\documentclass[11pt]{article}
\usepackage[margin=1in]{geometry}
\usepackage[utf8]{inputenc}      
\usepackage{amsmath}             
\usepackage{amssymb}             
\usepackage{mathtools}           
\usepackage{graphicx}            
\usepackage{subcaption}          
\usepackage{xcolor}              
\usepackage[colorlinks=true, linkcolor=blue, citecolor=blue, urlcolor=blue]{hyperref}
\usepackage[capitalise,nameinlink]{cleveref}
\usepackage{float}               
\usepackage{placeins}            
\usepackage{algorithm}
\usepackage{algorithmic}
\usepackage{tikz}                
\usetikzlibrary{tikzmark,decorations.pathreplacing,calc}
\usepackage{amsthm}              

\newcommand{\IEEEPARstart}[2]{#1#2}
\newenvironment{IEEEkeywords}{\par\vspace{0.5em}\noindent\textbf{Keywords---}\;}{\par\vspace{0.5em}}
\newcommand{\appendices}{\appendix\renewcommand{\thesection}{\Alph{section}}%
  \let\origsection\section
  \renewcommand{\section}[1]{\origsection{Appendix \thesection: ##1}}}

\graphicspath{{./}}

\newcommand{\R}{\mathbb{R}} 
\newcommand{\Radon}{\mathcal{R}} 
\newcommand{\Edge}{\mathcal{E}} 
\newcommand{\sample}{\mathbf{f}} 
\newcommand{\sinogram}{\mathbf{y}} 
\newcommand{\A}{\mathbf{A}} 
\newcommand{\BeamSet}{\Phi } 
\newcommand{\Indicator}{\mathcal{I}} 
\newcommand{\IIndicator}{\hat{\mathcal{I}}} 
\newcommand{\function}{\mathcal{F}} 
\newcommand{\Region}{\mathcal{D}}
\newcommand{\Boundary}{\mathcal{B}}
\newcommand{\measure}{\mu}
\newcommand{\beam}{\mathbf{u}}
\newcommand{\GP}{\mathcal{GP}}
\newcommand{\Kernel}{\mathbf{K}}
\newcommand{\CI}{\hat{\Sigma}}
\newcommand{\Obj}{\psi}
\newcommand{\ObjBatch}{\Psi}
\newcommand{\U}{\mathbf{U}}
\newcommand{\CoVar}{\Gamma}
\newcommand{\zscore}{\mathcal{Z}}
\newcommand{\sampleMean}{\mathbf{\bar{f}}}

\newcommand{\resolution}{N} 
\DeclareMathOperator*{\argmax}{arg\,max}
\DeclareMathOperator*{\argmin}{arg\,min}
\newtheorem{theorem}{Theorem}[section] 

\title{Edge-Aligned Beam Placement in Scanning Probe Tomography via Reconstruction-Free Sequential Design of Experiments}
\author{San~Dinh, Zichao~Wendy~Di, and~Matt~Menickelly%
\thanks{S. Dinh, Z. W. Di, and M. Menickelly are with the Mathematics and Computer Science Division, Argonne National Laboratory, Lemont, IL 60439 USA (e-mail: sdinh@anl.gov; wendydi@anl.gov; mmenickelly@anl.gov).}%
}
\date{\today}

\begin{document}

\maketitle

\begin{abstract}
In X-ray scanning probe tomography, reconstruction quality generally improves with larger numbers of projections. However, additional projections increase experiment costs, acquisition time, and the radiation dose imparted to the sample.
One mitigation to these trade-offs is to adopt a sequential design of experiments, in which each subsequent measurement is determined as a function of previously acquired data in order to maximize information gain. In scanning probe tomography, a widely used heuristic to maximize information is to align beams with the edges of the sample.
A key challenge, however, is that the true sample is unknown, so identifying edge-aligned beams typically requires reconstructing the sample based on available measurements.
This work proposes a novel sequential design method that identifies edge-aligned measurements directly from the sinogram, bypassing any reconstruction, thereby improving computational efficiency and reducing the experimental design's susceptibility to reconstruction errors.
Our method dynamically selects the next set of measurement beams by maximizing an acquisition function that balances exploration and exploitation over the domain of all possible measurements, improving reconstruction quality while reducing measurement redundancy.
\end{abstract}

\begin{IEEEkeywords}
Computed tomography, sequential design of experiments, Gaussian process, edge detection, Radon transform, adaptive sampling, Bayesian optimization.
\end{IEEEkeywords}

\section{Introduction}
\IEEEPARstart{S}{ynchrotron} X-ray tomography is a powerful non-destructive technique for quantitative, high-resolution visualization of a sample's internal structure. Leveraging the high brilliance, coherence, and tunable energy of synchrotron radiation, it achieves sub-micron spatial resolution and element-specific contrast far beyond conventional laboratory CT systems. However, full-angle acquisitions are typically time-consuming, requiring thousands of high-precision projections that limit experimental throughput and constrain the study of dynamic or radiation-sensitive specimens \cite{wang2025}. Moreover, repeated or prolonged X-ray exposure can cause radiation damage to delicate materials and biological samples, motivating dose-efficient, adaptive acquisition strategies such as sequential experimental design, which strategically selects informative measurements to maximize information gain while minimizing exposures \cite{brenner2007}.

Design of experiments (DoE) is the mathematical framework for selecting experimental conditions in a structured manner to estimate precisely factor effects on a response while using minimal resources. In scanning probe tomography, DoE corresponds to the strategic selection of measurement beams that maximize the information obtained about the sample for a given budget. 
Two principal approaches are widely used; optimization-based methods and Bayesian methods.
Optimization-based methods cast the design task as the maximization of an objective quantifying measurement informativeness, often a bi-level formulation \cite{ruthotto2018,haber2009} whose lower level represents the forward or reconstruction model and whose upper level determines the optimal measurement configuration.
These methods benefit from advances in numerical optimization but generally require differentiable models and sufficient training data.
Bayesian methods instead incorporate prior knowledge and uncertainty through probabilistic modeling \cite{burger2021}; a prior is placed on the unknown parameters and the posterior is updated as measurements are acquired, with design criteria commonly expressed through the posterior covariance, whose inverse relates to the Fisher information matrix \cite{kagan1999}. Minimizing posterior variance therefore corresponds to maximizing expected information, and posterior updates become tractable when the forward model is linear and the prior is Gaussian. Both classes of methods select measurements that maximize expected information under resource constraints, at substantial computational cost that grows rapidly with reconstruction resolution and measurement count.

Sequential DoE is a variation of DoE in which measurements are selected iteratively, with each new choice informed by previously acquired data \cite{huan2016,maaninen2023,go2025a}. It is typically implemented as a closed-loop procedure; an internal model evaluates the informativeness of candidate measurements given the current state of knowledge, the most informative candidate is selected, the corresponding data are collected, and the model is updated. This adaptive strategy 
can reduce the number of measurements required to achieve a desired accuracy. However, the need for efficient online updates imposes strict computational requirements at each iteration, amplifying the challenges already present in traditional DoE methods.

A widely used heuristic for experimental design in scanning probe tomography is \emph{edge alignment}, which asserts that measurements taken along object boundaries tend to be more informative than those taken elsewhere. For example, recent work employs an actor-critic network to learn a policy that selects measurement angles aligned with prominent edges in the reconstruction \cite{wang2024a,wang2025}; although edge alignment is not explicitly encoded in the reward function, the learned policy implicitly adopts this strategy and yields improved reconstruction quality. A limitation of such learned policies, however, is their dependence on the training data: trained on a library of simple basis shapes and assuming the sample to be a union of them, the policy generalizes well to samples composed of previously seen shapes but may degrade for the more complex geometries encountered in real samples. Edge information has also been incorporated into Bayesian formulations through approximate total variation priors \cite{barbano2022,helin2022}; as with other Bayesian DoE approaches, these formulations depend on a fixed linear forward model, typically a discrete Radon transform with a predetermined set of measurement beams, which makes evaluating the expected information gain for candidate measurements computationally expensive. Alternatively, the edge alignment principle has been embedded directly into acquisition functions for sequential design \cite{yang2023}, improving computational efficiency by avoiding the evaluation of complex statistical criteria. 
All such existing approaches require an initial reconstruction to estimate the location of the object's edges, adding implementation complexity, computational cost, and approximation error.

In this work, we introduce a sequential DoE framework for scanning probe tomography that identifies edge-aligned measurements (beams whose paths coincide with object boundaries) directly from the sinogram, thereby eliminating the need for an initial reconstruction. We establish mathematical properties of the Radon transform that link edges in the sample to characteristic features in the sinogram, enabling edge detection to be performed entirely in the measurement domain. A Gaussian process (GP) surrogate model is then constructed over the sinogram to interpolate edge information in unmeasured regions, and the GP's covariance structure is employed to discourage redundant measurements. 

The remainder of this paper is organized as follows. Section~\ref{sec: Background} provides necessary background on the Radon transform and the reconstruction methods used. Section~\ref{sec: Edge} analyzes how edge structure in the sample is translated to the sinogram and establishes the connection between edges in the image and sinogram domains. Section~\ref{sec: Methods} presents the proposed sequential DoE framework, the GP surrogate on the sinogram, and the edge-aligning acquisition functions for both beam- and angle-level selection. Section~\ref{sec: Computational Analysis} analyzes the computational complexity of our method and discusses implementation details. Section~\ref{sec: Numerical Sim} reports numerical experiments comparing the proposed methods to established baselines on multiple samples, including studies of the sensitivity of our method's performance to parameter selection and its robustness to noise. Finally, Section~\ref{sec: conclusion} concludes the manuscript and outlines directions for future work.

\section{Background} \label{sec: Background}
We employ the Radon transform as the forward model for X-ray scanning probe tomography.
Following the Lambert--Beer law for X-ray attenuation, the negative logarithm of the normalized beam intensity equals a line integral through the sample \cite{hansen2021}.
For a sample contained in the unit disk and characterized by the sample function $f(x):\R^2\to[0,\infty)$, the Radon transform at angle $\theta$ and distance $r$ is
\begin{equation} \label{eq:Radon_Transform}
    \Radon[f](\theta,r) = \int_{L_{\theta,r}} f(x)\,d\ell, \quad r\in[-1,1],
\end{equation}
where $L_{\theta,r}=\left\{x\in\R^2 \mid x_1\cos(\theta) + x_2\sin(\theta) = r\right\}$ denotes the X-ray beam path, parameterized by its angle with the horizontal axis and its distance from the origin. The restriction $r\in[-1,1]$ corresponds to the rotation plate that holds the sample; beams with $|r|>1$ do not intersect the object.

Discretizing the image on a $\resolution\times\resolution$ pixel grid and approximating the line integral by bilinear interpolation along the beam path yields a linear forward model.
Sampling at $n_\theta$ candidate angles and $n_r$ candidate offsets defines the measurement set $\BeamSet = \left\{(\theta_i, r_i)\right\}_{i=1}^{n_\theta n_r}$ and produces the discrete forward operator $\A\in\R^{n_\theta n_r \times \resolution^2}$, whose $i^{\text{th}}$ row encodes the contribution of every pixel to the line integral along $L_{\theta_i, r_i}$. Letting $\sample\in\R^{\resolution^2}$ denote the vectorized image and $\sinogram\in\R^{n_\theta n_r}$ the corresponding vector of measurements, the discrete Radon transform is
\begin{equation} \label{eq:discrete_Radon}
    \sinogram = \A\sample.
\end{equation}

The most commonly used inversion is the Filtered Back Projection (FBP) method \cite{bracewell1967, ramachandran1971}; the FBP is derived from the Fourier Slice Theorem \cite{bracewell1966} and hence assumes dense angular sampling. Because the sequential DoE proposed in this work intentionally produces sparse, non-uniform sinogram samples, we instead use the Simultaneous Algebraic Reconstruction Technique (SART) \cite{andersen1984, andersen1989}, which is better poised to account for the contribution of individual beams. 

\section{Edge Mapping Effect on the Radon Transform}
\label{sec: Edge}
Previous studies have shown that edge-aligned measurements provide richer information about the object of interest and lead to lower reconstruction error \cite{yang2023, wang2025, helin2022}. However, existing approaches typically require reconstructing the image first in order to determine the location of the object’s edges.
Our work demonstrates that edge-aligned measurements can instead be inferred directly from the sinogram. Thus, candidate measurements can be computed reliably and independently of the reconstruction method, without any loss in accuracy.

The object model needs to be modified to reflect the physical meaning of edges. Since the goal of tomographic reconstruction is to recover the internal structure of an object, the object can be assumed to consist of individual non-overlapping components. For example, they may correspond to distinct parts in a device or organs in a body. The non-overlapping requirement is a mathematical convenience ensuring the indicator functions below do not overlap; it does not restrict generality, since any overlapping configuration can be decomposed into non-overlapping components by treating each shared region as its own component.
To capture this notion mathematically, we introduce the smoothed positive indicator
\begin{equation} \label{eq: Positive Indicator}
    \Indicator[\function;\gamma](x) = 1-\exp\left(-\gamma \| \function(x) \|^2\right),
\end{equation}
where $\function:\R^2 \to \R$ is an arbitrary scalar-valued function 
and the parameter $\gamma > 0$ controls the sharpness of the transition between $0$ and $1$. In the tomography setting considered in this work, $\function$ corresponds to the sample function $f$. As $\gamma \to \infty$,  $\Indicator[\function;\gamma](x)$ approaches $1$ wherever $\function(x) \neq 0$ and $ \Indicator[\function;\gamma](x)$ approaches $0$ wherever $\function(x) = 0$, recovering the positive indicator function. Throughout we write $\Indicator[\function](x) := \lim_{\gamma\to\infty}\Indicator[\function;\gamma](x)$.

For each component of the object, the sample function is assumed to be the product of a positive indicator and a Lipschitz continuous function.
Lipschitz continuity captures the notion that the image intensity varies smoothly within each component and changes abruptly only when crossing into a different component.
In the sample space, let $f_i(x)$ denote a Lipschitz continuous function associated with the $i^\text{th}$ component, and let $\Region_i:\R^2\to\R_{\ge 0}$ denote a  function whose support coincides with the $i^\text{th}$ component (e.g., a non-negative distance-like function), so that $\Indicator[\Region_i](x)$ equals one inside the component and zero outside. The mathematical model of the object of interest composed of $n_f$ components is then given as
\begin{equation}
    f(x) = \sum_{i=1}^{n_f} \Indicator[\Region_i](x)f_i(x).
\end{equation}

We define the edge mapping of an image as a function that produces a new image with values approaching 1 at edges and approaching 0 in smoother regions.
In particular, we define the \emph{smoothed} edge-mapping function as
\begin{equation} \label{eq: edge map}
    \Edge[f;\alpha](x) = 1 - \exp \left(-\alpha\left\| \nabla f (x) \right\|^2\right),
\end{equation}
where the \emph{edge parameter} $\alpha > 0$ controls how sharply the edge transitions appear in the edge mapping.
Observe that in regions where $\Region_i(x) \neq 0$ for some $i$ (i.e., in the interior of a component), the norm of the gradient of the indicator satisfies $\left\| \nabla  \Indicator[\Region_i](x) \right\|^2 = 0$, and the norm of the gradient of each Lipschitz continuous function is bounded by a finite constant by definition, $\left\| \nabla f_i(x) \right\| \leq K_i$. Thus, in the interior of any component, $\left\|\nabla f(x)\right\|$ is bounded; thus, for any $\epsilon > 0$ there exists a sufficiently small $\alpha$ such that $\Edge[f;\alpha](x) < \epsilon$.
On the other hand, as $\gamma\to\infty$ in (\ref{eq: Positive Indicator}), the gradient of the indicator diverges on the component boundary, so the magnitude of $\nabla f$ becomes unbounded there. As a result, for any fixed $\alpha > 0$,
\begin{equation} \label{eq: edge limit}
    \lim_{\Region_i(x) \to 0} \Edge[f;\alpha](x) = 1.
\end{equation}
The limit in (\ref{eq: edge limit}) holds whenever $x$ approaches the boundary $\{x:\Region_i(x)=0\}$ of some component $i$. Intuitively, since the gradient inside the $i$th component is bounded by $K_i$ while the gradient across a component boundary is unbounded, the edge map (\ref{eq: edge map}) can be tuned via $\alpha$ to take the value $1$ along boundaries and $0$ elsewhere. Accordingly, we define the (\emph{idealized}) edge-mapping function as the limit
\begin{equation}
    \Edge[f](x) := \lim_{\gamma \to \infty} \Edge[f;\gamma^{-1}](x),
    \label{eq: edge map exact}
\end{equation}
where the edge parameter and indicator sharpness are coupled as $\alpha=\gamma^{-1}$: taking $\alpha\to0$ alone at fixed $\gamma$ would drive $\Edge[f;\alpha]$ to $0$ everywhere, whereas under the joint scaling $\alpha=\gamma^{-1}\to0$ the bounded interior contribution $\alpha\|\nabla f\|^2\to0$ is suppressed to $0$ while the divergent boundary contribution $\alpha\|\nabla f\|^2\sim\gamma\to\infty$ saturates at $1$; see Subsection \ref{subsec: convex edge radon} for the explicit limit construction.

To establish the connection between the edge-mapping function and the Radon transform, we now present and prove a key theorem. This theorem formalizes the relationship between the edge-mapping of an image and its corresponding sinogram, providing the mathematical foundation for the proposed edge-alignment strategy later introduced in Section \ref{sec: Methods}.

\begin{theorem}[Edge Radon Transform Correspondence] \label{theorem: edge radon}
Let $f(x)$ be a sample function composed of a finite number of non-overlapping components, each expressed as the product of a positive indicator and a Lipschitz continuous function. Let $\Edge[f](x)$ denote the edge-mapping function defined in (\ref{eq: edge map exact}), and let $\Radon[\Edge[f]](\theta,r)$ denote its Radon transform as defined in (\ref{eq:Radon_Transform}), where the path integral is interpreted as a Lebesgue integral. Then the following correspondence holds:
\begin{equation}
\label{eq:thm3point1}
    \lim_{\gamma \to \infty}\Indicator[\Radon[\Edge[f]];\gamma](\theta,r) = \Edge[\Radon[f]](\theta,r),
\end{equation}
with $\Indicator[\,\cdot\,;\gamma]$ as defined in (\ref{eq: Positive Indicator}).
\end{theorem}

In this section, we prove Theorem \ref{theorem: edge radon} by first considering the case of a single convex component (Subsection \ref{subsec: convex edge radon}; detailed proof deferred to Appendix~\ref{appendix: convex proof}) and then extending the result to the general case of multiple non-overlapping components (Subsection \ref{subsec: non-convex edge radon}).

\subsection{Single Convex Component} \label{subsec: convex edge radon}
For the case of a single convex object, both sides of the correspondence in Theorem~\ref{theorem: edge radon} evaluate to $1$ when the line $L_{\theta,r}$ contains a boundary segment of the object and to $0$ otherwise. The detailed derivation is given in Appendix~\ref{appendix: convex proof}; here we summarize the underlying intuition and illustrate the result on a representative example.

Intuitively, edges correspond to locations where the image jumps sharply in value, which in X-ray tomography occur at boundaries between materials. Accordingly, we assume the intensity varies smoothly within each component but jumps steeply across component boundaries---assumptions encoded, respectively, by the Lipschitz continuity and the positive indicator function.

The edge mapping function is constructed to take the value one at component boundaries and zero elsewhere. This contrast follows from the gradient structure of the sample function: the Lipschitz assumption keeps $\|\nabla f\|$ bounded in the interior of each component, while the sharp positive indicator makes $\|\nabla f\|$ divergent across component boundaries. Then, $\Edge[f;\alpha] = 1 - \exp(-\alpha\|\nabla f\|^2)$ from (\ref{eq: edge map}), evaluated in the joint limit $\alpha \to 0$ (see Appendix A), suppresses the bounded interior values to zero while saturating the divergent boundary values at one.

\begin{figure}[ht]
    \centering
    \begin{subfigure}[t]{0.19\columnwidth}
        \centering
        \includegraphics[width=\linewidth]{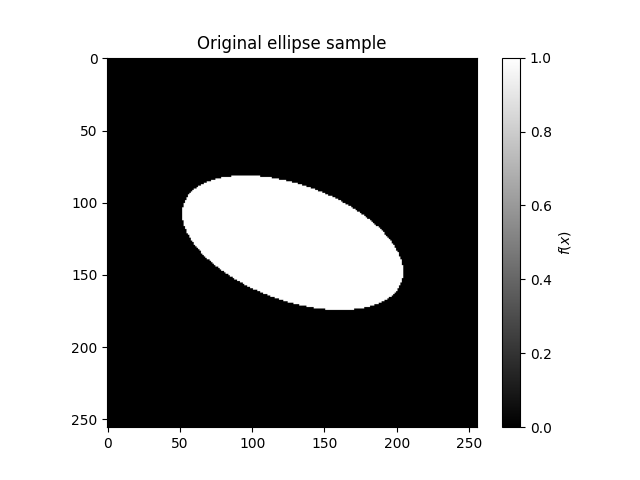}
        \caption{True sample}
        \label{subfig: edge radon convex sample}
    \end{subfigure}
    \hfill
    \begin{subfigure}[t]{0.19\columnwidth}
        \centering
        \includegraphics[width=\linewidth]{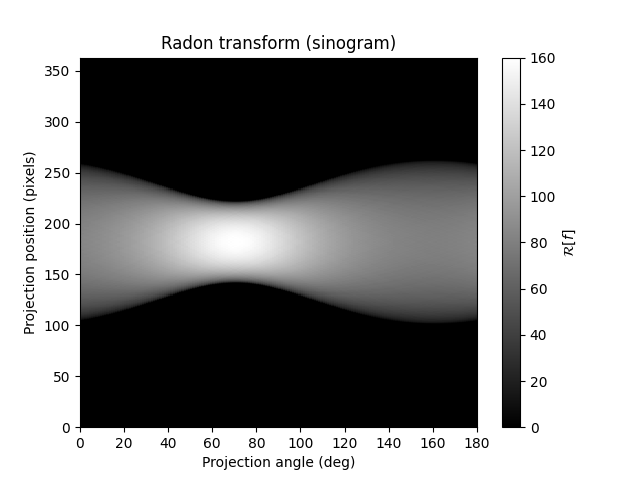}
        \caption{Sinogram of true sample}
        \label{subfig: edge radon sinogram of sample}
    \end{subfigure}
    \hfill
    \begin{subfigure}[t]{0.19\columnwidth}
        \centering
        \includegraphics[width=\linewidth]{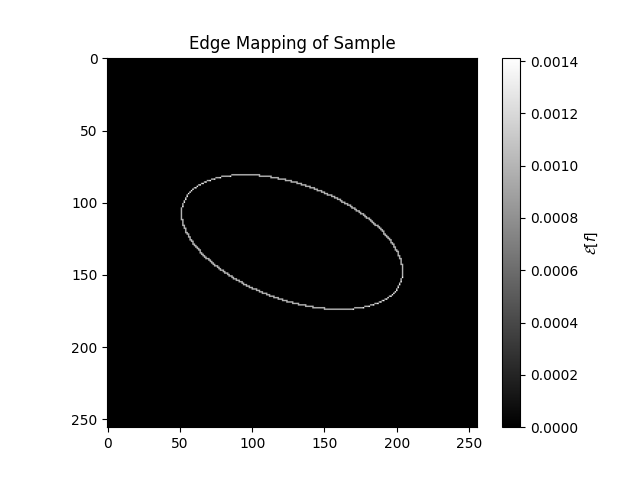}
        \caption{Edge-mapping function of sample}
        \label{subfig: edge radon edge of sample}
    \end{subfigure}
    \hfill
    \begin{subfigure}[t]{0.19\columnwidth}
        \centering
        \includegraphics[width=\linewidth]{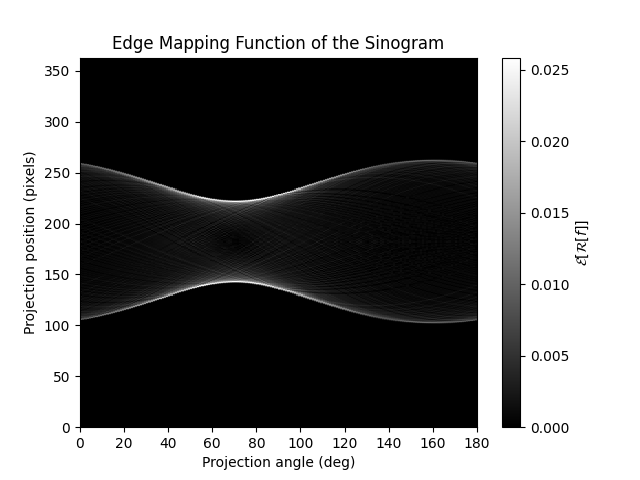}
        \caption{Edge mapping of sinogram}
        \label{subfig: edge radon edge of sinogram}
    \end{subfigure}
    \hfill
    \begin{subfigure}[t]{0.19\columnwidth}
        \centering
        \includegraphics[width=\linewidth]{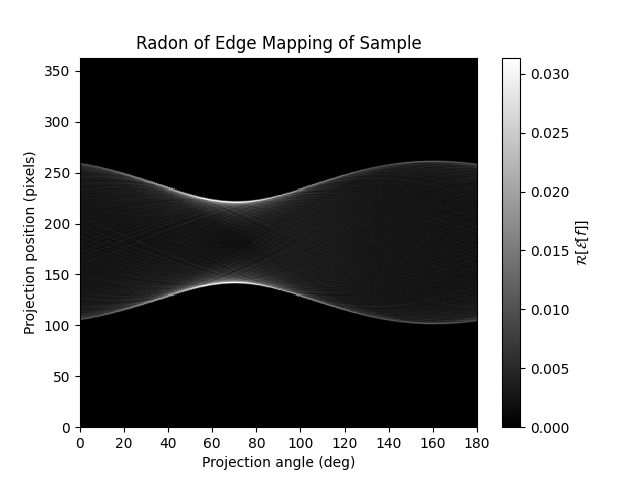}
        \caption{Radon transform of edge-mapping function}
        \label{subfig: edge radon radon of edge}
    \end{subfigure}
    \caption{Illustration of the equivalence in equation (\ref{eq: edge-radon equality}) using a convex test object.}
    \label{fig: edge radon convex}
\end{figure}

Specializing Theorem~\ref{theorem: edge radon} to a single convex object gives the identity derived in Appendix~\ref{appendix: convex proof}, equation~(\ref{eq: edge-radon equality}): the edge mapping of the Radon transform equals the positive indicator applied to the Radon transform of the edge mapping. Convexity is essential here. Because a convex object has a boundary made of straight faces, a line $L_{\theta,r}$ either contains an entire boundary face---so that $\Radon[\Edge[f]]$ accumulates that face's full length---or meets the boundary only at isolated points and contributes nothing. This binary alternative holds for \emph{every} line only when the object is convex, and the lines aligned with a face form a measure-zero set in $(\theta,r)$ that coincides exactly with the edges of the sinogram; a nonconvex object can violate it (a single line may contain several disjoint boundary segments), which is why Subsection~\ref{subsec: non-convex edge radon} decomposes the object into convex pieces. An illustrative example is given in Figure \ref{fig: edge radon convex}. In this example, the reference object is a simple ellipse, as shown in Figure \ref{fig: edge radon convex}(\subref{subfig: edge radon convex sample}), and its corresponding sinogram is shown in Figure \ref{fig: edge radon convex}(\subref{subfig: edge radon sinogram of sample}). The edge maps of Figure \ref{fig: edge radon convex}(\subref{subfig: edge radon convex sample}) and Figure \ref{fig: edge radon convex}(\subref{subfig: edge radon sinogram of sample}) are given in Figure \ref{fig: edge radon convex}(\subref{subfig: edge radon edge of sample}) and Figure \ref{fig: edge radon convex}(\subref{subfig: edge radon edge of sinogram}), respectively. The Radon transform of Figure \ref{fig: edge radon convex}(\subref{subfig: edge radon edge of sample}) is shown in Figure \ref{fig: edge radon convex}(\subref{subfig: edge radon radon of edge}). We observe that Figure \ref{fig: edge radon convex}(\subref{subfig: edge radon edge of sinogram}) and Figure \ref{fig: edge radon convex}(\subref{subfig: edge radon radon of edge}) exhibit a high degree of visual similarity. This example empirically demonstrates the relationship between the Radon transform and the edge mapping function.

\subsection{Extension to the Non-Convex Setting} \label{subsec: non-convex edge radon}
The convex-case correspondence extends to nonconvex sets: since any nonconvex shape decomposes into a union of non-overlapping convex polyhedra, we modify the positive indicator function as
\begin{equation} \label{eq: composite indicator}
    \IIndicator[Hx-l;\beta](x) =\sum_{j=1}^{n_P} \prod_{i=1}^{n_h} \frac{1}{1 + \exp(-\beta (H_{i,j} x - l_{i,j}))},
\end{equation}

where $n_P$ is the number of non-overlapping convex polyhedra in the decomposition, and $H_{i,j}$ and $l_{i,j}$ describe an $i$-th bounding half-plane (with boundary $\Boundary_{H,i,j}$) defining the $j$-th polyhedron. 
Because the polyhedra do not overlap, at most one term in the outer sum is nonzero at any point away from shared boundaries, so the composite indicator function $\IIndicator$ in (\ref{eq: composite indicator}) remains a proper indicator function, bounded by $1$. When two polyhedra overlap only at a shared boundary $H_0x-l_0=0$, the indicator function cancels on this boundary.
Since $\IIndicator$ remains constant (equal to $1$) along a shared boundary, $\nabla\IIndicator$ stays bounded there, so the divergence that produces $\Edge[f]=1$ in the convex case vanishes, and (\ref{eq: edge map}) yields zero. Thus, the edge-mapping function for a non-convex object is
\begin{equation}
\begin{aligned}
    \Edge[f](x) &= \lim_{\alpha \to 0} \Edge[f;\alpha](x) \\
    &= \begin{cases}
        1 & \text{if } x\in \Boundary_{H,i,j} \text{ for some } (i,j), \\
          & \quad \text{and } \Boundary_{H,i,j} \text{ is not shared} \\
          & \quad \text{between components}, \\
        0 & \text{otherwise.}
    \end{cases}
\end{aligned}
\end{equation}
By this construction,
Theorem \ref{theorem: edge radon} extends to the non-convex case.

\begin{figure}[H]
    \centering
    \begin{subfigure}[t]{0.19\columnwidth}
        \centering
        \includegraphics[width=\linewidth]{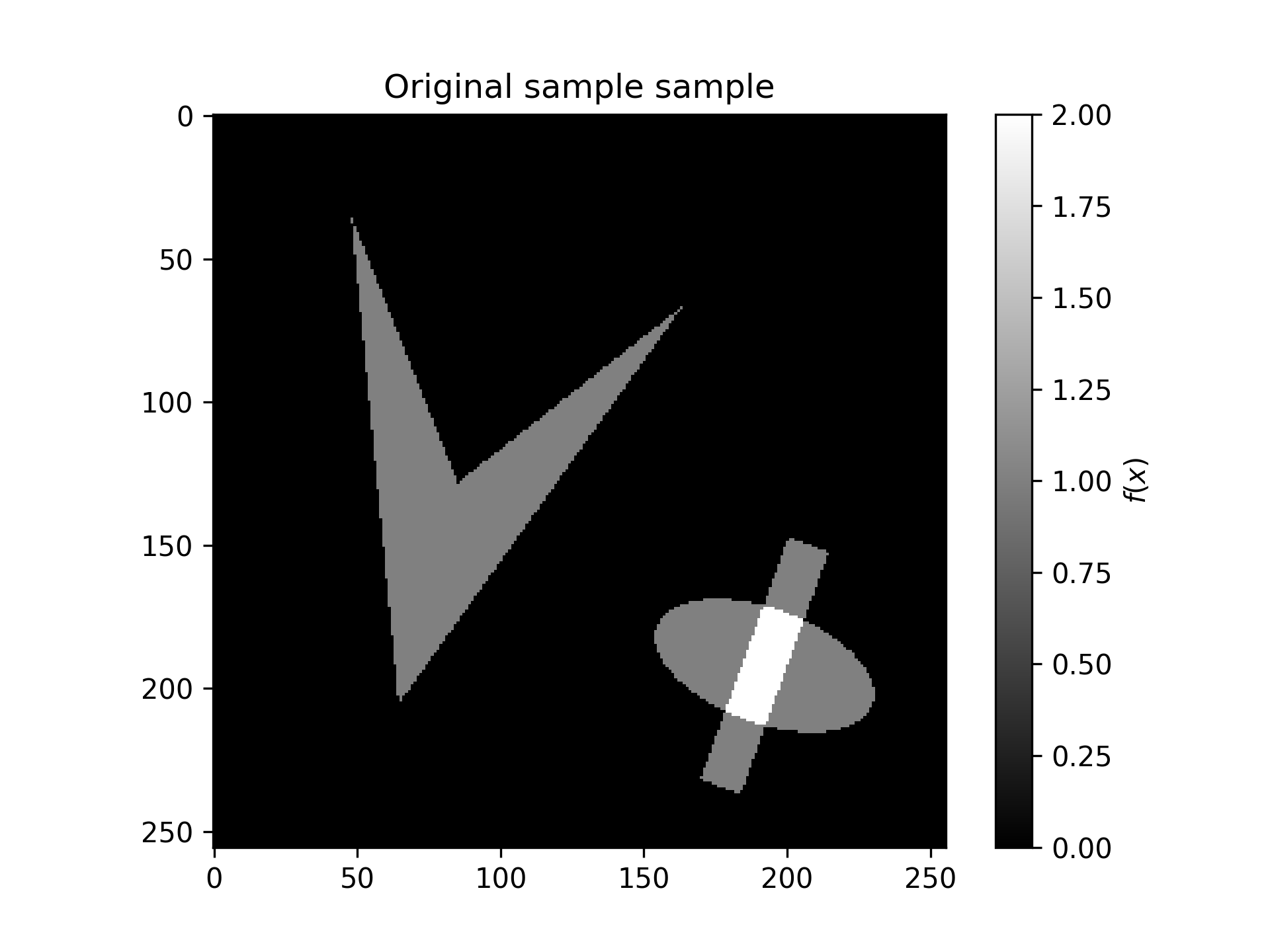}
        \caption{Generalized Object}
    \end{subfigure}
    \hfill
    \begin{subfigure}[t]{0.19\columnwidth}
        \centering
        \includegraphics[width=\linewidth]{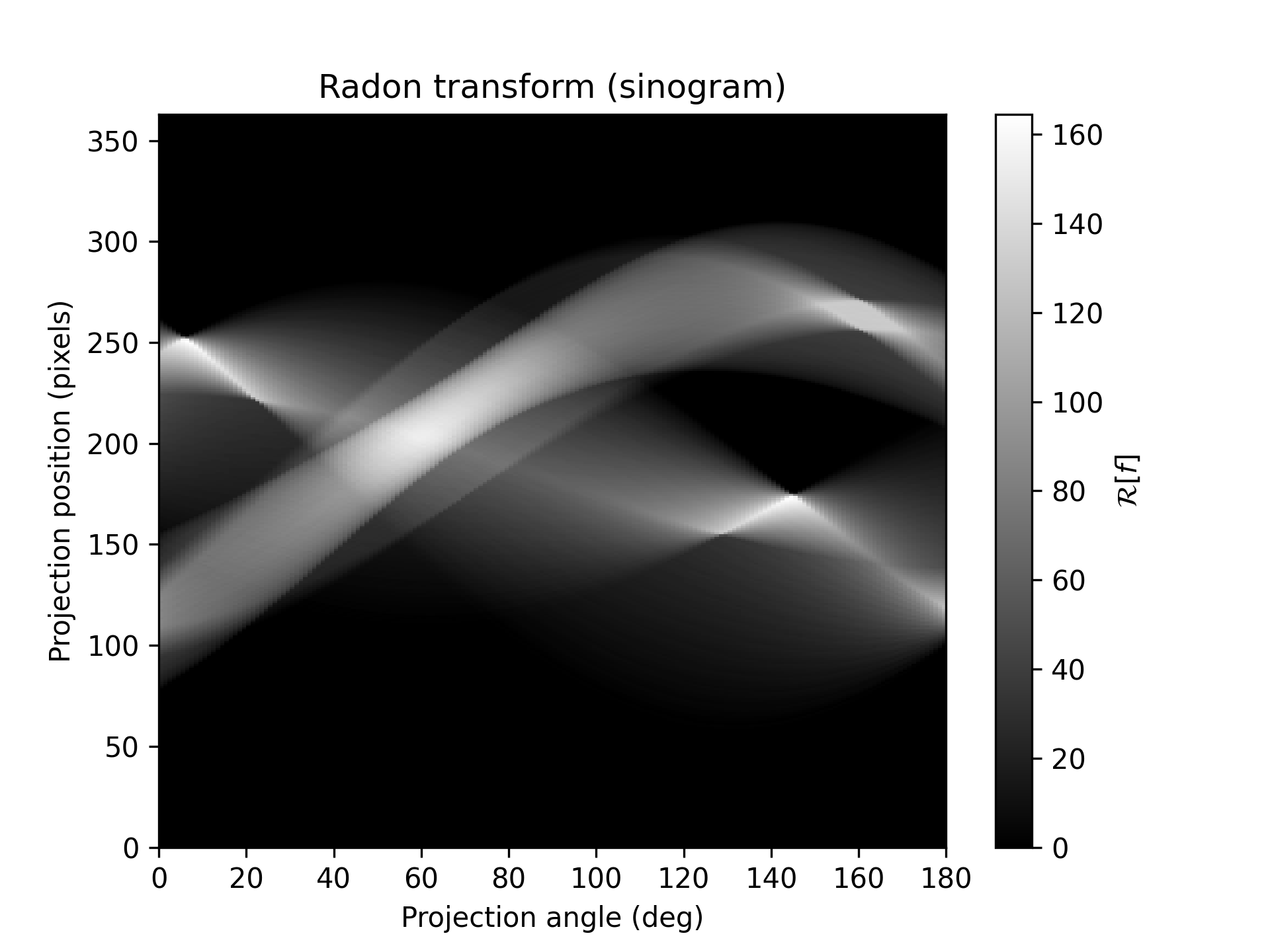}
        \caption{Sinogram of generalized object}
    \end{subfigure}
    \hfill
    \begin{subfigure}[t]{0.19\columnwidth}
        \centering
        \includegraphics[width=\linewidth]{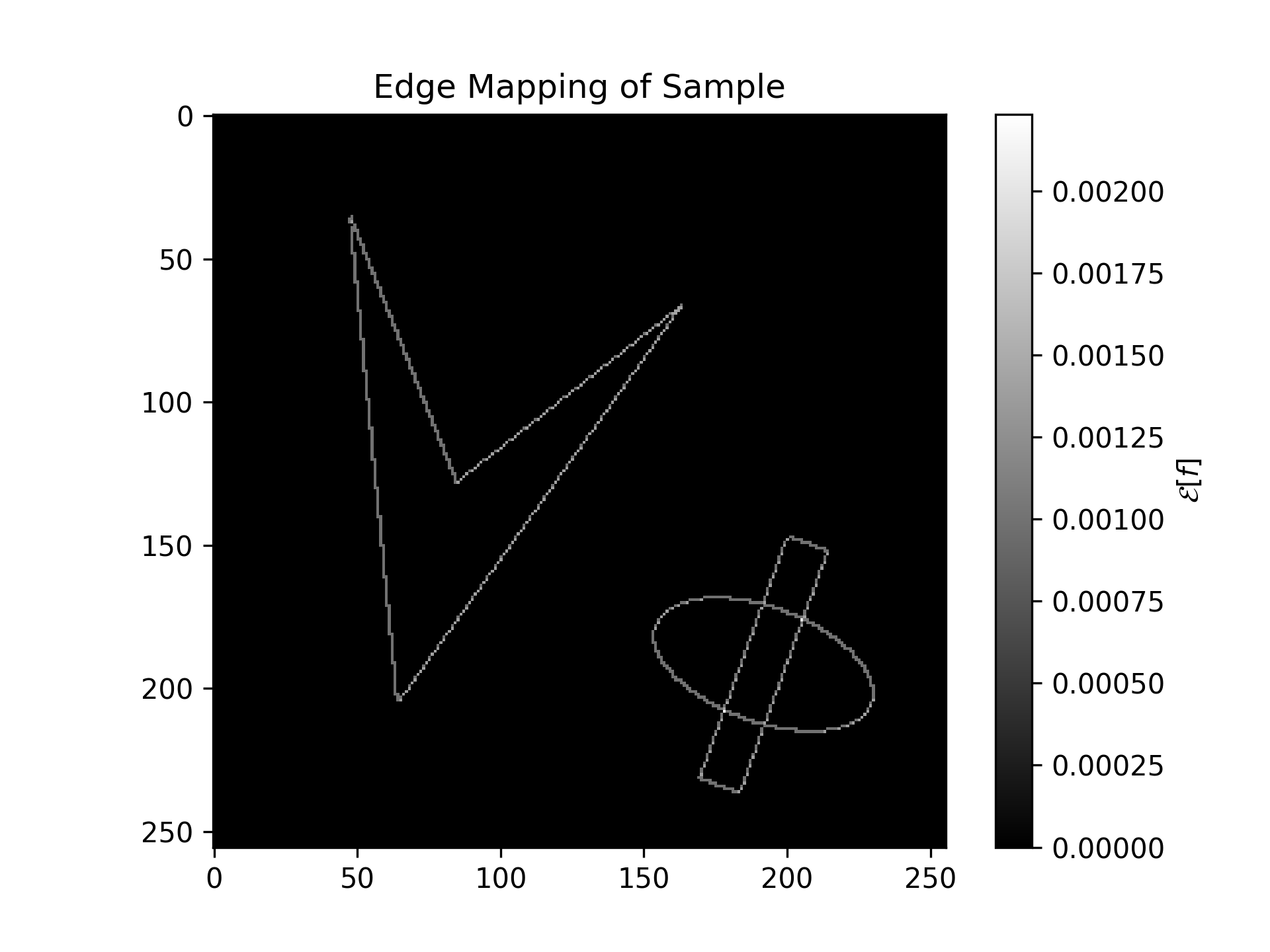}
        \caption{Edge mapping of object}
    \end{subfigure}
    \hfill
    \begin{subfigure}[t]{0.19\columnwidth}
        \centering
        \includegraphics[width=\linewidth]{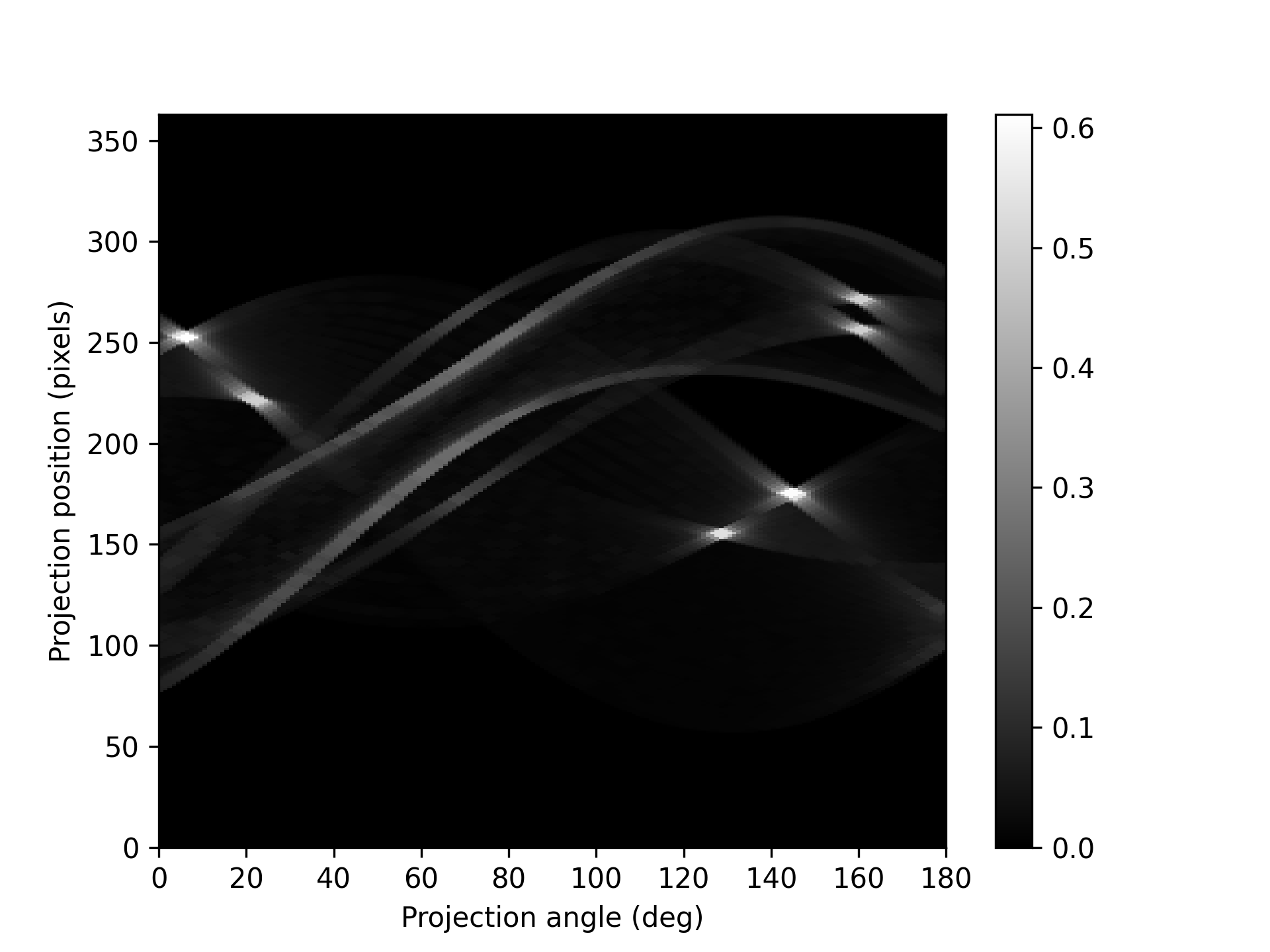}
        \caption{Edge mapping of sinogram}
    \end{subfigure}
    \hfill
    \begin{subfigure}[t]{0.19\columnwidth}
        \centering
        \includegraphics[width=\linewidth]{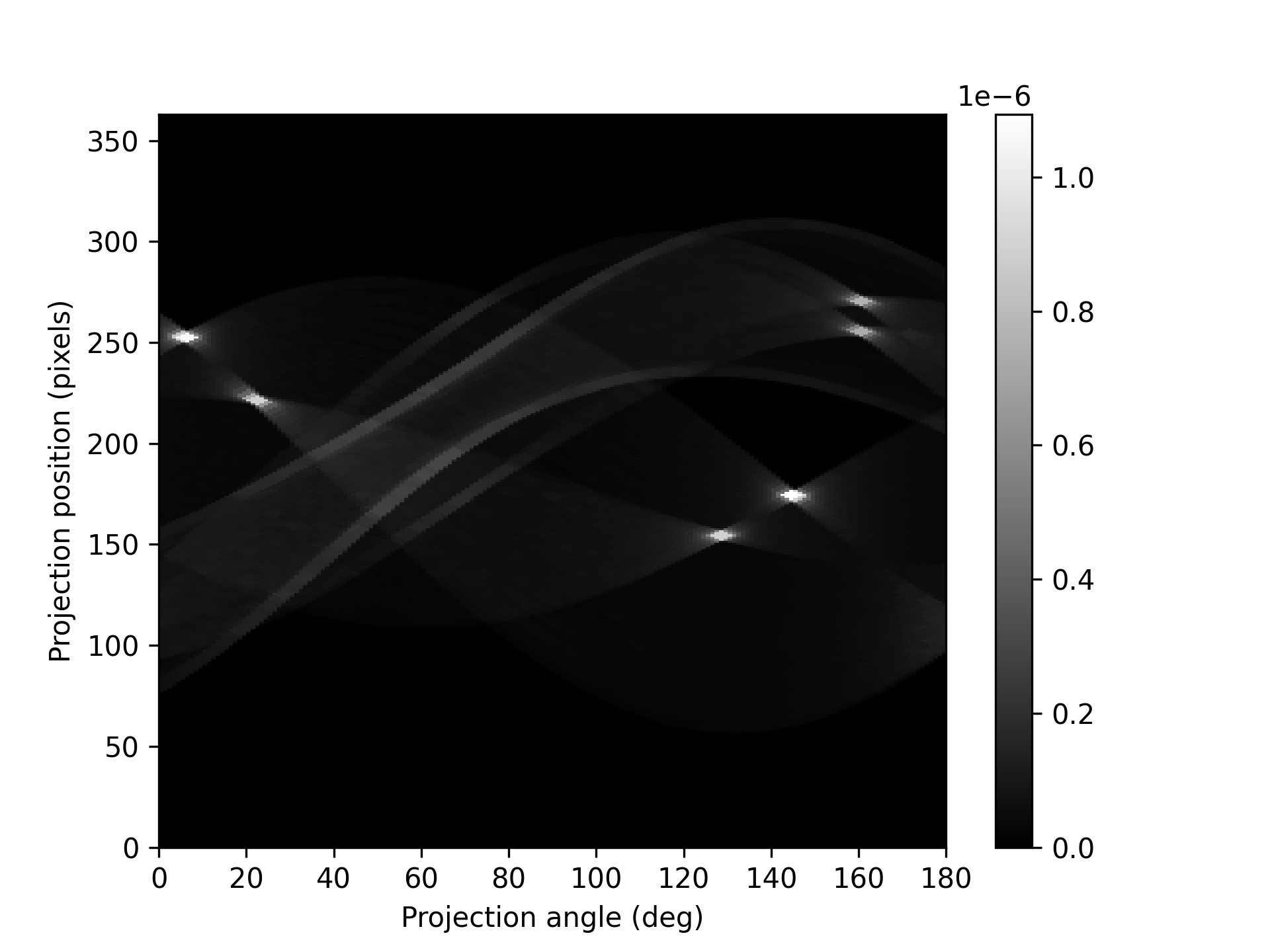}
        \caption{Radon transform of edge mapping}
    \end{subfigure}
    \caption{Edge mapping and Radon transform of generalized objects}
    \label{fig: edge radon generalized}
\end{figure}

By linearity of the Radon transform, the transform of a multi-component object (a union of non-overlapping convex components) equals the sum of the individual component transforms. The edge-mapping function $\Edge$, however, is nonlinear and does not distribute over this sum. Instead, since the edges of the component sinograms---each supported on a one-dimensional curve in the $(\theta,r)$ plane where that component's Radon transform is non-smooth---overlap only on a set of measure zero, the edge-mapping function~(\ref{eq: edge map exact})
equals $1$ wherever any single component sinogram has an edge---the union (pointwise maximum) of the component edge maps rather than their sum. Figure \ref{fig: edge radon generalized} illustrates this property for a generalized multi-component object.

\section{Methodology} \label{sec: Methods}
\subsection{Sequential Design of Experiments}

In this work, individual beams are selected sequentially, unlike previous X-ray imaging methods that optimized projection angles only for raster scans \cite{burger2021}. Our approach is a sequential DoE framework: measurements are acquired in batches, and each batch is informed by previous measurements. 
Let \(\BeamSet_k = \{ \beam_1, \beam_2, \ldots, \beam_{m_k} \}\) denote the set of all beams $(\theta, r)$ corresponding to previously acquired measurements at iteration $k$, where $m_k = |\BeamSet_k|$ is the number of acquired beams. 
Thus, \(\BeamSet_1 \subset \BeamSet_2 \subset \ldots \subset \BeamSet\) since the sequence only accumulates new measurements over time. 
Let the observed sinogram values for these beams be collected in the vector $Y_k(\BeamSet_k) = [y_1, y_2, \ldots, y_{m_k}]^\top$, where $y_i = \Radon[f](\beam_i)$ for each $\beam_i \in \BeamSet_k$.
The method in Algorithm \ref{algorithm: Adaptive Beam Selection} selects the next batch of measurement beams by maximizing an \emph{acquisition function} $\ObjBatch(\U, \BeamSet_k, Y_k(\BeamSet_k))$, which quantifies the believed informativeness of a candidate batch \(\U = \{\beam_1, \ldots, \beam_{n_b}\}\) given prior measurements.
At each iteration, $n_b$ new measurements are selected. In our experiments, we will take $n_b = n_r$, matching the number of measurements per angle in an evenly spaced raster scan.
Generally, an acquisition function \(\ObjBatch\) is designed to balance exploration -- identifying regions of high uncertainty -- with exploitation -- prioritizing beams anticipated to yield the most informative data.
A detailed formulation of a specific $\ObjBatch$ is presented in \Cref{sec: Acquisition}, and the optimal batch \(\U^*_k\) at iteration $k$ (line 7 of Algorithm \ref{algorithm: Adaptive Beam Selection}) is obtained using Algorithm \ref{algorithm: acquisition}.
The overall procedure is summarized in Algorithm \ref{algorithm: Adaptive Beam Selection}.

\begin{algorithm} [htb]
\caption{Adaptive Beam Selection for scanning probe tomography}
\label{algorithm: Adaptive Beam Selection}
\begin{algorithmic}[1]
\STATE \textbf{Input:} An acquisition function $\ObjBatch$ and the number of iterations $K_{\max}$.
\STATE \textbf{Initialization:} Define the initial set of measurements $\BeamSet_0$.
\FOR{$k = 0$ to $K_{\max}$}
    \STATE Perform or collect the measurements
    \STATE \hspace{1em} $Y_k(\BeamSet_k)= [y_1, y_2, \ldots, y_{m_k}]^\top, \text{ where } y_i = \Radon[f](\beam_i) \text{ for each } \beam_i \in \BeamSet_k.$
    \STATE Formulate an optimal design of experiments \COMMENT{\Cref{algorithm: baseline Bayesian angle,algorithm: acquisition} are used here for the design of experiment}
    \STATE \hspace{1em} $\U^*_k = \argmax\limits_{\substack{\U\subset\BeamSet\setminus\BeamSet_k \\ |\U|=n_b}} \ObjBatch(\U, \BeamSet_k, Y_k(\BeamSet_k)).$
    \STATE Update $\BeamSet_{k+1} = \BeamSet_k \cup \U^*_k.$
\ENDFOR
\STATE \textbf{End}
\end{algorithmic}
\end{algorithm}

\subsection{Edge-Aligning Beam and Angle Search}\label{sec: Acquisition}
The acquisition function we propose to employ in Algorithm \ref{algorithm: Adaptive Beam Selection} is guided by two well-established heuristics in scanning probe tomography: 1) projections intersecting regions of high image gradient (e.g., projections aligned with edges) typically yield the most information for reconstruction; and 2) measurements obtained along geometrically similar beams tend to be redundant, yielding limited  information gain.  
Unlike existing methods, our method applies the edge-alignment principle directly in the sinogram domain, eliminating the need for image reconstruction at each iteration of Algorithm \ref{algorithm: Adaptive Beam Selection}.

To construct an acquisition function $\ObjBatch$ that promotes edge-aligned beams, interpolation of the sparsely sampled sinogram is required.
To achieve this, a zero-mean GP prior is imposed on the sinogram, that is,  $\sinogram(\beam) \sim \GP(0, \Kernel(\beam, \beam'))$, where $\beam$ represents the beam parameters. 
In this formulation, the GP maps the beam parameters $\theta$ and $r$ to the corresponding sinogram values, with the covariance function $\Kernel$ chosen as the Matérn-$3/2$ kernel to capture smooth variations across the sinogram.
Assuming the observation noise of the Radon transform is independent and identically distributed with variance $\sigma^2$, the interpolated sinogram value at a measurement beam $\beam^*$ can be obtained using the posterior mean of the GP,
\begin{align}
    \hat{\sinogram}_k(\beam^*) &= \Kernel(\beam^*,\BeamSet_k)^\top\left[\Kernel(\BeamSet_k,\BeamSet_k) + \sigma^2 I\right]^{-1} Y_k(\BeamSet_k) \label{eq: GP mean}.
\end{align}

We additionally employ the posterior covariance of the GP
to determine the width of a confidence interval for the value of $\hat{\sinogram}_k(\beam)$ at each candidate beam $\beam\in\BeamSet$.
This confidence interval width, denoted $\CI_k(\beam)$, appears in the acquisition function $\ObjBatch$, and is intended to promote exploration in the measurement space, since smaller confidence intervals correspond to regions of the sinogram where more measurements are already available.
The width of a $(1-\delta)$ confidence interval around $\hat{\sinogram}_k(\beam)$ for any $\beam\in\BeamSet$ is computed
from the GP posterior covariance by evaluating the variance
\begin{equation}
    \begin{aligned}
        \hat{\sigma}_k(\beam)& = \Kernel(\beam,\beam) \\
        &- \Kernel(\beam,\BeamSet_k)^\top\left[\Kernel(\BeamSet_k,\BeamSet_k) + \sigma^2 I\right]^{-1}\Kernel(\beam,\BeamSet_k),
    \end{aligned}
\end{equation}
and then evaluating 
\begin{equation}\label{eq:ci_width}
    \CI_k(\beam) = 2\zscore(1-\delta/2)\sqrt{\hat{\sigma}_k(\beam)},
\end{equation}
where $\zscore(\cdot)$ denotes the inverse cdf of the standard normal distribution (i.e., the $z$-score).
In our tests, we let $\delta = 0.05$ (a 95\% confidence interval).

We now turn our attention to the edge-alignment term in our acquisition function, which we refer to as an \emph{edge-mapping function}. 
Let $\Delta \theta$ denote the difference between two consecutive angles in the (assumed uniformly discretized) set $\BeamSet$.
The perturbation of the input between two consecutive angles is represented by the vector $\Delta \beam_\theta = [\Delta \theta, 0]^\top$. 
Similarly, let $\Delta r$ denote the difference between two consecutive scanning positions along a fixed angle; the corresponding perturbation vector is denoted $\Delta \beam_r = [0, \Delta r]^\top$. 
The edge-mapping function, $\hat{\Edge}$, of the sinogram is defined by the magnitude of the gradient of the sinogram with respect to these two perturbations for each input $\beam$.
This function is an approximation of the edge-mapping function in (\ref{eq: edge map exact}).
This provides a finite-difference approximation of the gradient in the edge-mapping function,
\begin{align}
    \Delta \hat{\sinogram}_{k,\theta}(\beam) &=\frac{\hat{\sinogram}_k(\beam + \Delta \beam_\theta) - \hat{\sinogram}_k(\beam)}{\Delta \theta},  \\
    \Delta \hat{\sinogram}_{k,r}(\beam) &=\frac{\hat{\sinogram}_k(\beam + \Delta \beam_r) - \hat{\sinogram}_k(\beam)}{\Delta r}, \\
    \hat{\Edge}[\hat{\sinogram}_k;\alpha](\beam) &= 1 - \exp\Big(-\alpha\,\bigl(
        \Delta \hat{\sinogram}_{k,\theta}(\beam)^2 \nonumber \\
        &\qquad\quad + \Delta \hat{\sinogram}_{k,r}(\beam)^2
        \bigr)\Big),\label{eq:edge_mapping}
\end{align}
matching the squared-norm form of the smoothed edge mapping (\ref{eq: edge map}).

The acquisition function is then defined as the product of the confidence interval width $\CI_k(\beam)$, defined in \eqref{eq:ci_width}, and the edge-mapping term $\hat{\Edge}[\hat{\sinogram}_k;\alpha](\beam)$, defined in \eqref{eq:edge_mapping}, where $\hat{\sinogram}_k$ denotes the GP posterior mean of the sinogram given $Y_k(\BeamSet_k)$. The per-beam acquisition is
\begin{equation} \label{eq: beam acquisition}
        \Obj(\beam) := \CI_k(\beam)\,\hat{\Edge}[\hat{\sinogram}_k;\alpha](\beam),
        \qquad \beam \in \BeamSet.
\end{equation}

The acquisition in (\ref{eq: beam acquisition}) can be deployed at two granularities, depending on the scanning hardware: selecting individual beams, or selecting whole projection angles. Both variants share the GP surrogate, confidence-interval width, and edge-mapping function developed above, and differ only in their acquisition objective and selection rule.

\emph{Beam-level selection (independent-beam scanners).} \Cref{algorithm: Adaptive Beam Selection} requests, simultaneously, a batch of $n_b$ measurements. For a batch, we define the acquisition value as the sum of the individual beam acquisition values,
\begin{equation} \label{eq: batch acquisition}
    \max \ObjBatch(\U, \BeamSet_k, Y_k(\BeamSet_k)) := \max \sum_{i=1}^{n_b} \Obj(\beam_i), \quad \U = \{\beam_1,\dots,\beam_{n_b}\}.
\end{equation}
Given the simplicity of computing the acquisition function and the finiteness of the candidate set $\BeamSet$, an approximate solution to (\ref{eq: batch acquisition}) can be obtained by greedily selecting beams in descending order of $\Obj(\beam)$; this is a heuristic for (\ref{eq: batch acquisition}) rather than its exact solution.
A naive greedy selection, however, tends to cluster beams in regions of the sinogram with high values of $\Obj$, producing redundant measurements.
To prevent this, a minimum-distance threshold $\Delta_{\Obj}$ is imposed between sequentially selected beams.

\emph{Angle-level selection (conventional full-field tomography).} In conventional full-field tomography, each rotation acquires an entire projection (all $n_r$ beams at one angle) simultaneously, rather than individual beams. For this acquisition mode, an angle-search variant selects a single angle per iteration and queries the full projection at that angle; it is the natural counterpart to the Bayesian angle-selection baseline of Section \ref{subsec: Bayesian Angle sOED} and is used in the angle-distribution comparison of Section \ref{sec: Numerical Sim}. Whereas the beam-level acquisition (\ref{eq: beam acquisition}) scores individual points in the two-dimensional sinogram domain, the angle-level acquisition $\Obj_\theta(\theta)$ aggregates that score over all offsets at a fixed $\theta$:
\begin{equation} \label{eq: angle acquisition}
    \Obj_\theta(\theta) := \sum_{i=1}^{n_r}  \CI_k([\theta,r_i])\, \hat{\Edge}[\hat{\sinogram}_k;\alpha]([\theta,r_i]),
\end{equation}
and the variant selects the single angle that maximizes (\ref{eq: angle acquisition}), then queries all $n_r$ beams at that angle.

Both variants implement the batch-selection step at line 7 of Algorithm \ref{algorithm: Adaptive Beam Selection}, returning the batch $\U^*_k$ that the outer loop appends to $\BeamSet_k$, as detailed in Algorithm \ref{algorithm: acquisition}.

\begin{algorithm}[htb]
\caption{Edge-Aligning Selection: Beam and Angle Variants}
\label{algorithm: acquisition}
\begin{algorithmic}[1]
\STATE \textbf{At iteration $k$:}
\STATE From $\BeamSet_k$, $Y_k(\BeamSet_k)$, compute the GP posterior mean $\hat{\sinogram}_k$ \eqref{eq: GP mean}, the confidence-interval width $\CI_k$ \eqref{eq:ci_width}, and the edge mapping $\hat{\Edge}[\hat{\sinogram}_k;\alpha]$ \eqref{eq:edge_mapping}.
\IF{independent-beam (beam-level) variant}
    \STATE Construct the per-beam acquisition $\Obj(\beam)$ from \eqref{eq: beam acquisition}.
    \STATE Generate a sorted list $\bar{\U} = (\beam_{i_1},\beam_{i_2},\ldots,\beam_{i_{n_\theta n_r}})$ such that $\Obj(\beam_{i_1}) \ge \Obj(\beam_{i_2}) \ge \cdots \ge \Obj(\beam_{i_{n_\theta n_r}})$.
    \FOR{$j = 1$ to $n_b$}
        \STATE Set $\beam^*_j \gets$ the first element of $\bar{\U}$ and remove it from $\bar{\U}$.
        \STATE Remove every $\beam \in \bar{\U}$ with $\|\beam^*_j - \beam\| \leq \Delta_{\Obj}$.
    \ENDFOR
    \STATE \textbf{Return:}  $\U^*_k = \{\beam^*_1,\beam^*_2,\ldots,\beam^*_{n_b}\}$.
\ELSE
    \STATE Construct the per-angle acquisition $\Obj_\theta(\theta_j)$ from \eqref{eq: angle acquisition} for $j=1,\ldots,n_\theta$.
    \STATE Select the optimal angle $\theta^* = \argmax_{\theta_j} \Obj_\theta(\theta_j)$.
    \STATE \textbf{Return:}  $\U^*_k = \{(\theta^*, r_1), (\theta^*, r_2), \ldots, (\theta^*, r_{n_r})\}$.
\ENDIF
\end{algorithmic}
\end{algorithm}

\subsection{Baseline: Bayesian Sequential Design of Experiment} \label{subsec: Bayesian Angle sOED}
In the literature, sequential DoE is often guided either by Bayesian uncertainty quantification, which penalizes measurements with low information content, or by edge-aligning heuristics, which emphasize measurements that capture object boundaries. In contrast to the edge-aligning method presented above, this subsection adopts a baseline sequential strategy adapted from \cite{burger2021}, which selects measurement angles based on the D-optimality criterion.

A Gaussian prior is imposed on the discrete sample $\sample$, with the SART reconstruction from the initial measurements taken as the prior mean $\sampleMean_0$ and an initial covariance $\CoVar_0$ (e.g., a diagonal matrix). Let $\A_k\in\R^{m_k\times \resolution^2}$ denote the discrete Radon operator restricted to the rows associated with $\BeamSet_k$, where $m_k = |\BeamSet_k|$. Because the Radon transform is linear in the sample for this fixed measurement matrix, each acquired measurement updates the posterior through the standard Kalman/GP equations, yielding the posterior mean $\sampleMean_{k \mid k}$ and covariance $\CoVar_{k \mid k}$ at iteration $k$ under independent Gaussian measurement noise of covariance $\CoVar_e$; the full probabilistic derivation is provided in Appendix~\ref{appendix: bayesian baseline}. For any candidate measurement set $\BeamSet_i^c \subset \BeamSet$, let $\A(\BeamSet_i^c)$ denote the discrete Radon operator restricted to its rows and let $\CoVar^c_e$ denote the measurement-noise covariance; in the typical case of scanning all $n_r$ beam positions along a single angle $\theta_i$, the candidate set is $\BeamSet_i^c = \{[\theta_i,r_1],\ldots,[\theta_i,r_{n_r}]\}$.

As an optional step, we may fold the current edge information into the prior covariance via a precision-weighted update, following the covariance intersection framework~\cite{julier1997}. Let $\CoVar^E_k$ be a diagonal precision-like matrix whose entries are large where the current reconstruction shows strong edge support; here $\Edge$ is applied pixel-wise to the discrete reconstruction via finite differences in place of the continuous gradient in (\ref{eq: edge map}).
That is, we set
\begin{align}
    \CoVar^E_k &= \mathrm{diag}(\Edge(\sampleMean_{k \mid k})) \text{ and } \\
    \CoVar^c_{k|k} &= \left((1-w)\CoVar^{-1}_{k|k} + w\CoVar^E_k\right)^{-1}, \label{eq: edge-prior covar}
\end{align}
where $w \in [0,1]$ trades off the data-driven posterior precision against the edge-promoting precision.

Treating the current posterior as the prior for the next acquisition, the posterior predictive joint distribution over a candidate set $\BeamSet_i^c$ is derived in Appendix~\ref{appendix: bayesian baseline}. Although the realized measurement value is unknown until acquired, the resulting predictive posterior covariance for the candidate $\BeamSet_i^c$ can be computed in closed form:
\begin{equation} \label{eq: posterior covar candidate}
\begin{split}
    \CoVar^c_{k+1\mid k}(\BeamSet_i^c) = \CoVar^c_{k|k} - {}&\CoVar^c_{k|k} \A(\BeamSet_i^c)^\top \\
    &\times \left[\A(\BeamSet_i^c) \CoVar^c_{k|k} \A(\BeamSet_i^c)^\top + \CoVar^c_e\right]^{-1} \\
    &\times \A(\BeamSet_i^c) \CoVar^c_{k|k}.
\end{split}
\end{equation}
The D-optimality criterion measures the volume of the posterior covariance ellipsoid; a D-optimal design minimizes the determinant of the posterior covariance\footnote{Equivalently, a $D$-optimal design maximizes the determinant of the posterior precision/Fisher information.}. 
Using a Cholesky decomposition $\CoVar^c_{k+1\mid k}(\BeamSet_i^c) = L_{k+1|k}(\BeamSet_i^c) L_{k+1|k}(\BeamSet_i^c)^\top$, the log-determinant is computed as
\begin{equation}
\begin{split}
    D(\BeamSet_i^c) &:= \log \det \CoVar^c_{k+1\mid k}(\BeamSet_i^c) \\
    &= 2\sum_{j=1}^{\resolution^2} \log [L_{k+1|k}(\BeamSet_i^c)]_{jj},
\end{split}
\end{equation}
where $[L_{k+1|k}(\BeamSet_i^c)]_{jj}$ are the diagonal entries of the Cholesky factor.
The optimal candidate set is then selected by minimizing the criterion
\begin{equation}
    \argmin_{\BeamSet_i^c} D(\BeamSet_i^c) =:\BeamSet^*_{k+1}.
\end{equation}

\begin{algorithm} [H]
\caption{Baseline: Bayesian Angle Selection}
\label{algorithm: baseline Bayesian angle}
\begin{algorithmic}[1]
\STATE \textbf{At iteration $k$:}
\STATE Update $\sampleMean_{k \mid k}, \CoVar_{k|k}$ from $\sampleMean_{k|k-1}, \CoVar_{k|k-1}$ via the Kalman update with the new measurement $\sinogram_k$.
\STATE Compute the current edge-prior covariance $\CoVar^c_{k|k}$ from (\ref{eq: edge-prior covar}).
\FOR{$i = 1$ to $n_\theta$}
    \STATE Compute the discrete Radon transform at angle $\theta_i$, $\A(\BeamSet_i^c)$.
    \STATE Compute the predictive posterior covariance $\CoVar^c_{k+1\mid k}(\BeamSet_i^c)$ from (\ref{eq: posterior covar candidate}).
    \STATE Compute the log-determinant $d_i = \log\det\CoVar^c_{k+1\mid k}(\BeamSet_i^c)$.
\ENDFOR
\STATE Select the optimal angle index $i^* = \argmin_i d_i$.
\STATE Update the forward operator
\STATE \hspace{1em} $\A_{k+1} = \left[\begin{array}{c}
    \A_k \\
    \A(\BeamSet_{i^*}^c)
\end{array}\right]$.
\STATE Use the predicted posterior covariance under the chosen candidate as the prior for iteration $k+1$: $\sampleMean_{k+1|k}=\sampleMean_{k \mid k}$, $\CoVar_{k+1|k}=\CoVar^c_{k+1|k}(\BeamSet_{i^*}^c)$.
\end{algorithmic}
\end{algorithm}

\section{Computational Analysis} \label{sec: Computational Analysis}
In this subsection, the online computational times of Algorithms \ref{algorithm: acquisition} and \ref{algorithm: baseline Bayesian angle} are compared. The complexity of each algorithm is scaled with the image side length $\resolution$, the number of candidate angles $n_\theta$, the number of beams per candidate angle $n_r$, and the per-iteration batch size $n_b$. The cost being evaluated is the floating point operation count.

Let us begin by evaluating the computational cost of the Bayesian angle selection baseline. At each candidate angle, this baseline forms the predictive posterior covariance $\CoVar^c_{k+1|k}(\BeamSet_i^c)$ in (\ref{eq: posterior covar candidate}) and computes its log-determinant via a Cholesky decomposition.

For an $\resolution \times \resolution$ image, this $\resolution^2 \times \resolution^2$ Cholesky decomposition requires $\tfrac{1}{3} \resolution^6$ FLOPs per candidate angle, so this step requires $C_{Chol} = \tfrac{1}{3}\, n_\theta\, \resolution^6$ FLOPs in total.

For our proposed approach, the computational cost is dominated by GP interpolation. Let $n_0$ denote the initial number of acquired measurements. At iteration $k$, there are $(n_0 + k\, n_b)$ data points available.

The GP interpolation requires computing the inverse of the covariance matrix $[\Kernel(\BeamSet_k,\BeamSet_k) + \sigma^2 I]$ in equation (\ref{eq: GP mean}). For $m_k = n_0 + k\, n_b$ measurements, this matrix inversion requires
\begin{equation}
    C_{GP} = \frac{1}{3}m_k^3 = \frac{1}{3}(n_0 + k\, n_b)^3
\end{equation}
FLOPs. However, we can use a block (Woodbury / Schur-complement) update when adding $s=n_b$ new measurements.  Let $m_k = n_0 + k\,n_b$ denote the number of measurements available in iteration $k$.  If we maintain the inverse of
$K(\BeamSet_k,\BeamSet_k)+\sigma^2 I$ from the previous iteration, forming the inverse for the enlarged matrix with $m_{k+1}=m_k+s$ points via the block-inversion (Schur complement) requires the following dominant operations:
\begin{itemize}
    \item form $A^{-1}B$ where $A^{-1}\in\R^{m_k\times m_k}$ and $B\in\R^{m_k\times s}$: cost $O(m_k^2 s)$,
    \item form the $s\times s$ Schur complement $S = C - B^\top A^{-1}B$: cost $O(m_k s^2)$,
    \item invert the $s\times s$ matrix $S$: cost $O(s^3)$,
    \item assemble the full block inverse (matrix multiplications of sizes involving $m_k$ and $s$): cost $O(m_k^2 s + m_k s^2)$.
\end{itemize}
Thus the incremental cost to update the inverse when adding $s=n_b$ beams is
\begin{equation}
    C_{\text{update}} = O\!\left(m_k^2\,n_b + n_b^3\right).
\end{equation}

The GP mean prediction for all candidate beams requires matrix-vector multiplication with computational cost
\begin{equation}
    C_{pred} = n_\theta\, n_r\, m_k = n_\theta\, n_r\, (n_0 + k\, n_b).
\end{equation}

The edge mapping computation using finite differences requires evaluating the GP at neighboring points, adding
\begin{equation}
    C_{edge} = 2\, n_\theta\, n_r.
\end{equation}

Therefore, the total computational cost per iteration is:
\begin{equation}
    C_{total} = \left(m_k^2\,n_b + n_b^3\right) + n_\theta\, n_r\, (n_0 + k\, n_b) + 2\, n_\theta\, n_r.
\end{equation}

Comparing the scaling behavior, the Bayesian baseline scales as $O(n_\theta\, \resolution^6)$ with respect to image resolution, while the proposed GP approach scales as $O(m_k^2\, n_b)$ with respect to the number of acquired measurements (a naive from-scratch inversion would instead cost $O(m_k^3)$). In the sparse-measurement regime targeted by this work, $m_k$ remains far below the pixel count $\resolution^2$, so $m_k^2\, n_b \ll n_\theta\, \resolution^6$ and the proposed approach is more efficient.

\section{Numerical Experiments} \label{sec: Numerical Sim}
\subsection{Performance Across Samples}
The first of our numerical tests evaluates the performance of the proposed algorithm on five test samples, each exhibiting distinct structural characteristics. Measurement data are simulated using a discrete Radon transform applied to the ground truth image associated with each sample.
Reconstruction accuracy is quantified using the root mean square error (RMSE) between the Simultaneous Algebraic Reconstruction Technique (SART) reconstruction \cite{andersen1984} and the ground truth. For consistency across methods, SART is also used for the Bayesian baseline in place of the posterior mean.

All methods are evaluated at a resolution of $128\times128$ pixels ($\resolution = 128$); in all experiments we further set $n_r = n_b = \resolution = 128$, so the number of beams per angle, the per-iteration batch size, and the image side length coincide numerically. The samples are tested and compared against two baseline methods: an evenly spaced raster scan, consistent with conventional scanning probe tomography measurements, and a sequential Bayesian design of experiment, as described in subsection \ref{subsec: Bayesian Angle sOED}.
The evenly spaced raster scan baseline distributes $128$ beams uniformly along the detector range at each angle, with the angles evenly spaced over \([0, \pi)\); the per-angle beam count thus matches the image resolution and the number of candidate scanning positions per angle in the sequential beam selection algorithm.

For a fair comparison, the number of beams per iteration is kept identical across both baselines and all sequential approaches. The sequential methods of Section \ref{sec: Methods} are initialized with a five-angle evenly spaced raster scan, so the total number of beams at iteration \(k\) is \(n_r \times (k + 5)\): the five initial angles plus the \(k\) angles selected thereafter.

\subsubsection{Sample 1: Tilted Ellipse}
\begin{figure}[H] 
    \centering
    \begin{subfigure}[t]{0.24\linewidth}
        \centering
        \includegraphics[width=\linewidth]{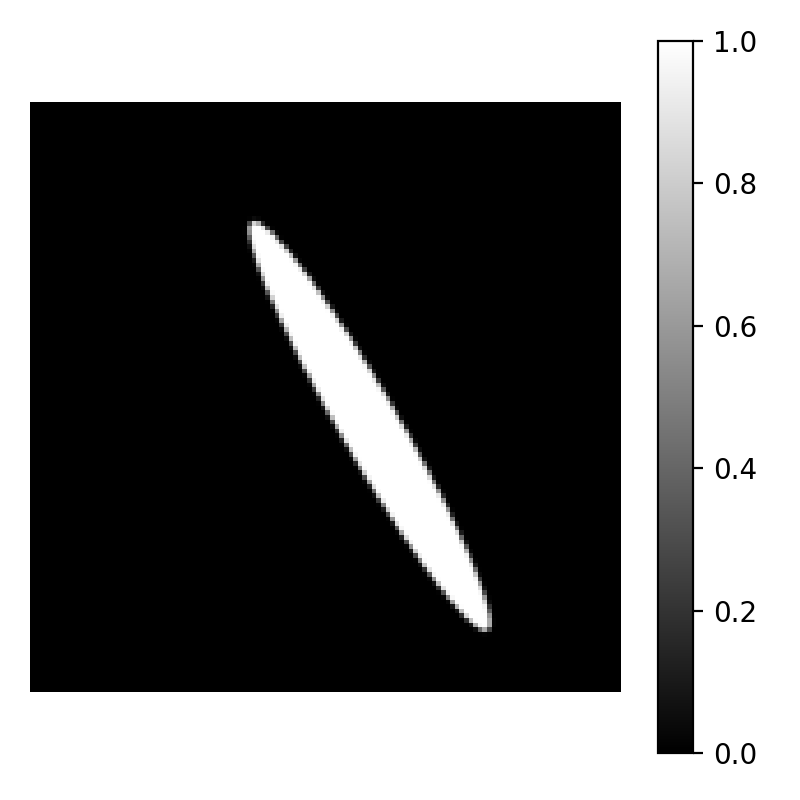}
        \caption{True Sample}
        \label{subfig: ground truth}
    \end{subfigure}
    \hfill
    \begin{subfigure}[t]{0.24\linewidth}
        \centering
        \includegraphics[width=\linewidth]{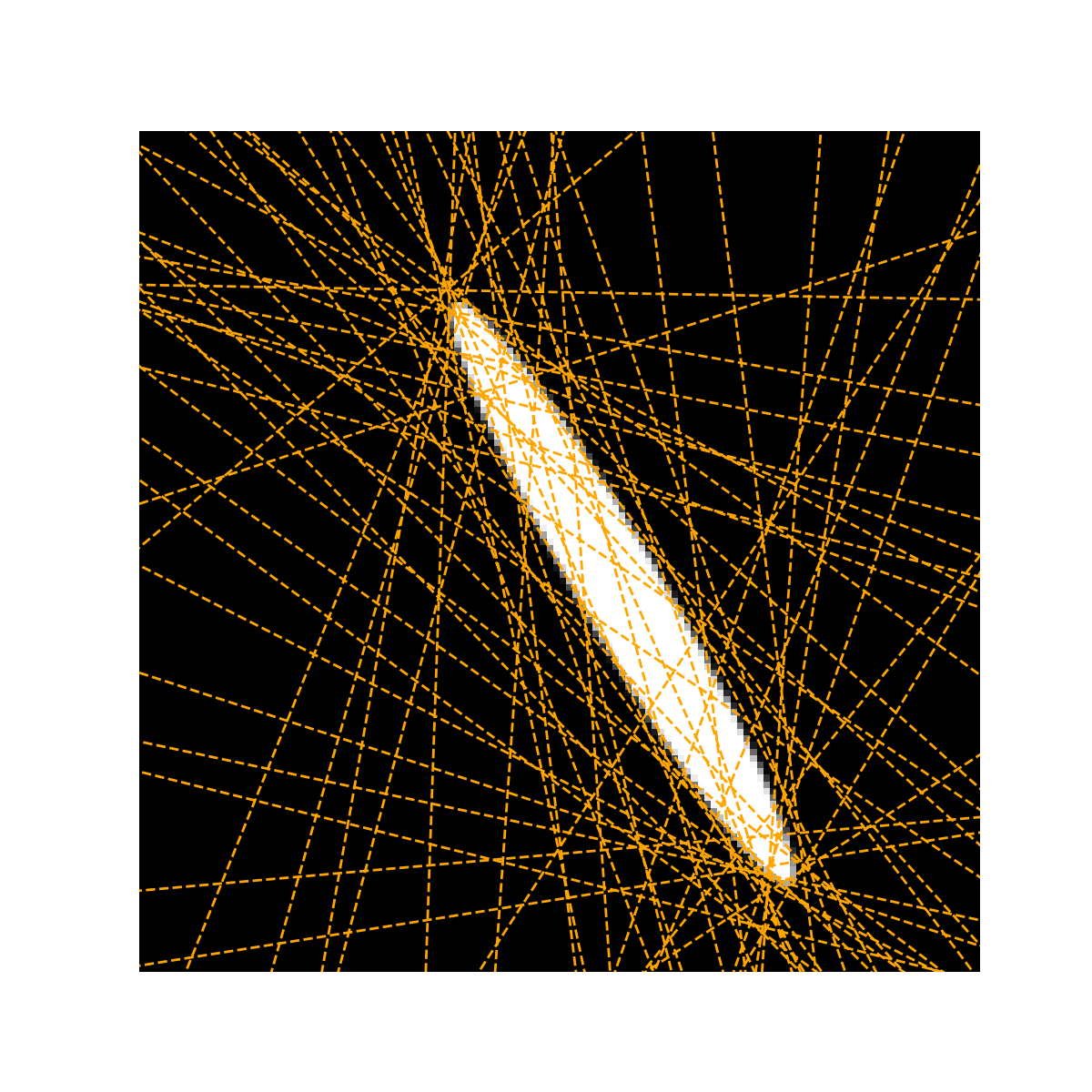}
        \caption{Beam selection at $10^{\text{th}}$ iteration}
        \label{subfig: ellipse beam}
    \end{subfigure}
    \hfill
    \begin{subfigure}[t]{0.24\linewidth}
        \centering
        \includegraphics[width=\linewidth]{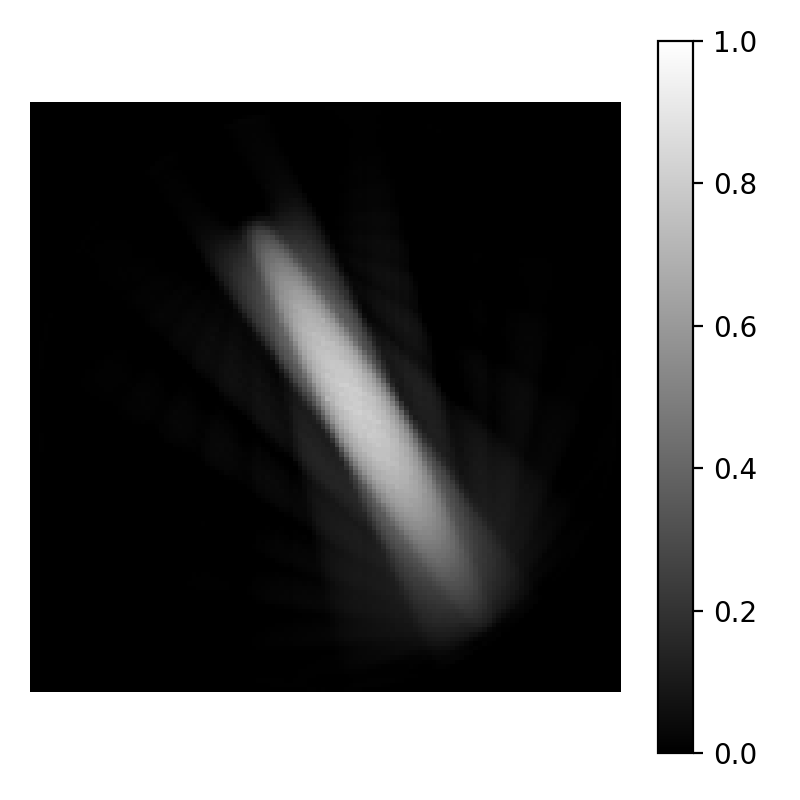}
        \caption{Reconstruction at $10^{\text{th}}$ iteration of evenly spaced raster scan baseline}
        \label{subfig: ellipse baseline reconstruction}
    \end{subfigure}
    \hfill
    \begin{subfigure}[t]{0.24\linewidth}
        \centering
        \includegraphics[width=\linewidth]{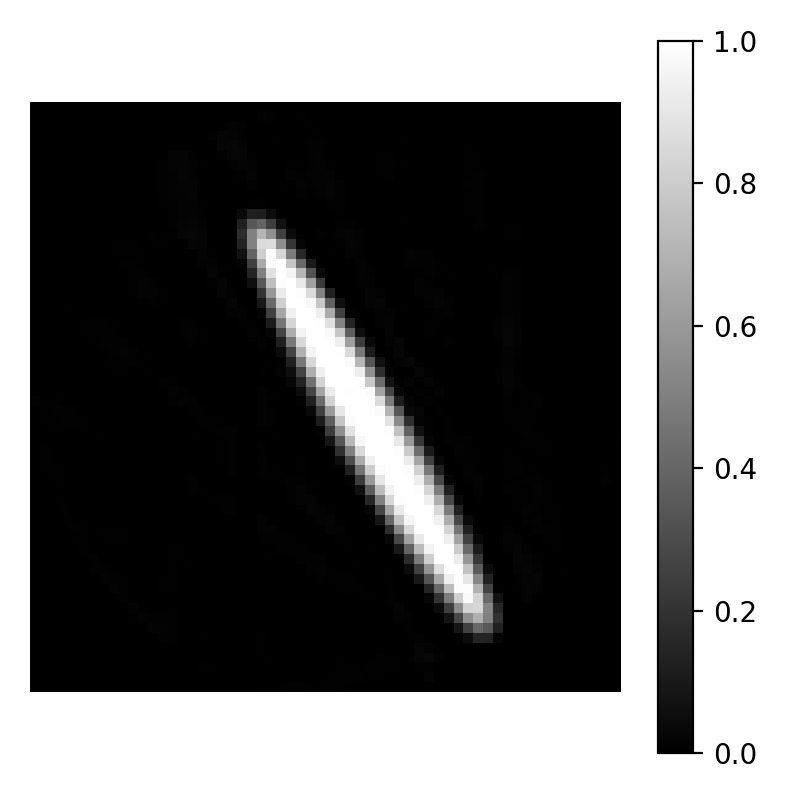}
        \caption{Reconstruction at $10^{\text{th}}$ iteration of beam selection}
        \label{subfig: ellipse beam reconstruction}
    \end{subfigure}
    \caption{Sample 1: Tilted Ellipse Sample}
    \label{fig: sample ellipse}
\end{figure}

The first sample is a thin ellipse with its major axis tilted at \(33^\circ\), as shown in Figure \ref{fig: sample ellipse}(\subref{subfig: ground truth}). It serves as a proof-of-concept for the edge-alignment component: because the sample is simple and highly anisotropic, most of its spatial variation lies along a narrow boundary, and a projection is informative largely when its angle aligns with that boundary. A correct angle-selection strategy should therefore place a higher density of measurements near the tilted major axis, providing a visually unambiguous check before the more complex geometries that follow, where informative angles are less apparent.

\begin{figure}[H]
    \centering
    \includegraphics[width=\linewidth]{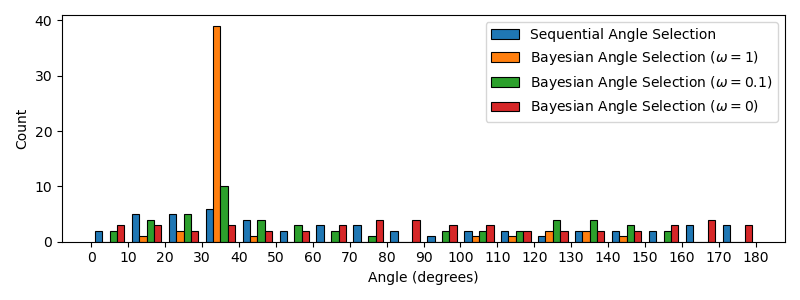}
    \caption{Angle distribution of different sequential angle selection methods}
    \label{fig: ellipse angle distribution}
\end{figure}

The edge–aligning behavior of both angle–selection strategies is evaluated using the tilted ellipse sample. Specifically, we compare the Bayesian baseline in Algorithm \ref{algorithm: baseline Bayesian angle} and the proposed reconstruction–free acquisition strategy in Algorithm \ref{algorithm: acquisition} (angle variant). To assess edge alignment, the angle distributions over the first 50 sequentially selected angles (excluding the initial angles) are collected and shown as a histogram in Figure \ref{fig: ellipse angle distribution}. The histogram overlays four angle-selection distributions: the Bayesian baseline at edge weights $w=0$, $w=0.1$, and $w=1$, and the proposed angle selection of Algorithm \ref{algorithm: acquisition} (angle variant). The horizontal axis denotes the selected angle and the vertical axis denotes the count of selections. Recall that the parameter $w$ in (\ref{eq: edge-prior covar}) controls the degree to which edge information influences the covariance fusion in the Bayesian baseline.

When the edge–weight parameter is set to $w=0$, the Bayesian update does not incorporate any edge information. In this setting the angle-selection criterion depends only on the measurement geometry $\A(\BeamSet_i^c)$ and not on the values of the acquired measurements; consequently the resulting angle trajectory is independent of the sample being scanned. This is reflected in Figure \ref{fig: ellipse angle distribution}, where the selected angles are approximately uniformly distributed over $[0^\circ, 180^\circ)$.

At the opposite extreme, when $w=1$ no uncertainty due to the forward Radon transform is included in the covariance update. Since the past measurement covariance term also produces a penalty against repeated measurements, setting $w=1$ removes this penalty entirely. In this case the algorithm repeatedly selects the angle that maximizes edge strength and becomes trapped there, leading to little or no improvement after a small number of iterations. This is evident in Figure \ref{fig: ellipse angle distribution}, where the angles are tightly concentrated near the tilt angle of the ellipse.

Based on these observations an intermediate value of $w=0.1$ is used for the Bayesian baseline in all subsequent sample tests. This choice provides a balance in which angles aligned with edges are prioritized while the covariance of the forward model is still retained to discourage repeated measurements. For each subsequent sample, the RMSE-vs-iteration plot then overlays four curves: the evenly spaced raster scan, the Bayesian baseline at $w=0.1$, the proposed angle selection (Algorithm \ref{algorithm: acquisition}, angle variant), and the proposed beam selection (Algorithm \ref{algorithm: acquisition}, beam variant); the horizontal axis denotes the iteration index and the vertical axis denotes reconstruction RMSE.

\begin{figure}[H]
    \centering
    \includegraphics[width=\linewidth]{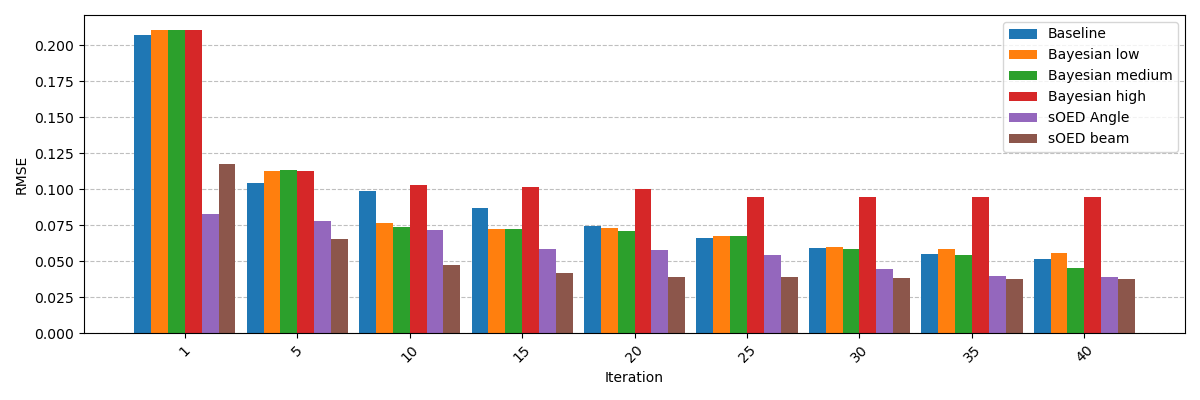}
    \caption{Comparison of RMSE between the baselines and the sequential design of experiment methods}
    \label{fig: ellipse RMSE}
\end{figure}

For beam selection, each individual beam is selected independently and assigned its own projection angle; therefore, the angle distribution directly characterizes the beam selection behavior. The edge–alignment behavior of the proposed method can also be visually confirmed by plotting the spatial locations of the selected beams, as shown in Figure \ref{fig: sample ellipse}(\subref{subfig: ellipse beam}).
The overall reconstruction performance of the different design–of–experiment approaches is summarized in Figure \ref{fig: ellipse RMSE}. This figure overlays the RMSE-vs-iteration curves for the evenly spaced raster scan, the Bayesian baseline at edge weights $w \in \{0, 0.1, 1\}$, the proposed angle selection, and the proposed beam selection.
 
The proposed beam and angle selection strategies achieve lower RMSE than both baselines, particularly during early iterations when only a limited number of projections are available. The Bayesian baseline with a large edge weight parameter $w$ exhibits the poorest reconstruction performance because measurements are concentrated almost exclusively around the major axis of the tilted ellipse. In this regime the sample covariance is dominated by the edge term in \eqref{eq: edge-prior covar}, which drives repeated measurements of the same angular neighborhood. When $w$ is moderately tuned, the Bayesian baseline outperforms the evenly spaced raster scan baseline at early iterations, since it can preferentially select angles aligned with the major axis of the object. Importantly, the Bayesian baseline does not remain trapped at those angles. The sample covariance still penalizes redundant acquisitions, which leads, over time, to a more uniform spread of measurements and ultimately yields better performance than the raster scan baseline.

The proposed angle selection method outperforms all Bayesian angle baselines. The primary reason is that the Bayesian edge alignment term depends on intermediate reconstructions, and at low measurement counts such reconstructions can yield incorrect edge estimates. The proposed method does not suffer from this failure mode since edge alignment is computed directly from sinogram interpolation rather than from a reconstruction. Furthermore, the proposed acquisition function achieves a more stable balance between edge alignment and global exploration, without requiring sensitive parameter tuning such as for $w$ in the Bayesian baselines.

The best performance is obtained with the proposed beam selection strategy. When measurements are scarce, this method provides the greatest improvement because beams can be concentrated in regions of higher structural complexity, yielding more informative data and therefore improved reconstruction accuracy at low measurement budgets.

As the iteration count increases, both baselines gradually improve and eventually approach the performance of the proposed methods. This is consistent with the convergence behavior of SART reconstructions: given sufficiently many measurements, a Kaczmarz-type iterative method converges toward the ground truth.

\subsubsection{Sample 2: Shepp-Logan Phantom}
\begin{figure}[H]
    \centering
    \begin{subfigure}[t]{0.24\linewidth}
        \centering
        \includegraphics[width=\linewidth]{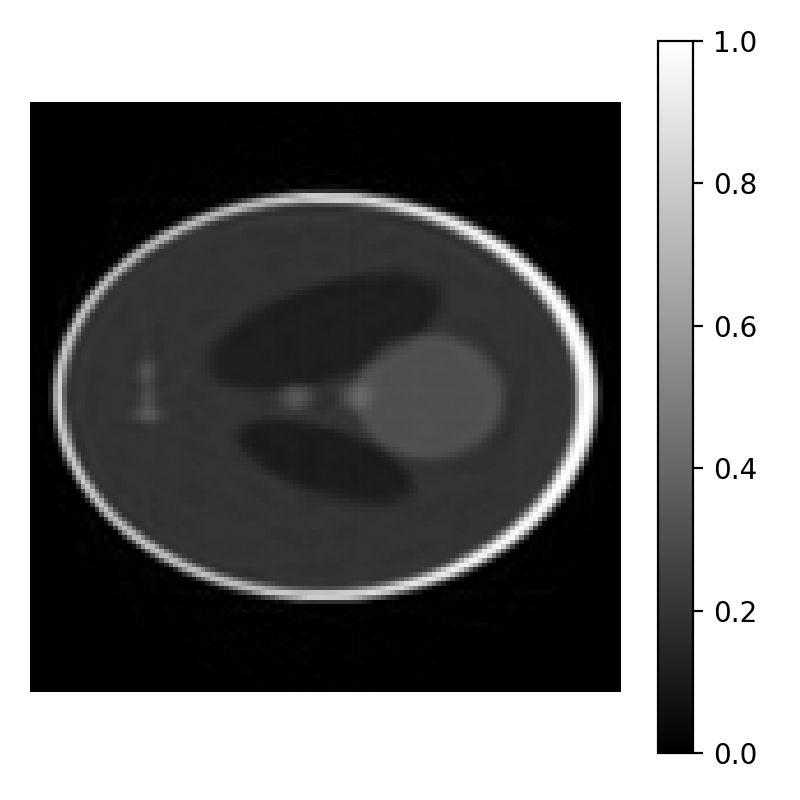}
        \caption{True Sample} 
        \label{subfig: phantom true}
    \end{subfigure}
    \hfil
    \begin{subfigure}[t]{0.24\linewidth}
        \centering
        \includegraphics[width=\linewidth]{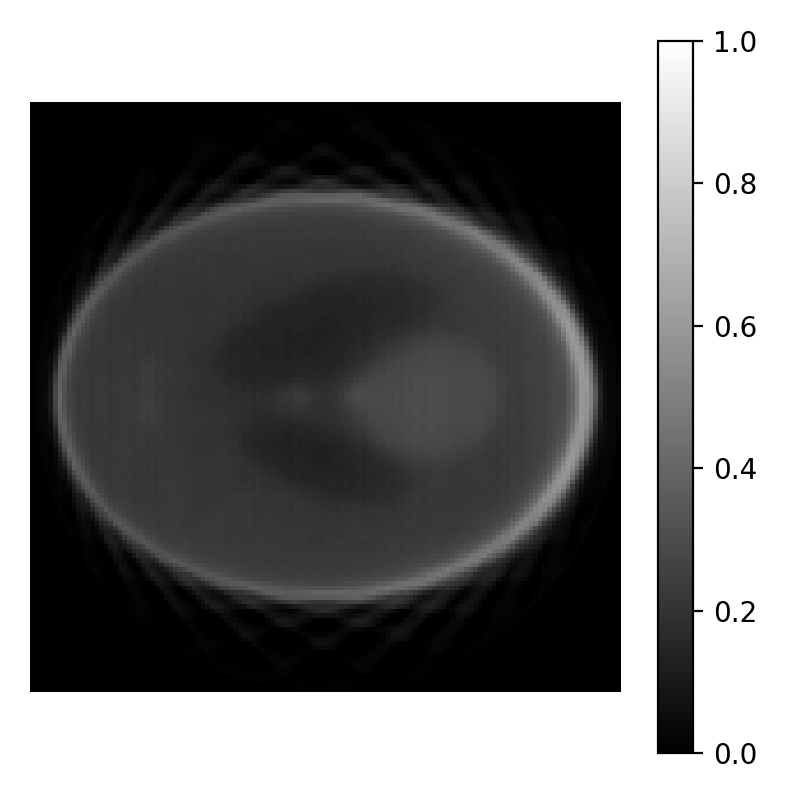}
        \caption{Raster scan baseline reconstruction} 
        \label{subfig: phantom baseline reconstruction}
    \end{subfigure}
    \hfil
    \begin{subfigure}[t]{0.24\linewidth}
        \centering
        \includegraphics[width=\linewidth]{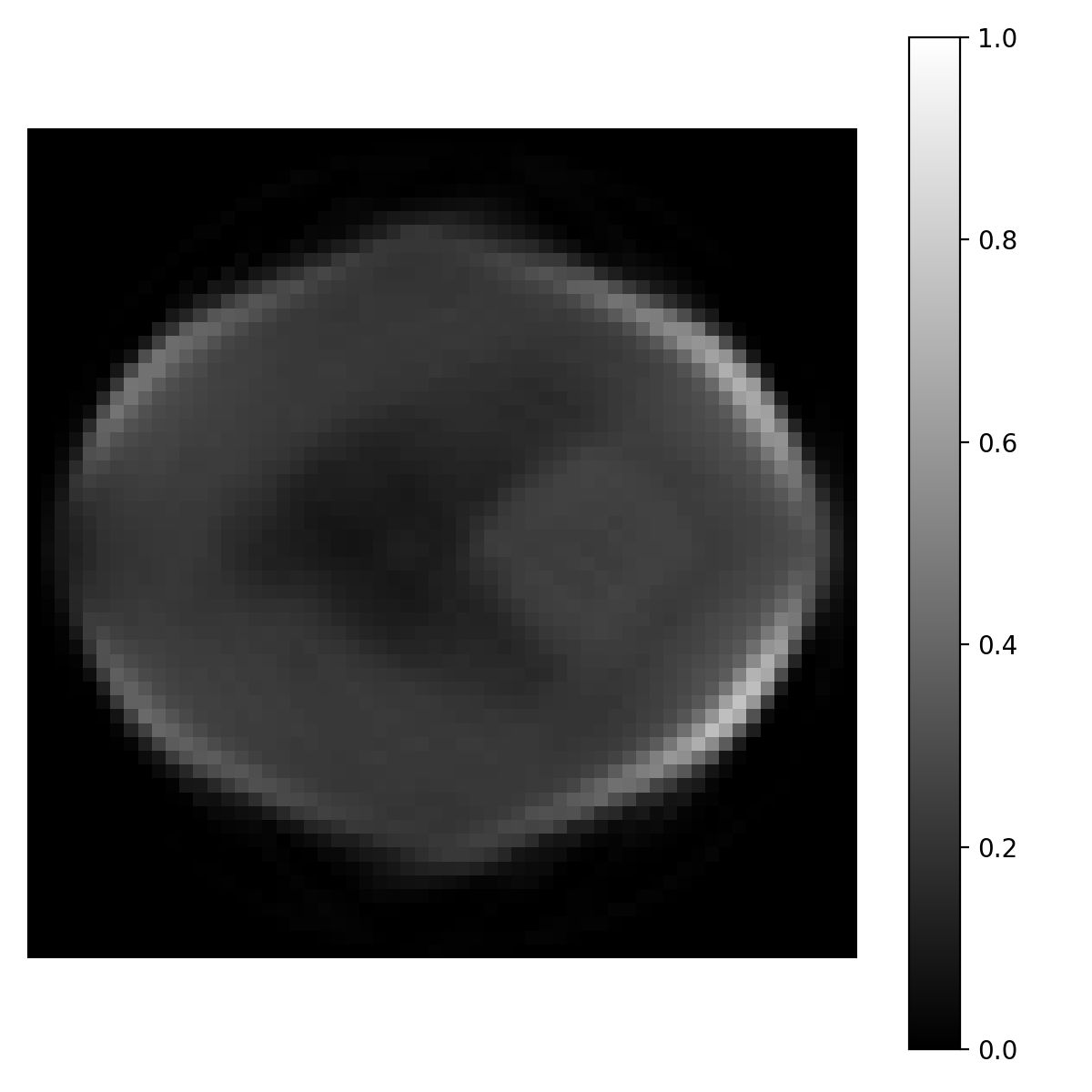}
        \caption{Bayesian baseline reconstruction} 
        \label{subfig: phantom Bayesian reconstruction}
    \end{subfigure}
    \hfil
    \begin{subfigure}[t]{0.24\linewidth}
        \centering
        \includegraphics[width=\linewidth]{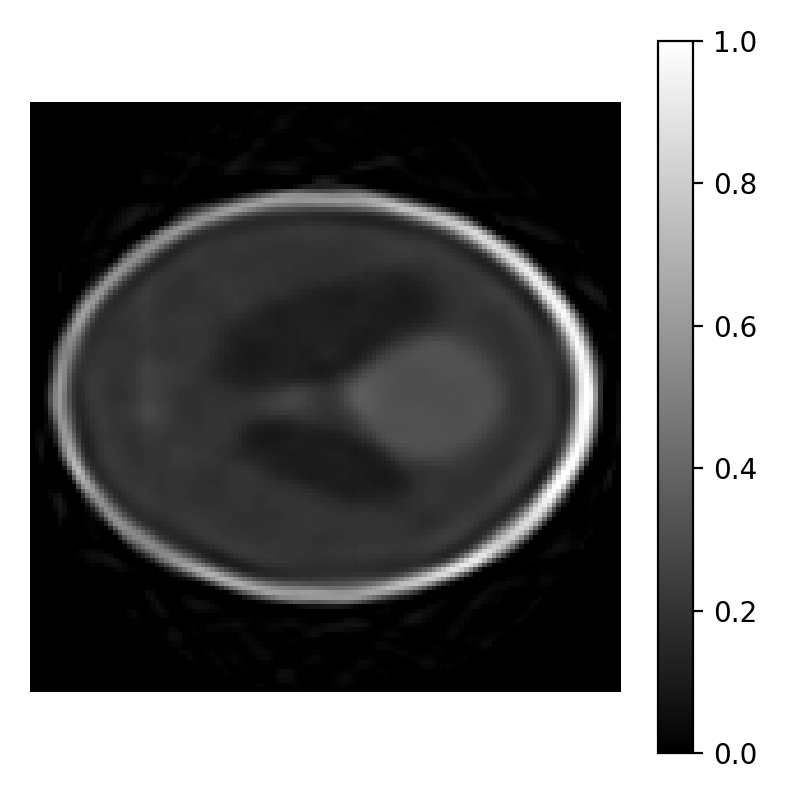}
        \caption{Beam selection reconstruction} 
        \label{subfig: phantom beam reconstruction}
    \end{subfigure}
    \caption{Shepp-Logan phantom Sample}
    \label{fig: phantom sample}
\end{figure}

To further demonstrate that the proposed approach can place measurements along sample boundaries without relying on any intermediate reconstruction, we consider the Shepp-Logan phantom, with the ground–truth image shown in Figure \ref{fig: phantom sample}. The Shepp-Logan phantom consists of high–contrast, piecewise–constant ellipsoidal regions with sharp and well–separated boundaries, so edge locations are unambiguous. Its analytical structure produces strong, localized features in the sinogram domain, which makes it a well–established benchmark for evaluating whether a beam–selection strategy correctly prioritizes edge–aligned measurements.

\begin{figure}[H]
    \centering
    \includegraphics[width=\linewidth]{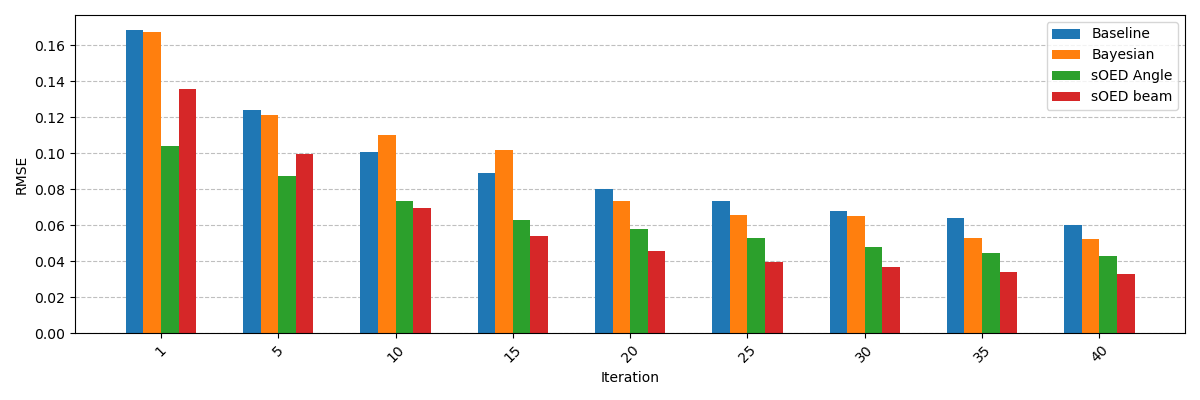}
    \caption{Comparison of RMSE between design of experiment methods on the phantom sample}
    \label{fig: phantom RMSE}
\end{figure}

The RMSE comparison in Figure \ref{fig: phantom RMSE} reflects the trends observed for the tilted-ellipse sample. The proposed sinogram-based sequential design, including both beam- and angle-selection variants, reduces RMSE substantially faster than the evenly spaced raster scan and the Bayesian baseline, particularly in the low-measurement (early-iteration) regime. Overall, the Bayesian baseline converges more effectively than the evenly spaced raster scan, except for a brief interval between iterations 6 and 15. This interval highlights one of the challenges associated with tuning the weight $w$ in the Bayesian baseline, as the Bayesian angle selection briefly prioritizes the tilted direction corresponding to the two darkest inner ellipses of the Shepp-Logan phantom. This behavior is evident in the reconstruction shown in Figure \ref{fig: phantom sample}(\subref{subfig: phantom Bayesian reconstruction}), where the reconstructed image is skewed along the orientation of these two ellipses.

In contrast, the proposed angle selection method does not exhibit this bias because it determines edge-alignment directly from the interpolated sinogram rather than relying on intermediate reconstructions. As a result, it identifies informative measurement directions without being misled by incomplete early reconstructions.

\begin{figure}[H] 
    \centering
    \begin{subfigure}[t]{0.19\linewidth}
        \centering
        \includegraphics[width=\linewidth]{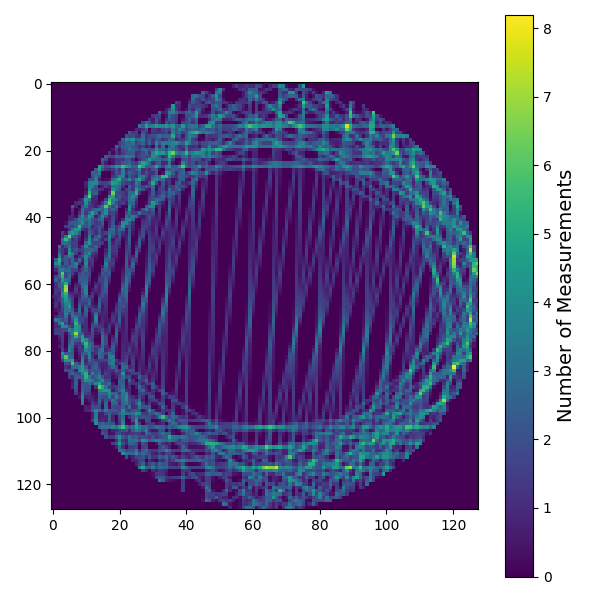}
        \caption{Iteration 1}
        \label{subfig: phantom beam distribution iter 1}
    \end{subfigure}
    \hfill
    \begin{subfigure}[t]{0.19\linewidth}
        \centering
        \includegraphics[width=\linewidth]{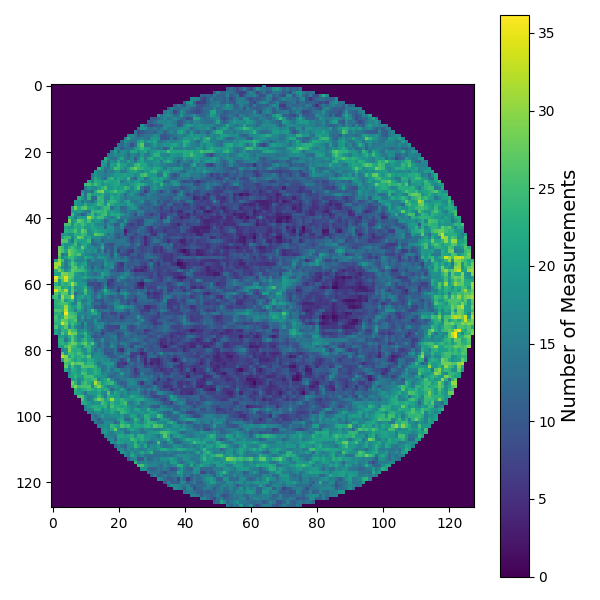}
        \caption{Iteration 10}
        \label{subfig: phantom beam distribution iter 10}
    \end{subfigure}
    \hfill
    \begin{subfigure}[t]{0.19\linewidth}
        \centering
        \includegraphics[width=\linewidth]{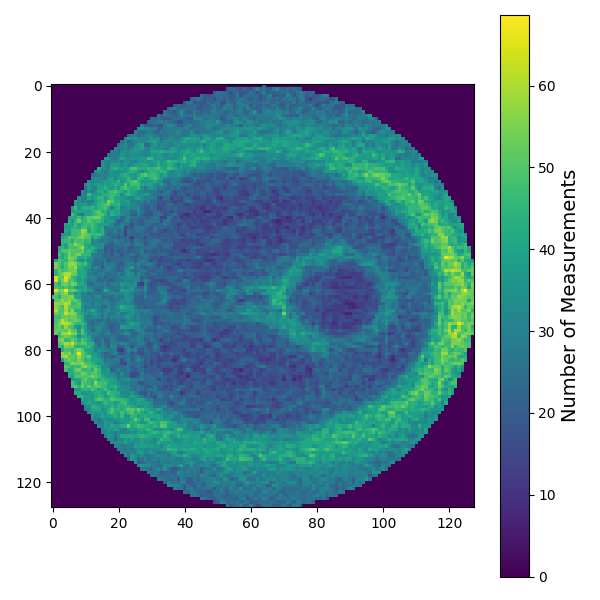}
        \caption{Iteration 20}
        \label{subfig: phantom beam distribution iter 20}
    \end{subfigure}
    \hfill
    \begin{subfigure}[t]{0.19\linewidth}
        \centering
        \includegraphics[width=\linewidth]{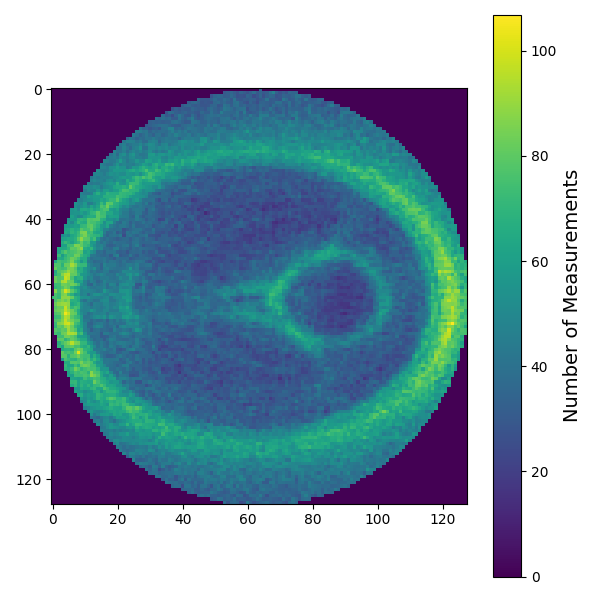}
        \caption{Iteration 30}
        \label{subfig: phantom beam distribution iter 30}
    \end{subfigure}
    \hfill
    \begin{subfigure}[t]{0.19\linewidth}
        \centering
        \includegraphics[width=\linewidth]{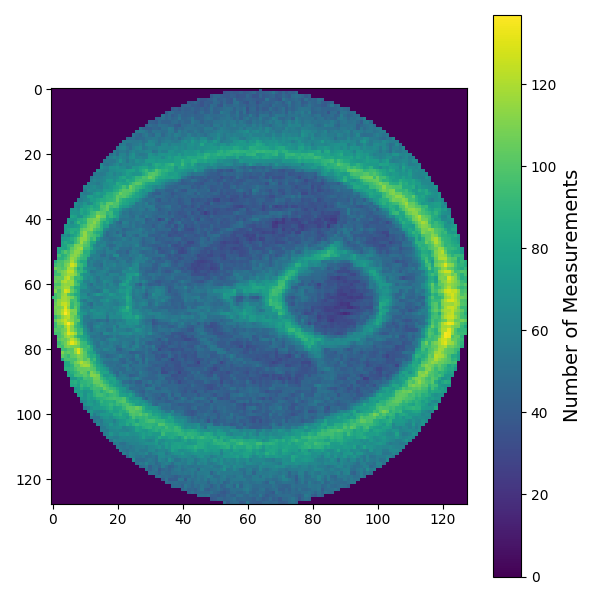}
        \caption{Iteration 40}
        \label{subfig: phantom beam distribution iter 40}
    \end{subfigure}
    \caption{Beam distribution on Shepp-Logan phantom sample}
    \label{fig: phantom beam distribution}
\end{figure}

For the proposed beam selection method, the reconstruction error decreases more rapidly than in the other approaches, particularly at early iterations, because the algorithm prioritizes placing beams along the edges of the sample. By acquiring measurements in regions where structural information is most informative, the method achieves lower RMSE at the same measurement budget. This behavior is illustrated in Figure \ref{fig: phantom beam distribution}, where the two-dimensional histogram of beam placements indicates the number of beams intersecting each pixel in the sample image. Regions with high counts coincide with feature boundaries, confirming that the acquisition strategy preferentially samples edges to improve reconstruction efficiency.

Figure \ref{fig: phantom beam distribution} also provides insight into why the angle selection method outperforms the beam selection method in the first few iterations. In early iterations, beam placement concentrates primarily on the outer boundary of the Shepp-Logan phantom, as seen in Figure \ref{fig: phantom beam distribution}(\subref{subfig: phantom beam distribution iter 1}). As more beams are added in subsequent iterations, the beam selection algorithm shifts the sampling to the interior of the phantom, with additional emphasis on the edges of individual ellipses. As a result, beam selection overtakes angle selection by the tenth iteration and remains superior thereafter.

\begin{figure}[H] 
    \centering
    \begin{subfigure}[t]{0.24\linewidth}
        \centering
        \includegraphics[width=\linewidth]{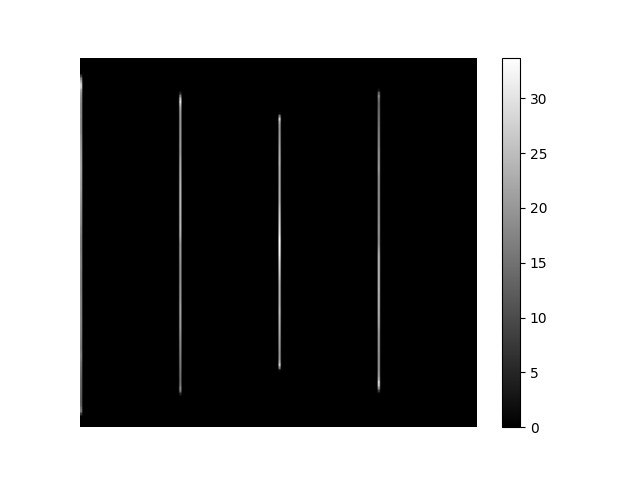}
        \caption{Collected measurements of the sinogram} 
    \end{subfigure}
    \hfill
    \begin{subfigure}[t]{0.24\linewidth}
        \centering
        \includegraphics[width=\linewidth]{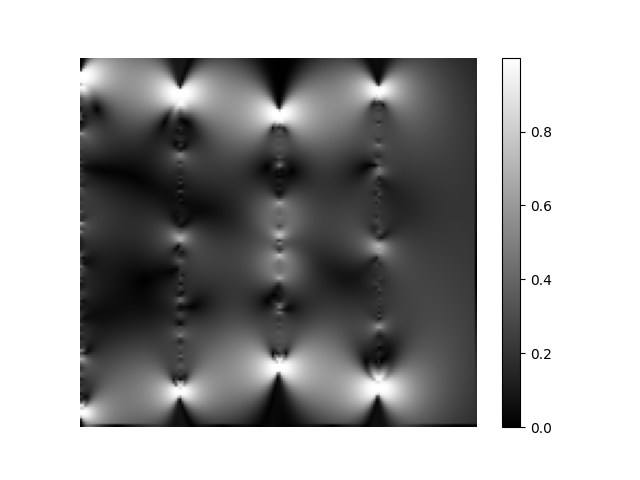}
        \caption{Edge-mapping component of the acquisition function} 
    \end{subfigure}
    \hfill
    \begin{subfigure}[t]{0.24\linewidth}
        \centering
        \includegraphics[width=\linewidth]{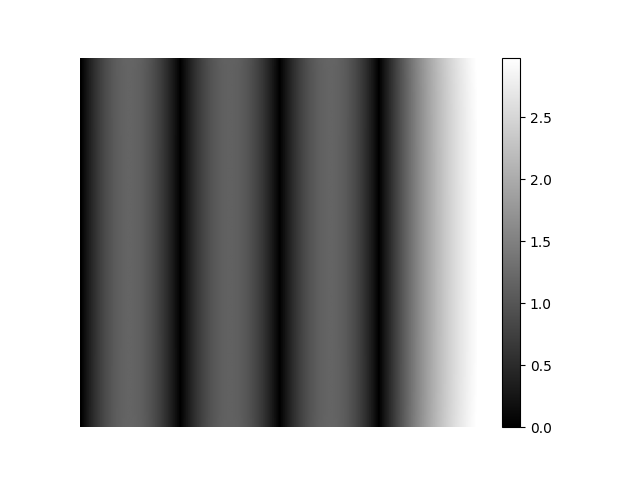}
        \caption{Confidence interval component  of the acquisition function} 
    \end{subfigure}
    \hfill
    \begin{subfigure}[t]{0.24\linewidth}
        \centering
        \includegraphics[width=\linewidth]{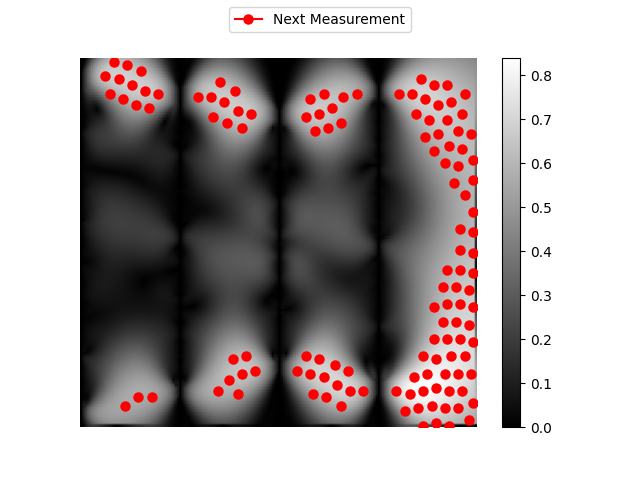}
        \caption{Acquisition function} 
    \end{subfigure}
    \caption{Beam selection iteration 1 on Shepp-Logan phantom sample}
    \label{fig: sinogram acquiistion phantom 1}
\end{figure}

\begin{figure}[H] 
    \centering
    \begin{subfigure}[t]{0.24\linewidth}
        \centering
        \includegraphics[width=\linewidth]{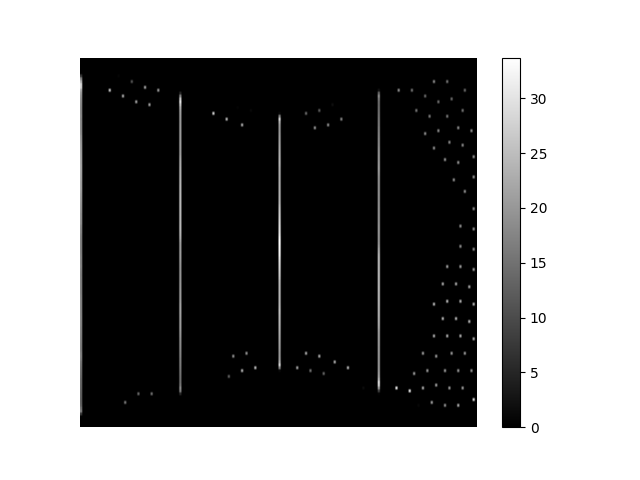}
        \caption{Collected measurements of the sinogram} 
    \end{subfigure}
    \hfill
    \begin{subfigure}[t]{0.24\linewidth}
        \centering
        \includegraphics[width=\linewidth]{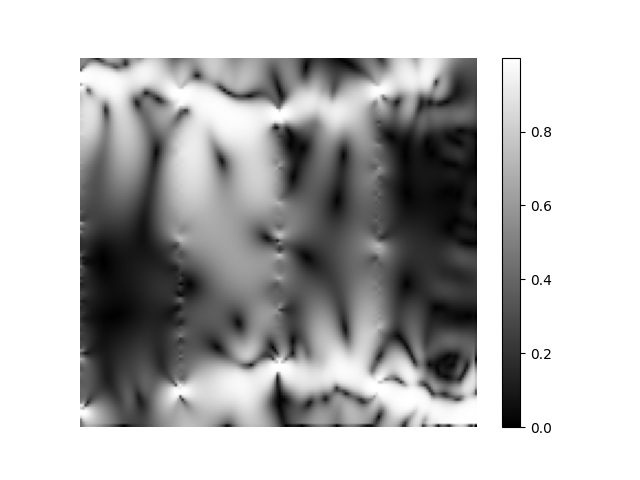}
        \caption{Edge-mapping component of the acquisition function} 
    \end{subfigure}
    \hfill
    \begin{subfigure}[t]{0.24\linewidth}
        \centering
        \includegraphics[width=\linewidth]{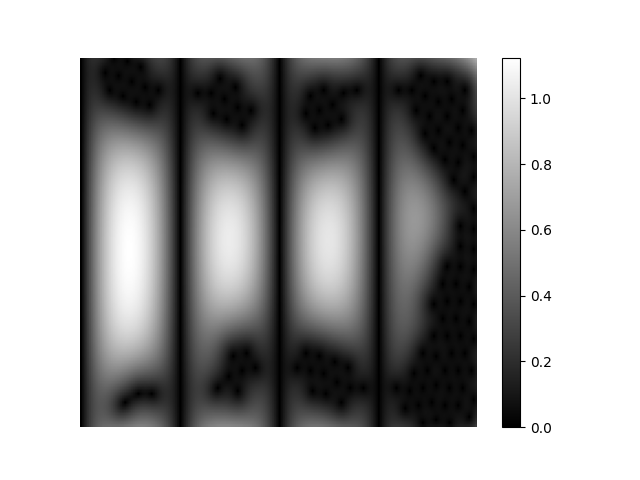}
        \caption{Confidence interval component  of the acquisition function} 
    \end{subfigure}
    \hfill
    \begin{subfigure}[t]{0.24\linewidth}
        \centering
        \includegraphics[width=\linewidth]{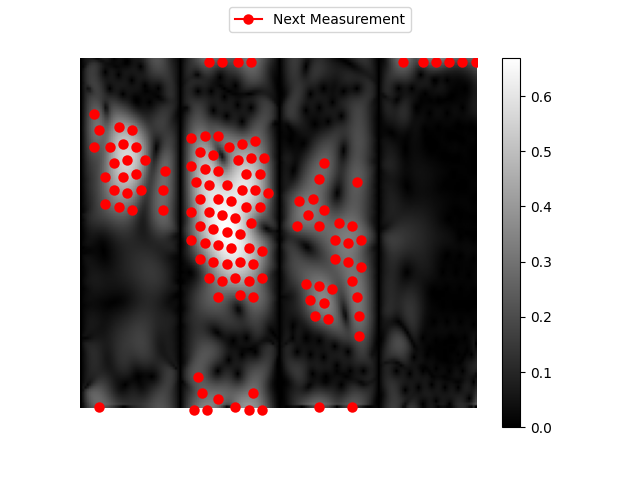}
        \caption{Acquisition function} 
    \end{subfigure}
    \caption{Beam selection iteration 2 on Shepp-Logan phantom sample}
    \label{fig: sinogram acquiistion phantom 2}
\end{figure}

\begin{figure}[H] 
    \centering
    \begin{subfigure}[t]{0.24\linewidth}
        \centering
        \includegraphics[width=\linewidth]{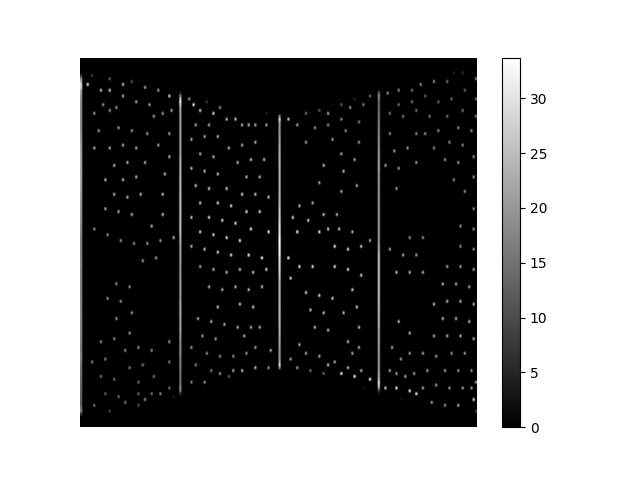}
        \caption{Collected measurements of the sinogram} 
    \end{subfigure}
    \hfill
    \begin{subfigure}[t]{0.24\linewidth}
        \centering
        \includegraphics[width=\linewidth]{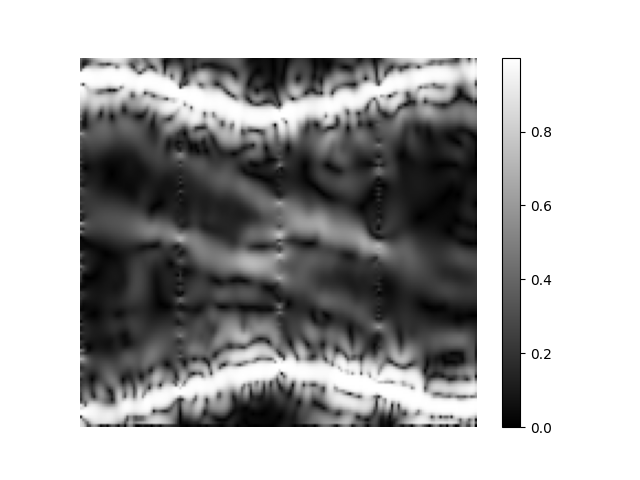}
        \caption{Edge-mapping component of the acquisition function} 
    \end{subfigure}
    \hfill
    \begin{subfigure}[t]{0.24\linewidth}
        \centering
        \includegraphics[width=\linewidth]{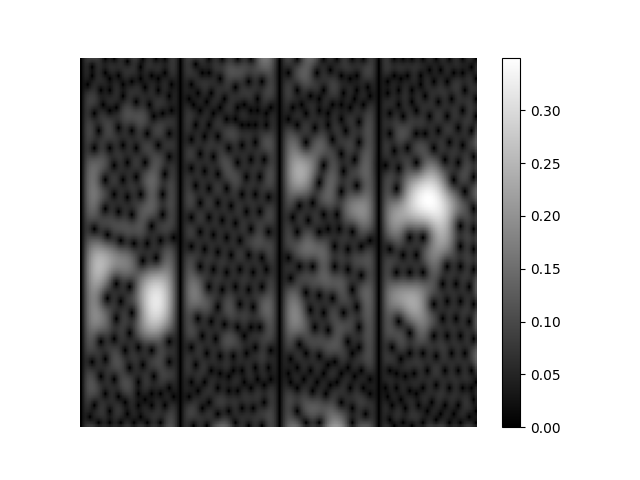}
        \caption{Confidence interval component  of the acquisition function} 
    \end{subfigure}
    \hfill
    \begin{subfigure}[t]{0.24\linewidth}
        \centering
        \includegraphics[width=\linewidth]{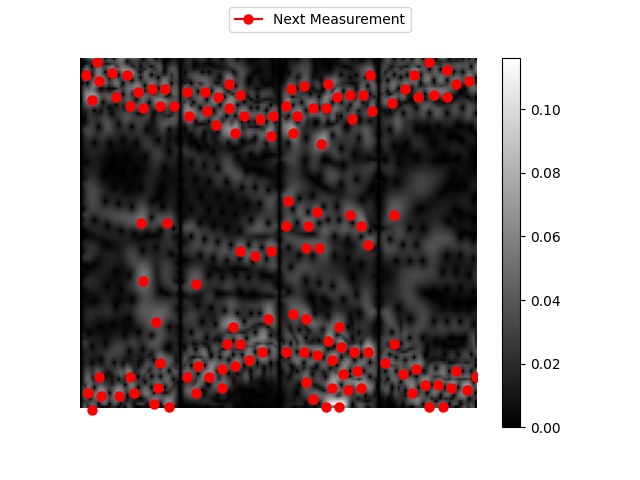}
        \caption{Acquisition function} 
    \end{subfigure}
    \caption{Beam selection iteration 5 on Shepp-Logan phantom sample}
    \label{fig: sinogram acquiistion phantom 3}
\end{figure}

\begin{figure}[H] 
    \centering
    \begin{subfigure}[t]{0.24\linewidth}
        \centering
        \includegraphics[width=\linewidth]{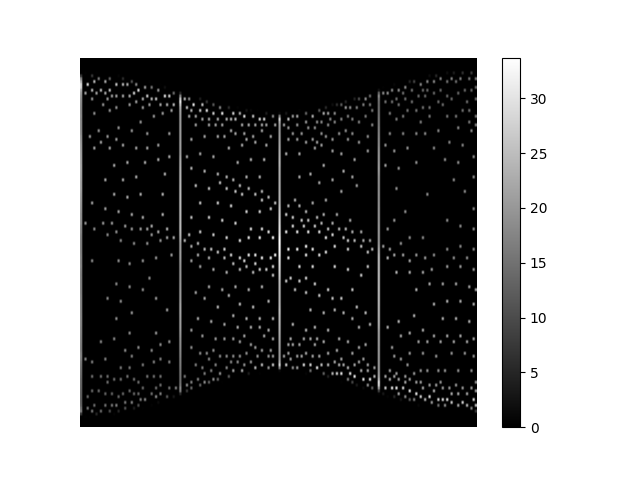}
        \caption{Collected measurements of the sinogram}
    \end{subfigure}
    \hfill
    \begin{subfigure}[t]{0.24\linewidth}
        \centering
        \includegraphics[width=\linewidth]{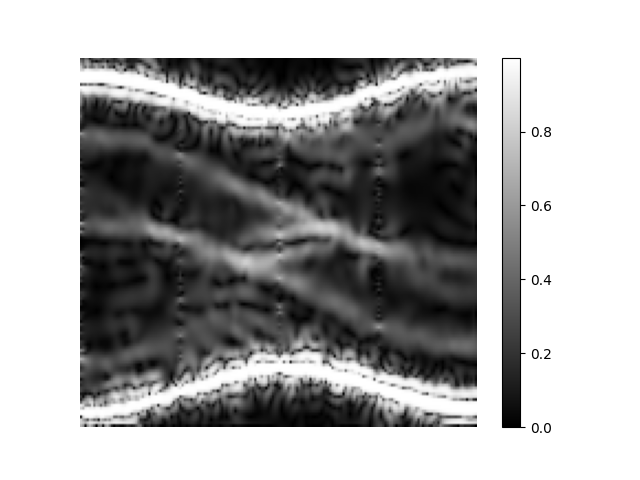}
        \caption{Edge-mapping component of the acquisition function} 
    \end{subfigure}
    \hfill
    \begin{subfigure}[t]{0.24\linewidth}
        \centering
        \includegraphics[width=\linewidth]{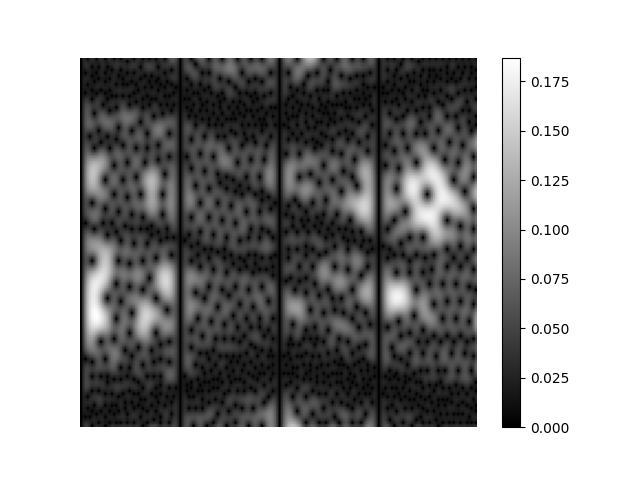}
        \caption{Confidence interval component  of the acquisition function} 
    \end{subfigure}
    \hfill
    \begin{subfigure}[t]{0.24\linewidth}
        \centering
        \includegraphics[width=\linewidth]{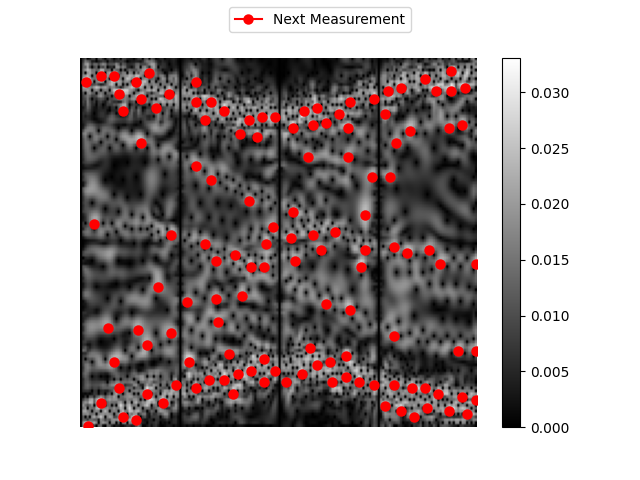}
        \caption{Acquisition function} 
    \end{subfigure}
    \caption{Beam selection iteration 10 on Shepp-Logan phantom sample}
    \label{fig: sinogram acquiistion phantom 4}
\end{figure}

The proposed beam-selection method places beams along the phantom boundaries without ever reconstructing an image; decisions rely solely on sinogram information. The acquisition-function evaluation (Algorithm \ref{algorithm: Adaptive Beam Selection}) is visualized at iterations 1, 2, 5, and 10 in Figures \ref{fig: sinogram acquiistion phantom 1}-\ref{fig: sinogram acquiistion phantom 4}. Each panel shows the components forming the early-iteration acquisition score: (i) the observed sinogram measurements, (ii) the edge-mapping term from the GP predictive mean, (iii) the confidence-interval (CI) width from the posterior variance, and (iv) their pointwise product, the final acquisition value. Operating directly in sinogram space, the method selects new beams without repeated reconstruction; the GP interpolant supplies both a smooth estimate for exploitation and a principled uncertainty measure for exploration.

This design has several implementation consequences. First, GP interpolation acts as an implicit denoiser, making the edge-mapping term more robust to noise under low-dose conditions. Second, the CI term down-weights previously sampled regions, which—together with the minimum-distance constraint $\Delta_{\Obj}$—prevents beam clustering. Third, the kernel choice (here, Matérn-3/2) and its hyperparameters set the interpolation scale, balancing local edge sensitivity against global smoothness; these are tuned offline. Finally, online update formulas (e.g., Woodbury / Schur complement) make GP updates efficient for the batch sizes considered, rendering exhaustive scoring over the finite candidate set tractable.

\begin{figure}[H] 
    \centering
    \begin{subfigure}[t]{0.45\linewidth}
        \centering
        \includegraphics[width=\linewidth]{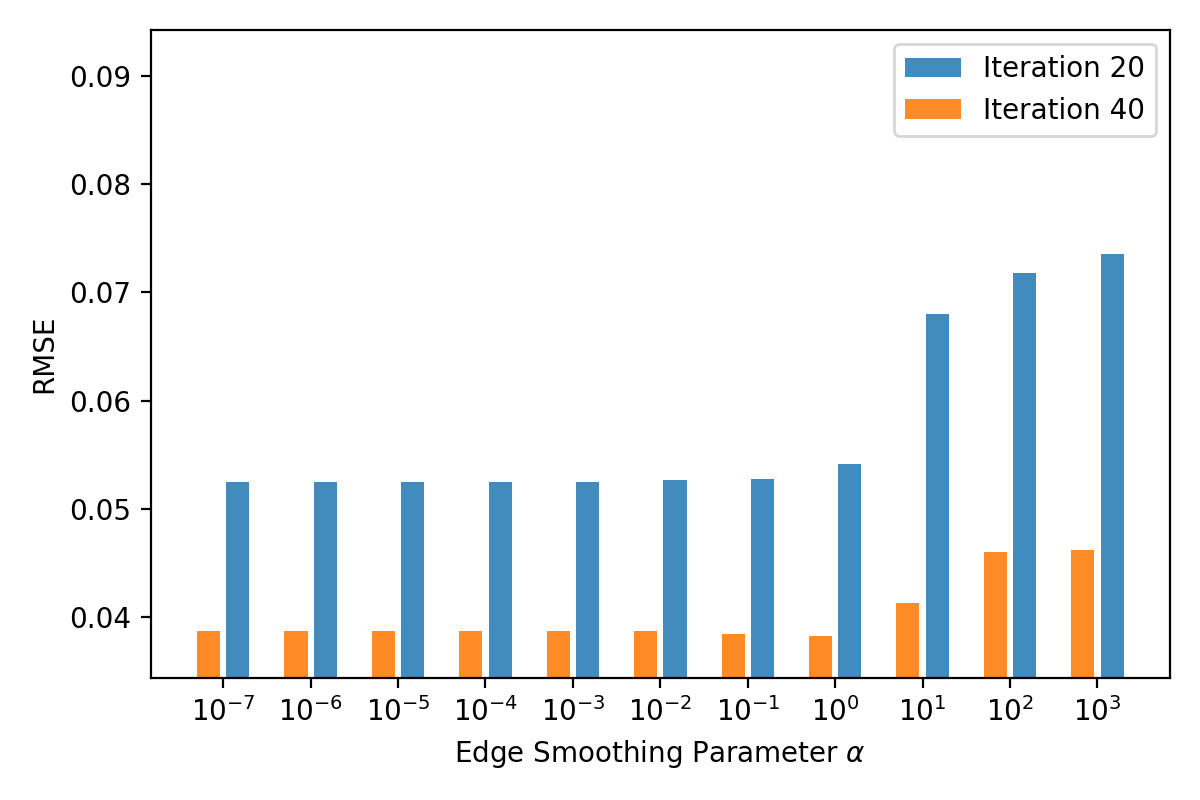}
        \caption{Beam selection parameter tuning}
        \label{subfig: beam parameter tuning}
    \end{subfigure}
    \hspace{0.1cm}
    \begin{subfigure}[t]{0.45\linewidth}
        \centering
        \includegraphics[width=\linewidth]{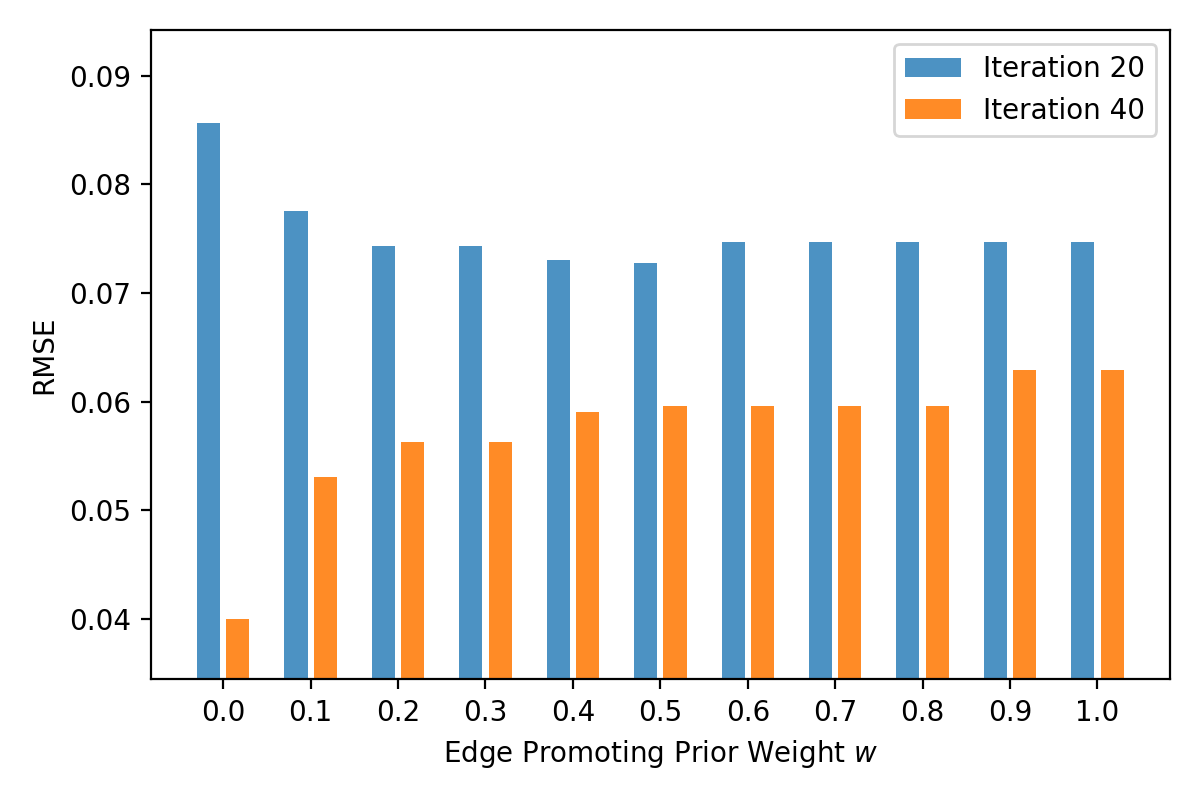}
        \caption{Bayesian baseline parameter tuning}
        \label{subfig: bayesian parameter tuning}
    \end{subfigure}
    \caption{Effect of parameter tuning on reconstruction RMSE at iteration 20 and 40}
    \label{fig: parameter tuning}
\end{figure}

As demonstrated in the previous examples, the performance of the Bayesian baseline depends strongly on the tuning of the edge weight parameter $w$, which must balance edge alignment against forward model uncertainty. This tradeoff is difficult to control in practice. In contrast, the proposed sinogram based beam selection method is substantially less sensitive to parameter tuning. To evaluate this difference, a case study was conducted on the current phantom sample to assess the effect of parameter choices on both the proposed method and the Bayesian baseline. The results are summarized in Figure \ref{fig: parameter tuning}; each subplot reports two RMSE-vs-parameter curves, one at iteration 20 and one at iteration 40.

For the Bayesian baseline, the edge weight parameter $w$ is varied from $0$ to $1$. As shown in Figure \ref{fig: parameter tuning}(\subref{subfig: bayesian parameter tuning}), large values of $w$ yield improved short term performance due to strong edge alignment. However, long term performance deteriorates because repeated measurements are insufficiently penalized. In contrast, at lower values of $w$, the method exhibits improved long term performance through better exploration of forward model uncertainty, but its short term accuracy suffers because edge alignment is underemphasized.

The proposed sinogram based beam selection method, shown in Figure \ref{fig: parameter tuning}(\subref{subfig: beam parameter tuning}), maintains stable performance across a wide range of parameter choices. Across a broad range of finite values of $\alpha$, the sinogram edge map reliably captures the true edge structure of the sample, consistent with the analysis in Section \ref{sec: Edge}. Consequently, the method is largely insensitive to the choice of $\alpha$, and performance remains consistent even when $\alpha$ varies across several orders of magnitude. Performance degrades only when $\alpha$ becomes too large, at which point the edge mapping no longer reflects the true edge locations. 

\subsubsection{Sample 3: Torso Cross-Section}
\begin{figure}[H]
    \centering
    \begin{subfigure}[t]{0.19\linewidth}
        \centering
        \includegraphics[width=\linewidth]{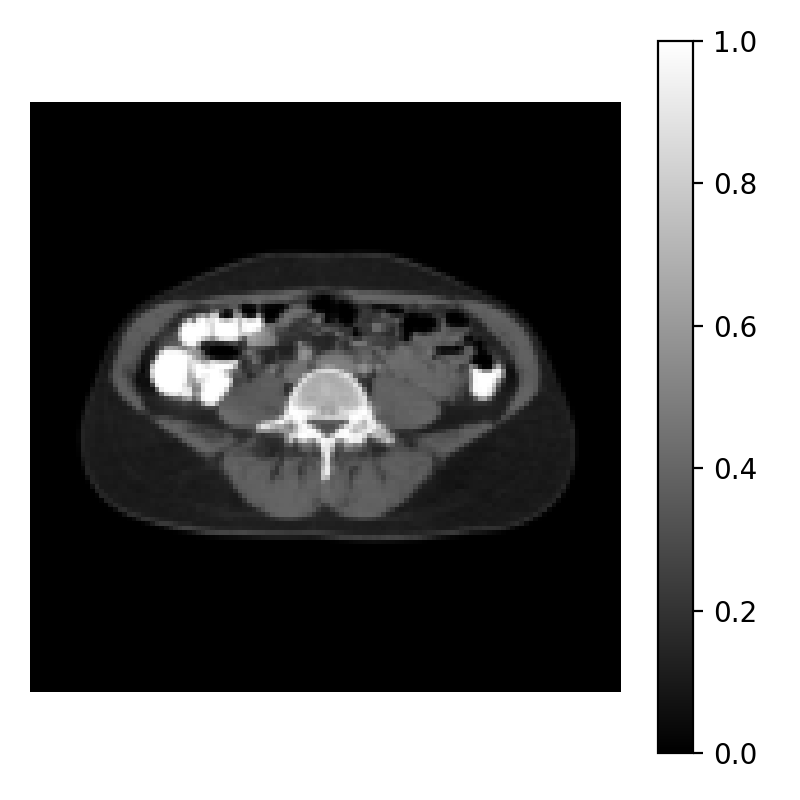}
        \caption{True sample} 
        \label{subfig: torso sample real}
    \end{subfigure}
    \hfil
    \begin{subfigure}[t]{0.19\linewidth}
        \centering
        \includegraphics[width=\linewidth]{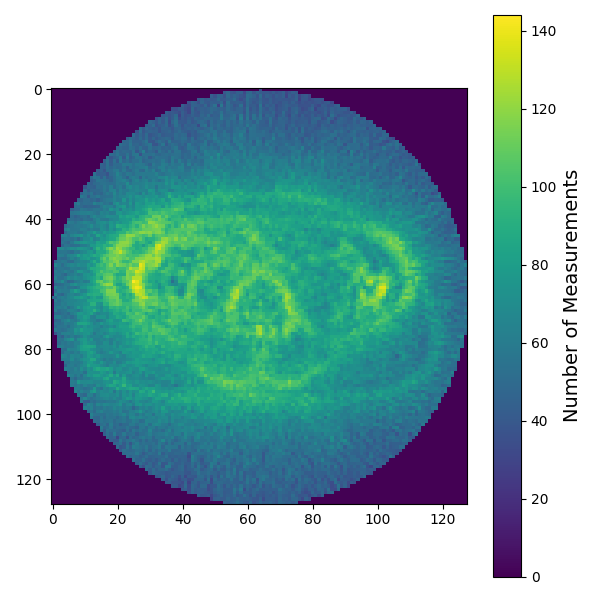}
        \caption{Beam distribution} 
        \label{subfig: torso beam distribution}
    \end{subfigure}
    \hfil
    \begin{subfigure}[t]{0.19\linewidth}
        \centering
        \includegraphics[width=\linewidth]{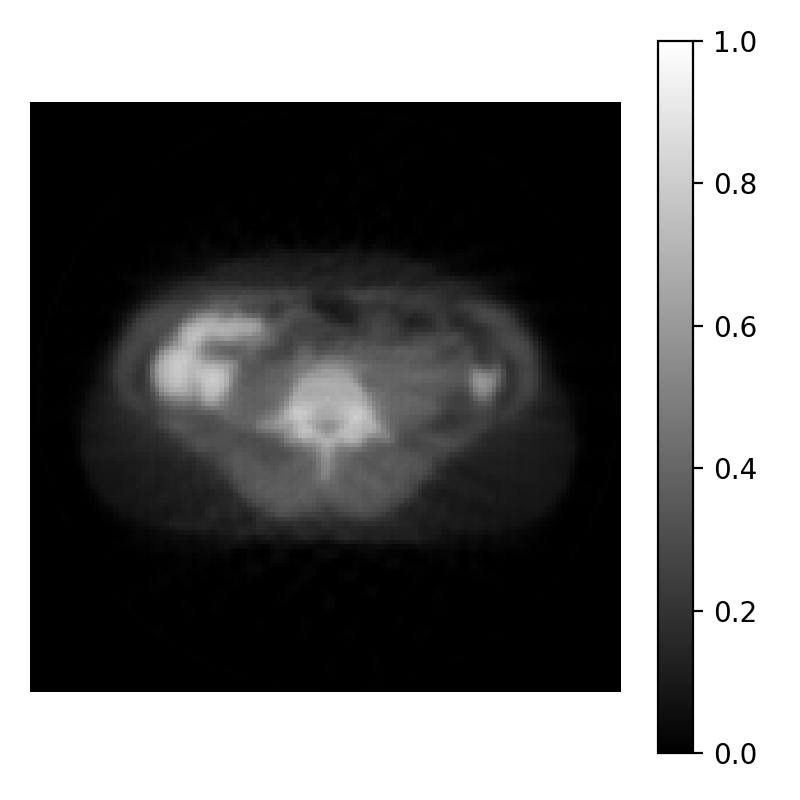}
        \caption{Raster scan baseline reconstruction} 
    \end{subfigure}
    \hfil
    \begin{subfigure}[t]{0.19\linewidth}
        \centering
        \includegraphics[width=\linewidth]{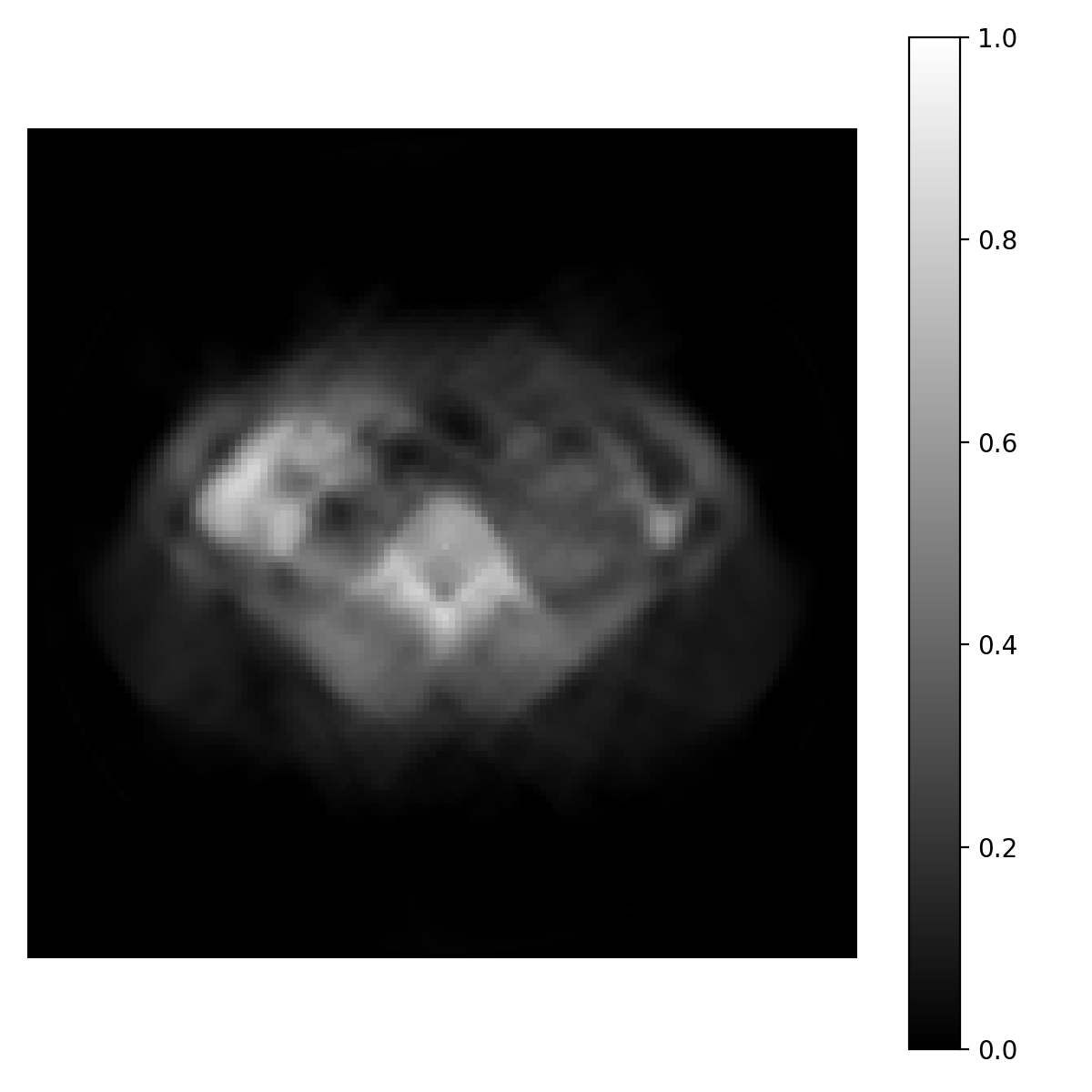}
        \caption{Bayesian baseline reconstruction} 
    \end{subfigure}
    \hfil
    \begin{subfigure}[t]{0.19\linewidth}
        \centering
        \includegraphics[width=\linewidth]{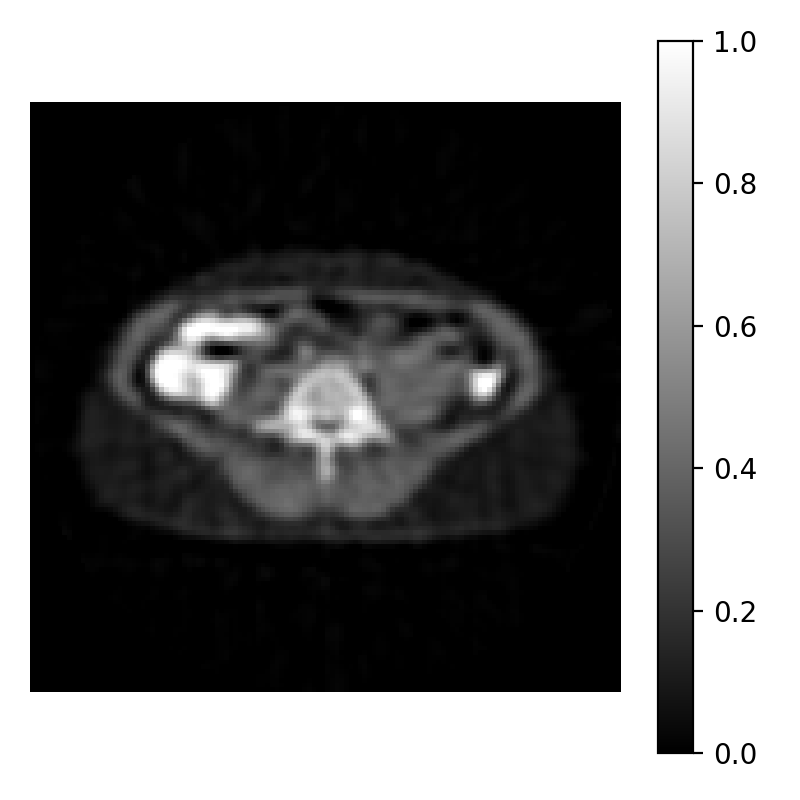}
        \caption{Beam selection reconstruction} 
    \end{subfigure}
    \caption{Torso sample}
    \label{fig: torso sample}
\end{figure}

\begin{figure}[H]
    \centering
    \includegraphics[width=\linewidth]{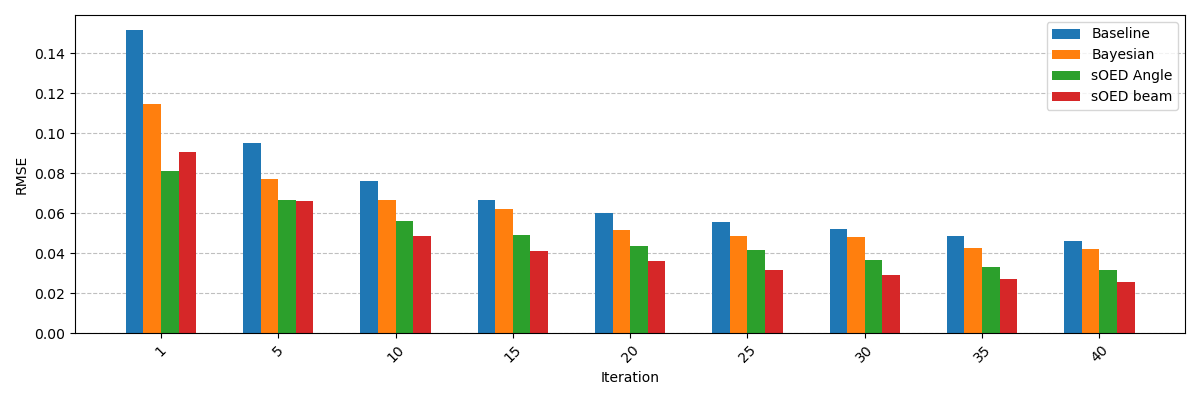}
    \caption{Comparison of RMSE between design of experiment methods on the torso sample}
    \label{fig: torso RMSE}
\end{figure}

The torso cross-section shown in Figure \ref{fig: torso sample}(\subref{subfig: torso sample real}) represents a more realistic and anatomically complex test case compared to the idealized Shepp-Logan phantom. Despite the increased structural complexity and varying contrast, the proposed sinogram-based beam selection exhibits the same qualitative behavior: beams concentrate along prominent anatomical boundaries (Figure \ref{fig: torso sample}(\subref{subfig: torso beam distribution})) and achieve the lowest RMSE at higher measurement budgets (Figure~\ref{fig: torso RMSE}). The proposed angle selection, having fewer degrees of freedom in beam placement, does not match the beam selection but still outperforms both the Bayesian and raster baselines; the Bayesian baseline in turn edges out the raster scan by prioritizing angles aligned with prominent boundaries. Overall, the trends match the phantom experiments: the proposed beam selection achieves improved reconstruction accuracy at the same measurement budget.

\subsubsection{Sample 4: Biological Sample} \label{subsec: bio sample}
\begin{figure}[H]
    \centering
    \begin{subfigure}[t]{0.19\linewidth}
        \centering
        \includegraphics[width=\linewidth]{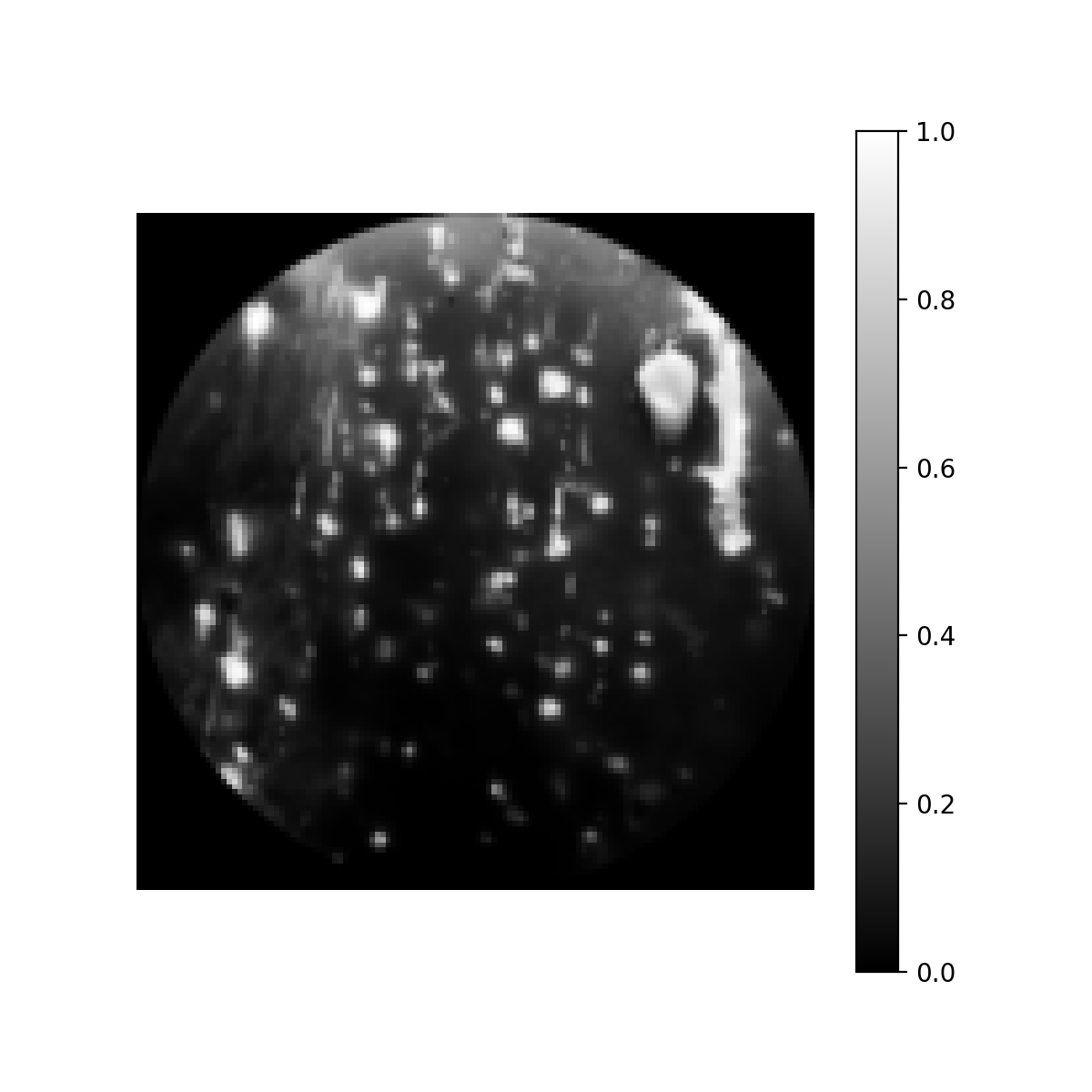}
        \caption{True sample} 
        \label{subfig: bio true}
    \end{subfigure}
    \hfil
    \begin{subfigure}[t]{0.19\linewidth}
        \centering
        \includegraphics[width=\linewidth]{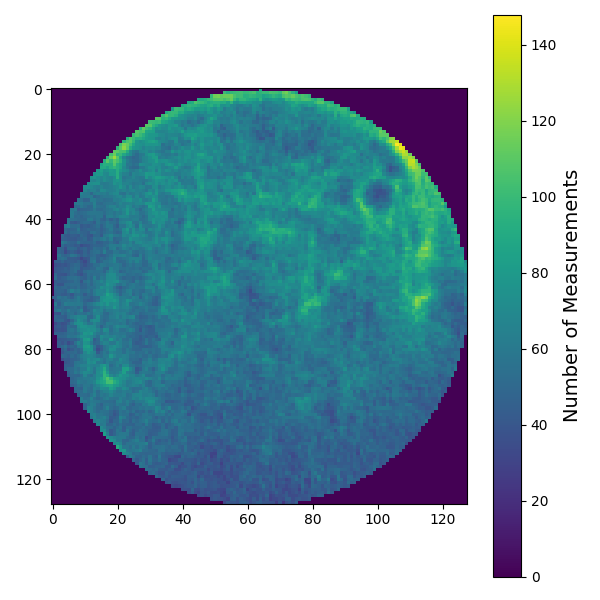}
        \caption{Beam distribution} 
        \label{subfig: bio beam distribution}
    \end{subfigure}
    \hfil
    \begin{subfigure}[t]{0.19\linewidth}
        \centering
        \includegraphics[width=\linewidth]{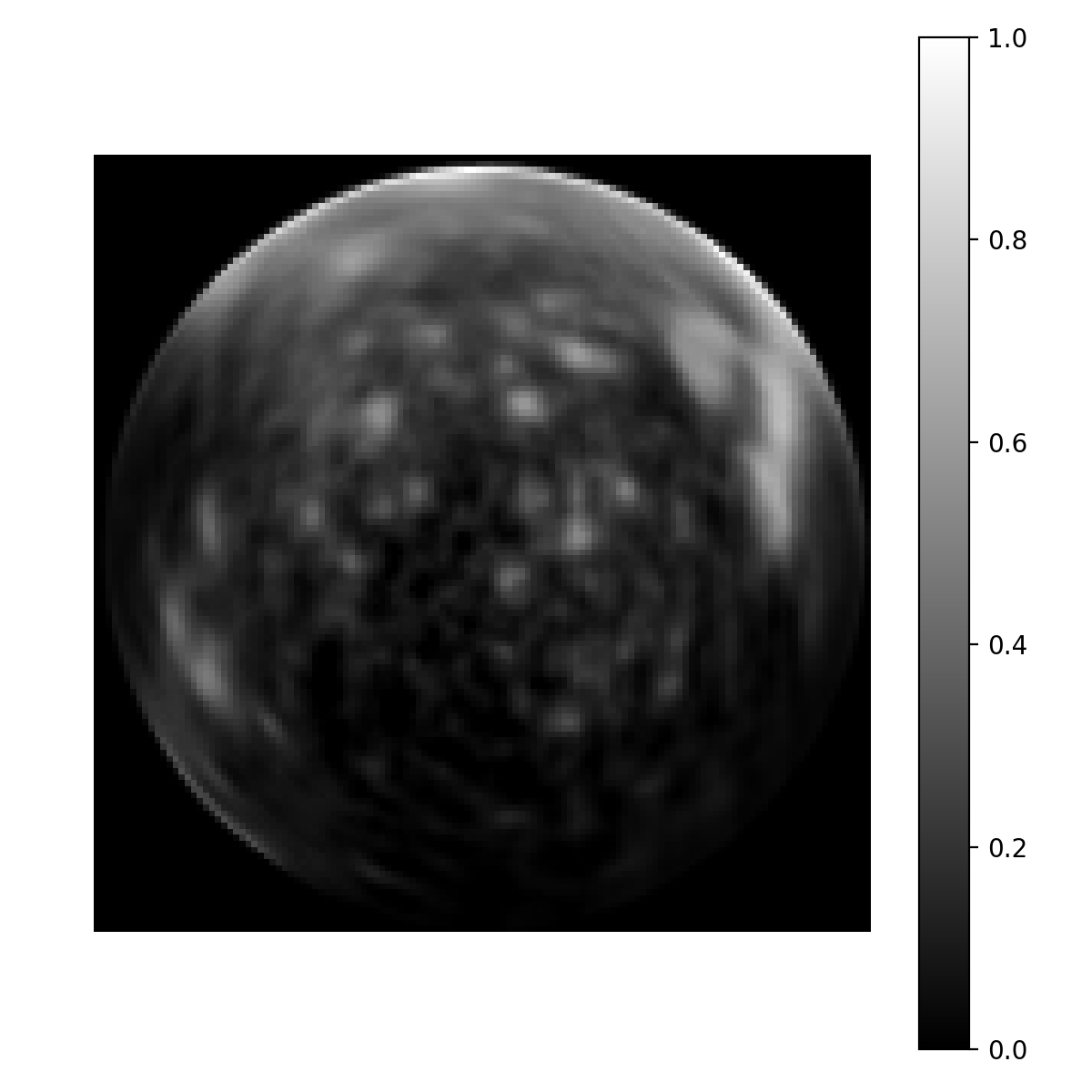}
        \caption{Raster scan baseline reconstruction} 
        \label{subfig: bio raster reconstruction}
    \end{subfigure}
    \hfil
    \begin{subfigure}[t]{0.19\linewidth}
        \centering
        \includegraphics[width=\linewidth]{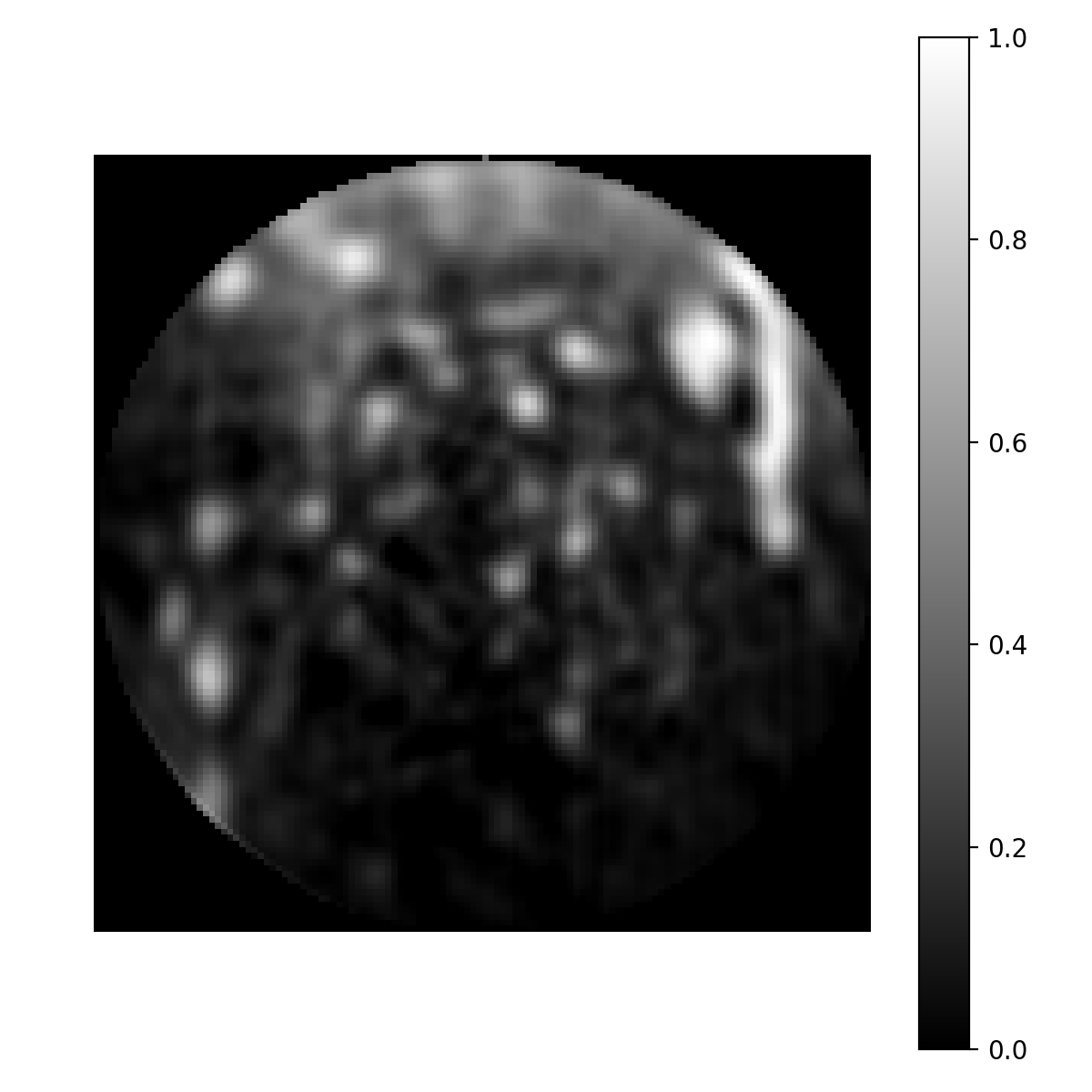}
        \caption{Bayesian baseline reconstruction} 
        \label{subfig: bio bayesian reconstruction}
    \end{subfigure}
    \hfil
    \begin{subfigure}[t]{0.19\linewidth}
        \centering
        \includegraphics[width=\linewidth]{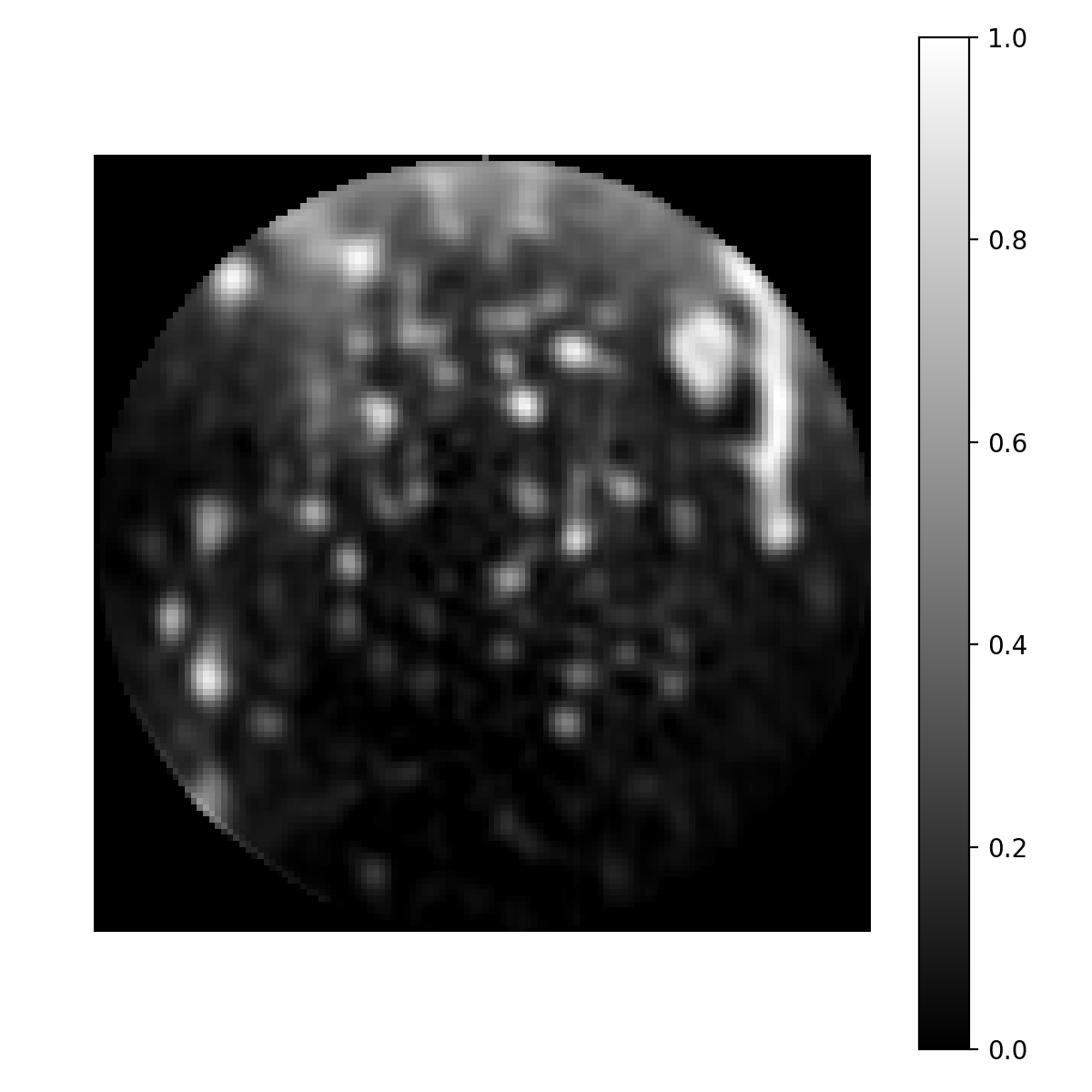}
        \caption{Beam selection reconstruction} 
        \label{subfig: bio beam beam reconstruction}
    \end{subfigure}
    \caption{Biological sample}
    \label{fig: bio sample}
\end{figure}

The next sample under consideration is a biological specimen, as shown in Figure \ref{fig: bio sample}(\subref{subfig: bio true}). This sample presents a unique challenge due to its sparsely distributed structural features. Because the edge promoting component of the proposed algorithm depends on interpolating the sinogram to identify edge locations, it becomes more difficult to accurately infer edge information when the true sample contains only sparse edges. This is reflected in the beam distribution histogram in Figure \ref{fig: bio sample}(\subref{subfig: bio beam distribution}), which captures primarily the large scale features of the sample, such as those in the upper region, particularly the upper right. In contrast, the finer scale features in the lower portion of the sample are not strongly emphasized in the beam distribution. Nevertheless, the proposed method still allocates a relatively uniform set of beams to explore this region.

\begin{figure}[H]
    \centering
    \includegraphics[width=\linewidth]{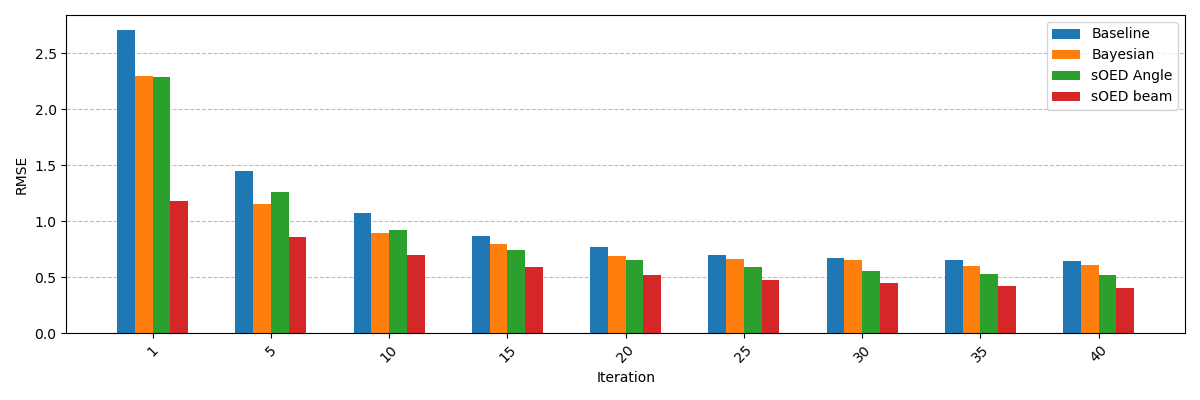}
    \caption{Comparison of RMSE between design of experiment methods on the biological sample}
    \label{fig: bio RMSE}
\end{figure}

Figure \ref{fig: bio RMSE} shows that the proposed beam selection remains superior to both baselines, particularly in the low measurement regime, and that the Bayesian baseline again outperforms the raster scan by prioritizing the more prominent edges. Between iterations 4 and 12 the Bayesian baseline slightly exceeds the proposed angle selection: with sparsely distributed edges, the sinogram-based edge mapping cannot capture all edge locations from only a few measurements, so the angle selection may miss some edge-aligned directions early on. As more measurements accrue, however, it ultimately surpasses the Bayesian baseline by balancing edge alignment with global exploration.
Overall, the proposed beam selection strategy consistently achieves the lowest RMSE across all iterations, demonstrating its effectiveness even in challenging scenarios characterized by sparse structural features.

\subsubsection{Sample 5: Siemens Star}
\begin{figure}[H]
    \centering
    \begin{subfigure}[t]{0.19\linewidth}
        \centering
        \includegraphics[width=\linewidth]{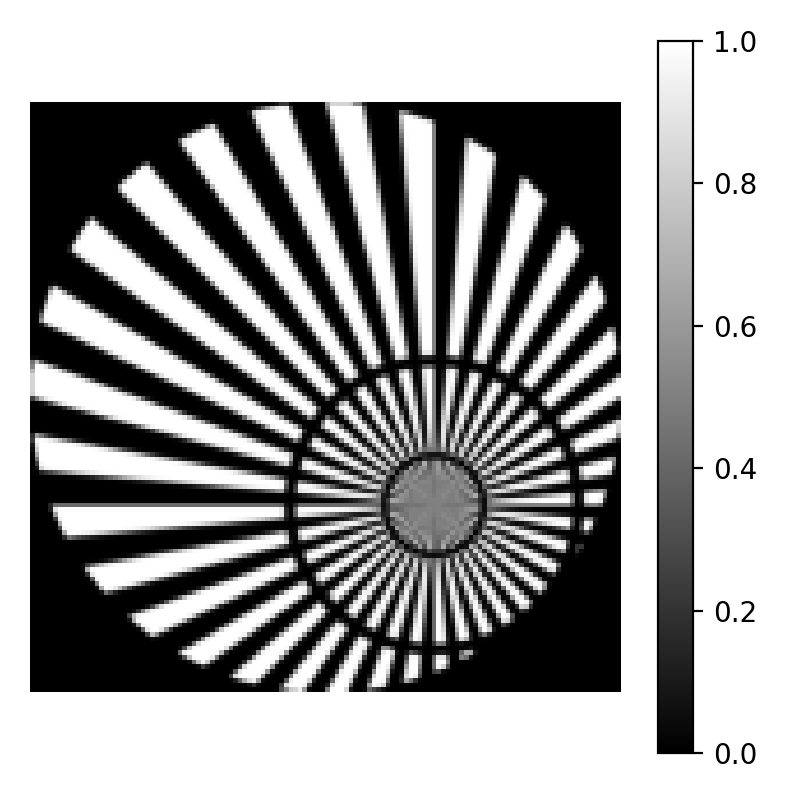}
        \caption{True sample} 
        \label{subfig: flower real}
    \end{subfigure}
    \hfil
    \begin{subfigure}[t]{0.19\linewidth}
        \centering
        \includegraphics[width=\linewidth]{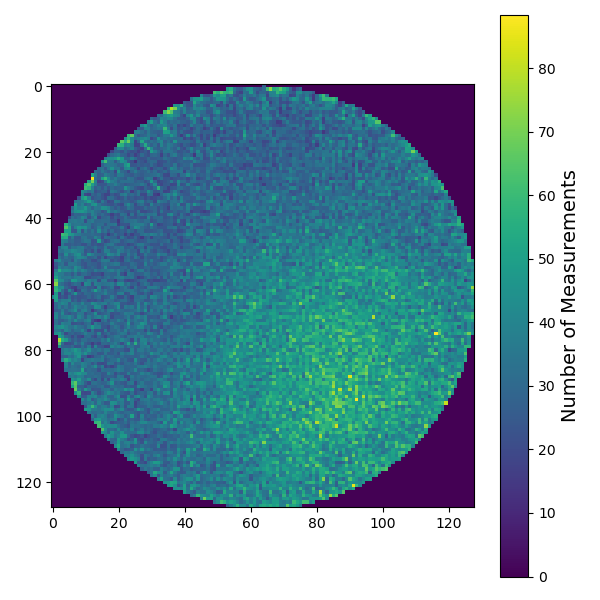}
        \caption{Beam distribution}
        \label{subfig: flower beam distribution}
    \end{subfigure}
    \hfil
    \begin{subfigure}[t]{0.19\linewidth}
        \centering
        \includegraphics[width=\linewidth]{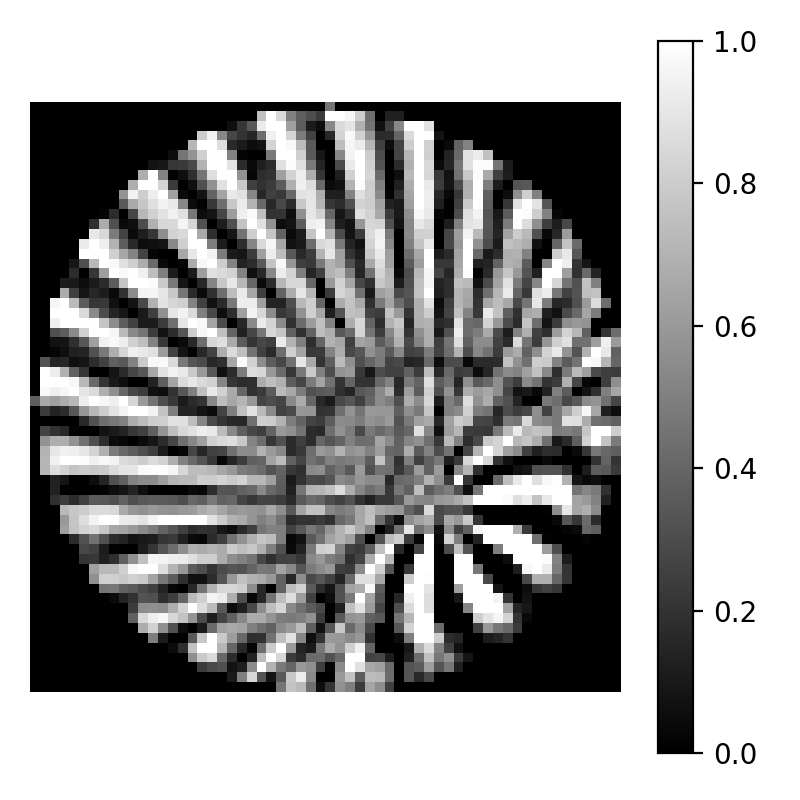}
        \caption{Raster scan baseline reconstruction} 
        \label{subfig: flower raster reconstruction}
    \end{subfigure}
    \hfil
    \begin{subfigure}[t]{0.19\linewidth}
        \centering
        \includegraphics[width=\linewidth]{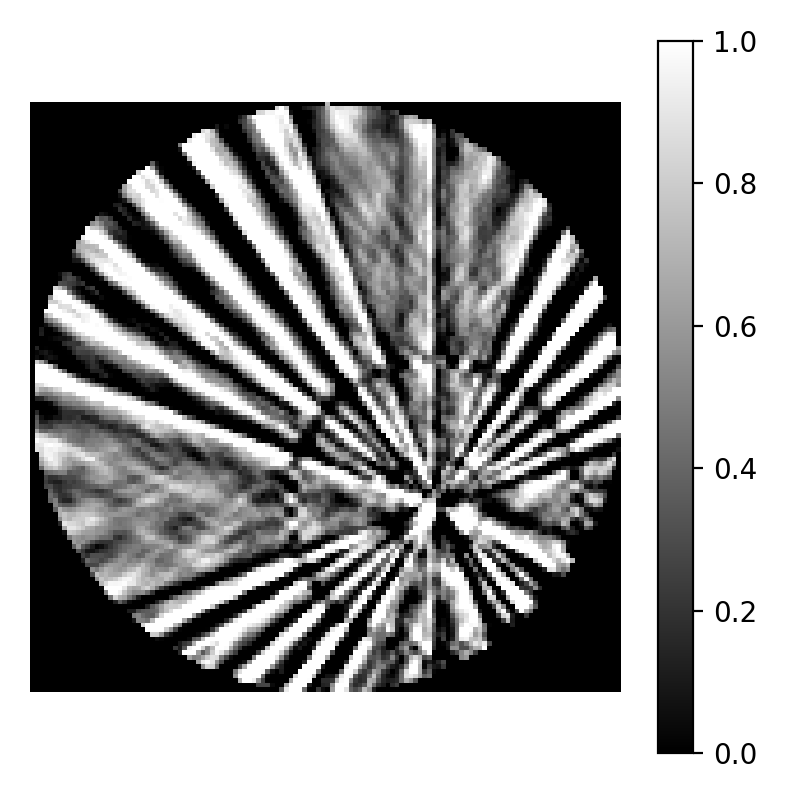}
        \caption{Bayesian baseline reconstruction} 
        \label{subfig: flower bayesian reconstruction}
    \end{subfigure}
    \hfil
    \begin{subfigure}[t]{0.19\linewidth}
        \centering
        \includegraphics[width=\linewidth]{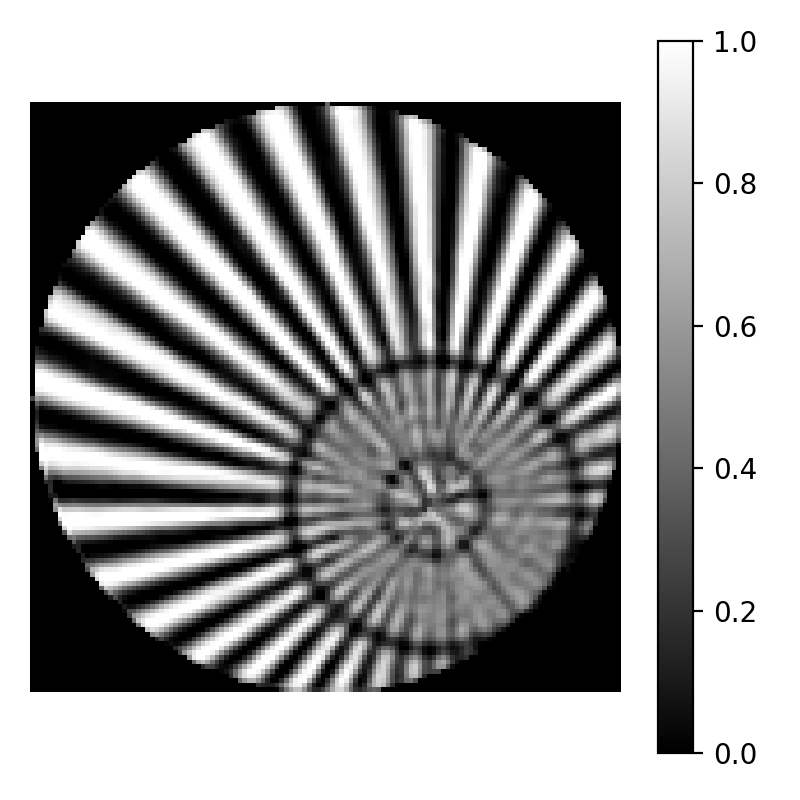}
        \caption{Beam selection reconstruction}
        \label{subfig: flower beam reconstruction}
    \end{subfigure}
    \caption{Siemens star sample}
    \label{fig: flower sample}
\end{figure}

The Siemens star shown in Figure \ref{fig: flower sample}(\subref{subfig: flower real}) contains high spatial-frequency content: 72 thin radial spokes that produce rapid angular variations in the sinogram, with the center intentionally offset toward the lower-right rather than the geometric center of the image. This pattern is especially challenging for evenly spaced raster scans with few projection angles. Figure \ref{fig: flower sample}(\subref{subfig: flower raster reconstruction}) shows the reconstruction from $25$ evenly spaced angles (the $20^{\text{th}}$ iteration of each sequential approach initialized with $5$ raster angles); the SART reconstruction exhibits noticeable artifacts near the star’s center, arising from the mismatch between the center of rotation and the star’s geometric center, compounded by the high spatial frequency of the spokes.

\begin{figure}[H]
    \centering
    \includegraphics[width=\linewidth]{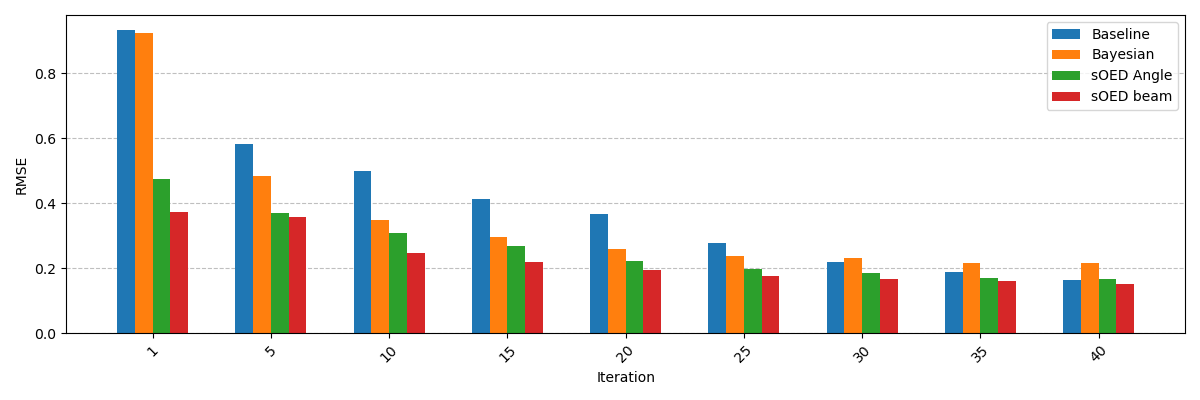}
    \caption{Comparison of RMSE between design of experiment methods on the Siemens star sample}
    \label{fig: flower RMSE}
\end{figure}

As shown in \Cref{fig: flower RMSE}, the proposed sinogram-based beam selection achieves a markedly faster reduction in RMSE compared with both the evenly spaced raster scan and the Bayesian baseline in the low-measurement regime. The beam distribution at the $20^\text{th}$ iteration, shown in Figure \ref{fig: flower sample}(\subref{subfig: flower beam distribution}), indicates that the beam selection algorithm is unable to resolve individual spokes of the Siemens star. This behavior is expected because, similar to the case discussed in \Cref{subsec: bio sample}, the algorithm cannot reliably interpolate edge information when the edge distribution is highly dense and changes rapidly between neighboring measurements. Nevertheless, the proposed method successfully places a higher density of beams near the center of the Siemens star, with a roughly uniform angular distribution around this region. This targeted placement leads to improved reconstruction quality using the same number of measurements, as illustrated in Figure \ref{fig: flower sample}(\subref{subfig: flower beam reconstruction}).

\subsection{Scaling of Beam Selection Algorithm}
\begin{figure}[H]
    \centering
    \begin{subfigure}[c]{0.45\columnwidth}
        \centering
        \includegraphics[width=\linewidth]{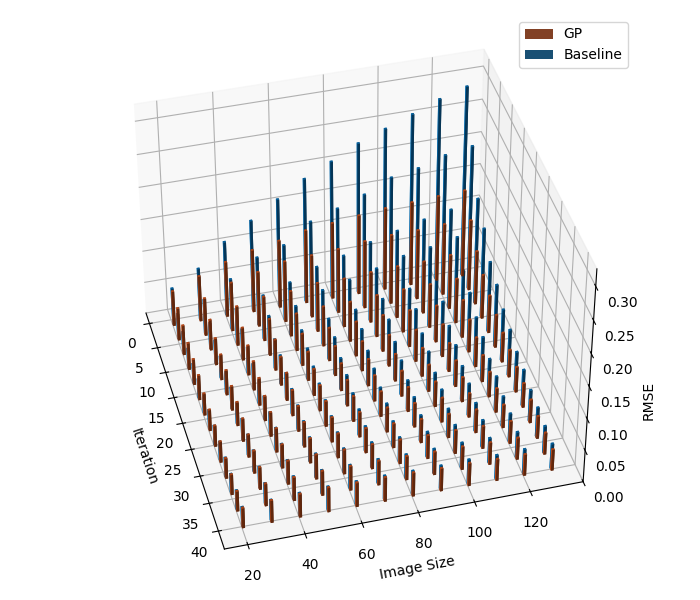}
    \end{subfigure}%
    \hfill
    \begin{subfigure}[t]{0.45\columnwidth}
        \centering
        \begin{subfigure}[t]{\linewidth}
            \centering
            \includegraphics[width=\linewidth]{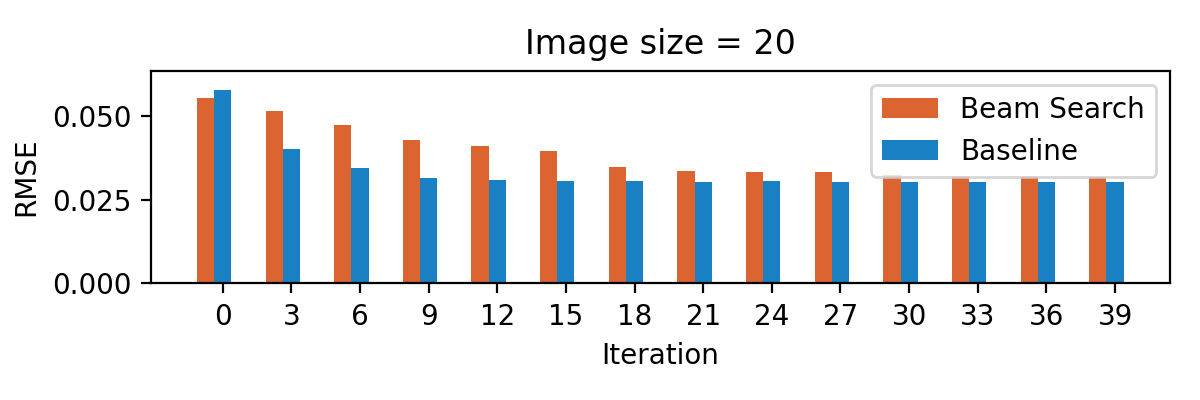}
        \end{subfigure}
        \vspace{0.5em}
        \begin{subfigure}[t]{\linewidth}
            \centering
            \includegraphics[width=\linewidth]{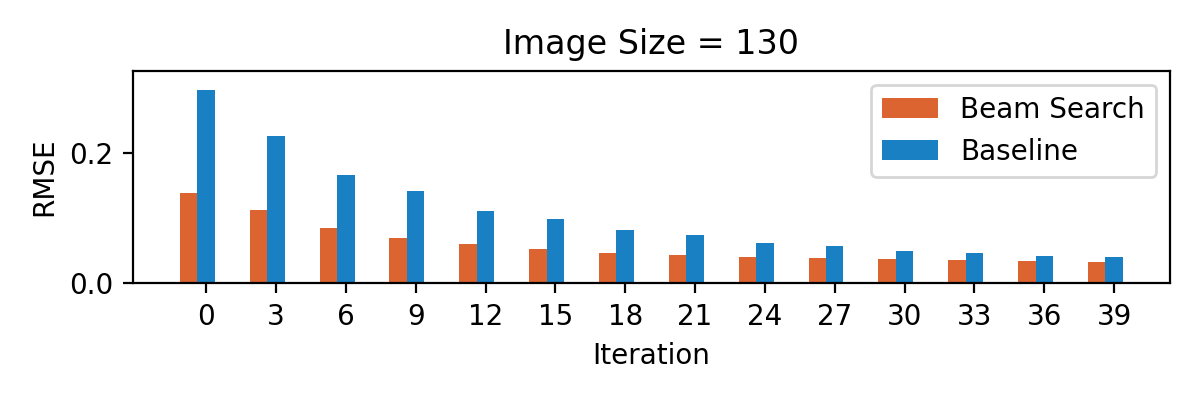}
        \end{subfigure}
    \end{subfigure}
    \caption{Scaling effect of beam selection sequential design of experiment.}
    \label{fig: phantom scale}
\end{figure}

The proposed beam–selection algorithm generally performs better at higher image resolutions. Since the number of beams per angle is tied to the pixel count ($n_r = \resolution$), increasing $\resolution$ enlarges the per-iteration measurement budget, broadening instantaneous sinogram coverage and making it more likely that boundary-intersecting rays are sampled early. Finer discretization also sharpens the GP surrogate and finite–difference edge metric: smaller $\Delta\theta$ and $\Delta r$ reduce gradient approximation error and yield narrower, more distinct edge peaks, strengthening the edge–alignment term in the acquisition function. Together these effects accelerate early-iteration RMSE reduction at higher resolutions (see Figure \ref{fig: phantom scale}). The trade–off is cost: as the number of acquired beams $m_k$ grows, GP updates and confidence–interval evaluations scale cubically naively—quadratically in $m_k$ with the incremental Schur–complement update of Section \ref{sec: Computational Analysis}—and very high resolution can oversmooth small features unless the kernel lengthscales and minimum–distance parameter $\Delta_{\Obj}$ are carefully tuned.

\begin{figure}[H]
    \centering
    \begin{subfigure}[t]{0.45\linewidth}
        \centering
        \includegraphics[width=\linewidth]{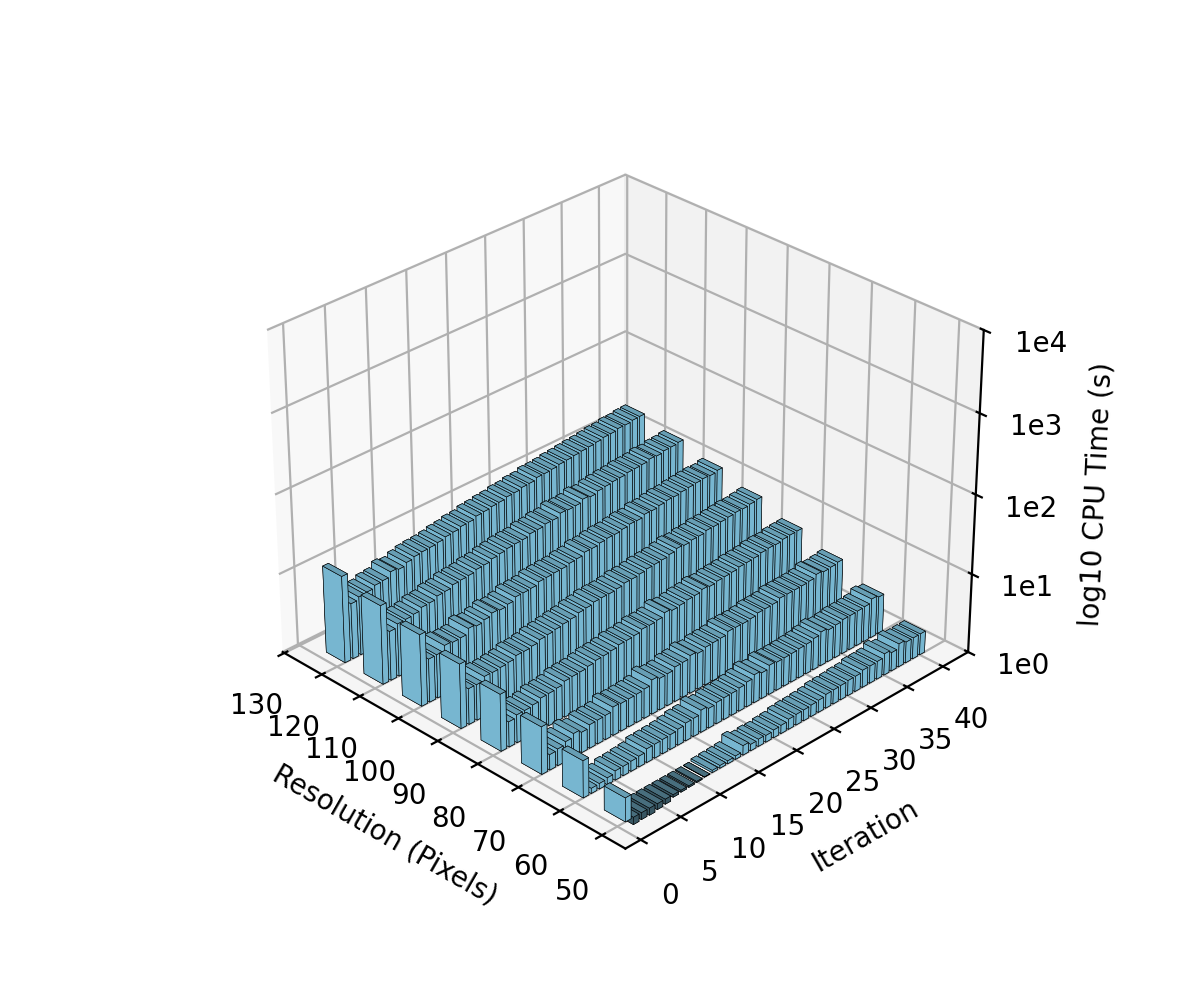}
        \caption{Computational time of proposed approach} 
    \end{subfigure}
    \hfil
    \begin{subfigure}[t]{0.45\linewidth}
        \centering
        \includegraphics[width=\linewidth]{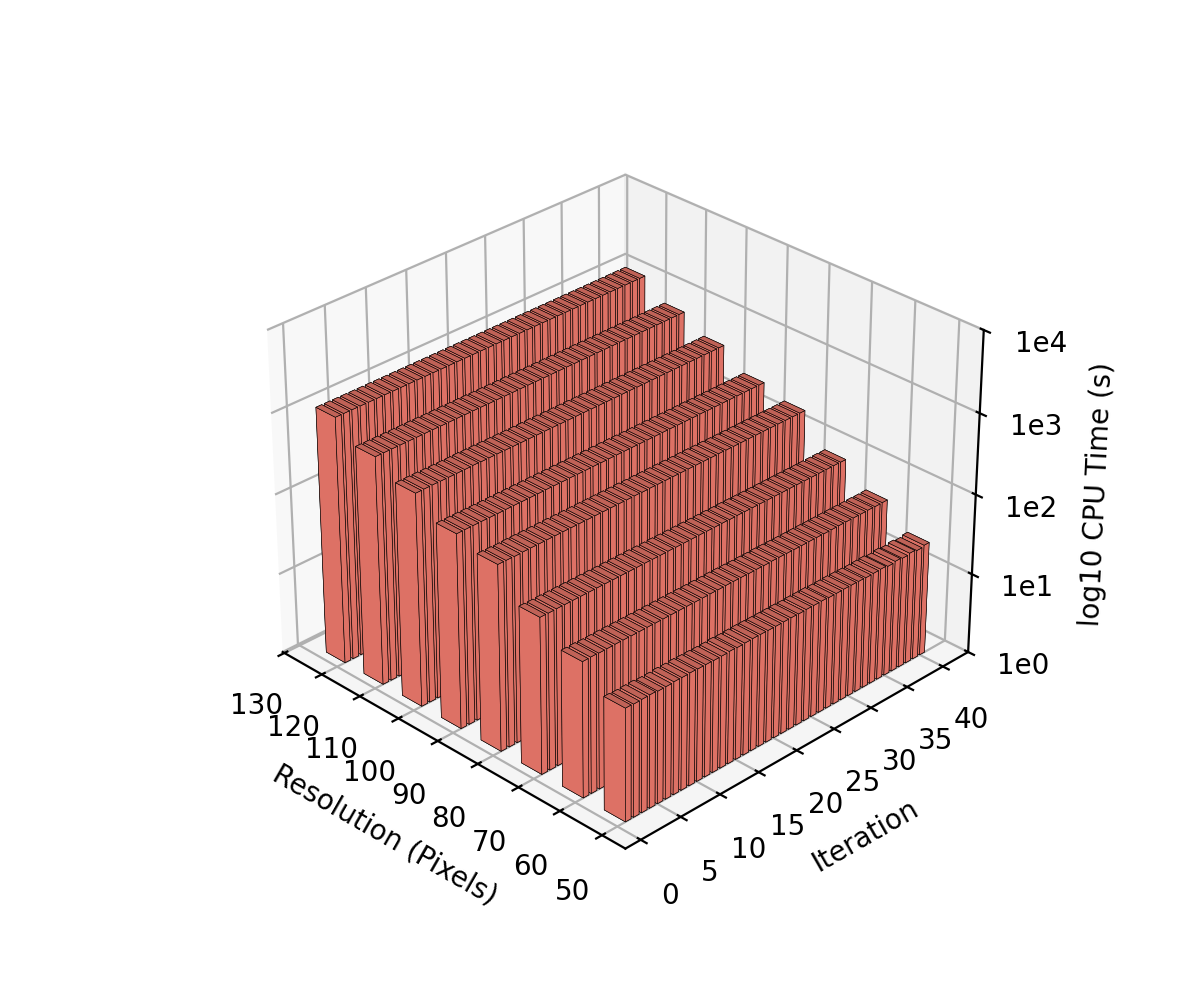}
        \caption{Computational time of Bayesian approach} 
    \end{subfigure}
    \caption{Computational time comparison}
    \label{fig: CPU time}
\end{figure}

The online computational cost per iteration of the proposed sinogram–GP approach is substantially lower than that of the Bayesian baseline, as shown in Figure \ref{fig: CPU time}. By leveraging incremental GP updates and inexpensive predictions over a finite candidate set, the per-iteration cost grows primarily with the current number of acquired beams (using efficient block updates) and the effort required to score candidates. In contrast, the Bayesian baseline requires repeated large Cholesky factorizations or determinant evaluations of high-dimensional posterior covariances, which scale with image resolution. In practice, this translates to orders-of-magnitude faster runtime per iteration for the proposed method, making it significantly more suitable for interactive or real-time beam selection.

\subsection{Noise Effect} 
\begin{figure}[H]
    \centering
    \begin{subfigure}[c]{0.45\columnwidth}
        \centering
        \includegraphics[width=\linewidth]{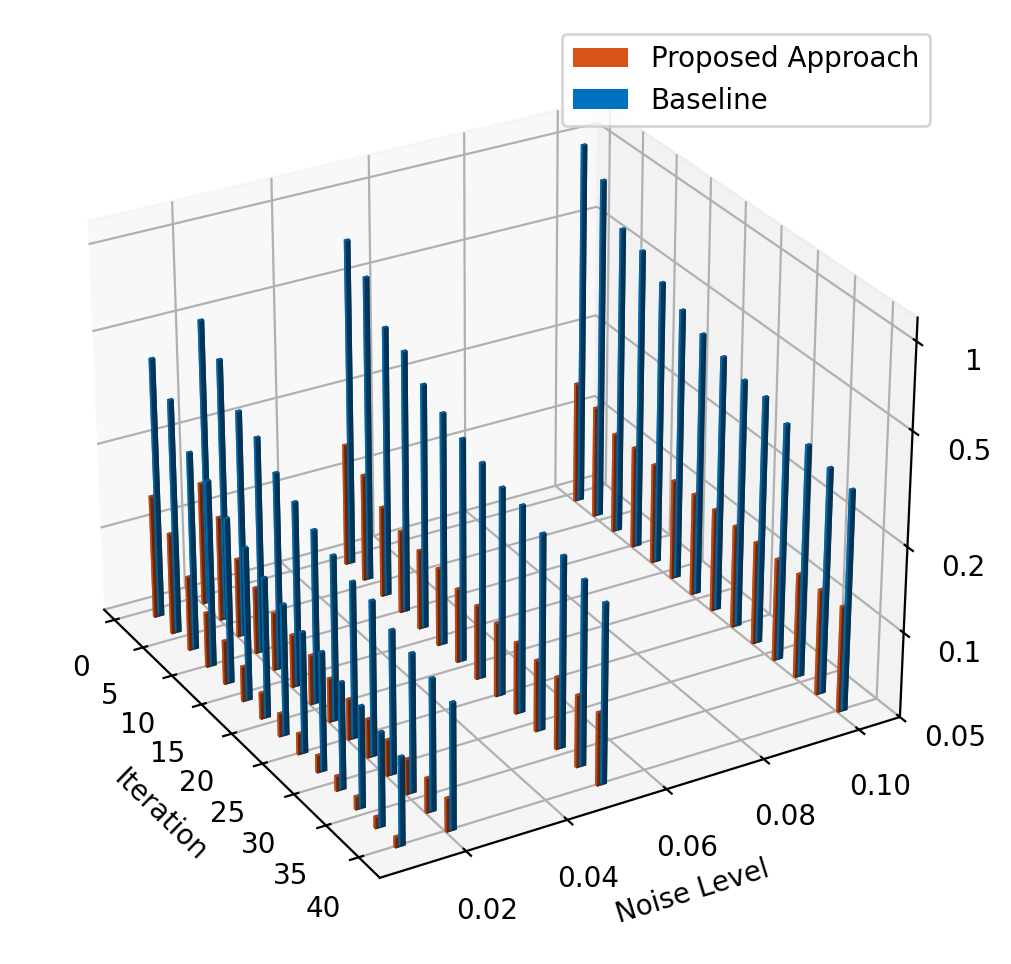}
    \end{subfigure}%
    \hfill
    \begin{subfigure}[t]{0.45\columnwidth}
        \centering
        \begin{subfigure}[t]{\linewidth}
            \centering
            \includegraphics[width=\linewidth]{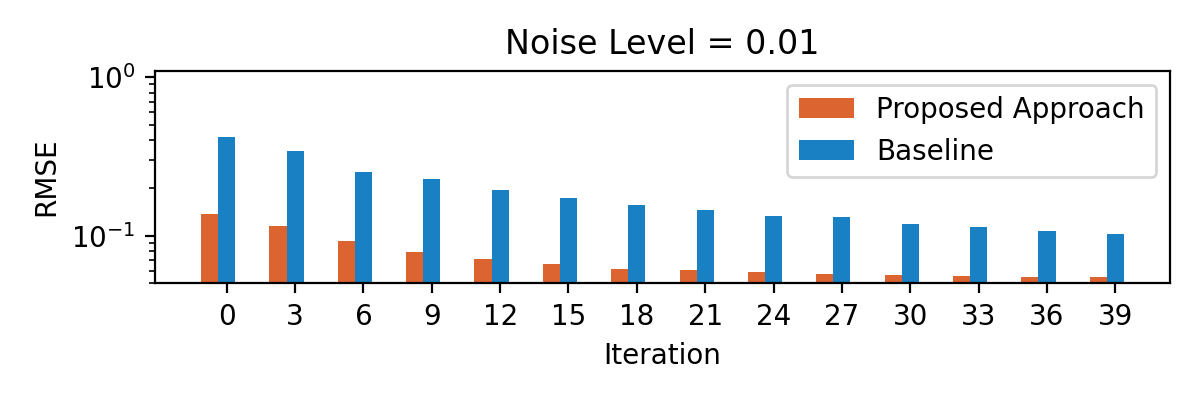}
        \end{subfigure}
        \vspace{0.5em}
        \begin{subfigure}[t]{\linewidth}
            \centering
            \includegraphics[width=\linewidth]{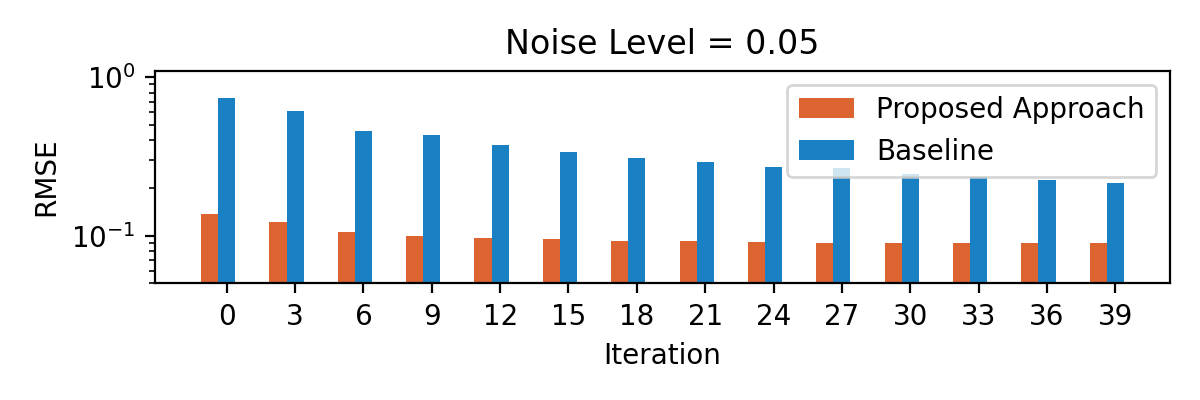}
        \end{subfigure}
    \end{subfigure}
    \caption{Robustness of the proposed beam-selection method to additive measurement noise.}
    \label{fig: Noise}
\end{figure}

An additional benefit of the proposed framework, derived from having imposed a GP on the sinogram, is that the predictive mean of the GP can act as a filter for measurement noise, as shown in Equation (\ref{eq: GP mean}). 
To demonstrate this, measurement noise is added and the noise level is defined as a fraction of the maximum noise-free signal along a single beam through the sample. As shown in Figure \ref{fig: Noise}, the reconstruction obtained from the proposed approach remains more accurate than the baseline even when the noise level reaches $5\%$. 


\section{Conclusion} \label{sec: conclusion}
This work presents a novel adaptive beam selection framework for scanning probe tomography that leverages edge information directly from the sinogram, eliminating the need for intermediate image reconstruction. 
By modeling the sinogram as a Gaussian process and designing an acquisition function that balances uncertainty with edge alignment, the proposed method efficiently identifies informative beams and reduces measurement redundancy. 
Numerical experiments demonstrate that this approach achieves improved reconstruction quality with the same number of measurements as conventional uniform sampling strategies. Although independent beam positioning introduces additional experimental overhead, the results underscore the method’s potential benefits and motivate future advances in adaptive scanning hardware.
\section*{Acknowledgment}
This work was supported in part by the U.S. Department of Energy, Office of Science,
Office of Advanced Scientific Computing Research Applied Mathematics Contract No. DE-AC02-06CH11357.

\bibliographystyle{unsrt}
\bibliography{Zotero_library_full.bib}

\clearpage
\section*{Supplementary Material}
\noindent The appendices below contain supplementary material for the manuscript. Equations in the appendices are numbered $(\mathrm{S}\cdot)$.

\setcounter{equation}{0}
\renewcommand{\theequation}{S\arabic{equation}} 
\renewcommand{\theHequation}{S\arabic{equation}} 
\appendices
\section{Proof of Theorem~\ref{theorem: edge radon} for a Single Convex Component}
\label{appendix: convex proof}

Consider the simplest case where the sample is a convex object. Since any convex region in Euclidean space can be approximated by a convex polyhedron \cite{bronshteyn1975}, denote this polyhedron by $H x \ge l$, where $H \in \R^{n_h \times 2}$ is the matrix representing the directions of the $n_h$ bounding half-planes, and $x\in\mathbb{R}^2$, as in the previous section. Let $H_i$ and $l_i$ be the $i^{\text{th}}$ row of $H$ and $l$ respectively. The positive indicator $\Indicator$ defined in (\ref{eq: Positive Indicator}) can be approximated by
\begin{equation} \label{eq: positive indicator approx}
    \IIndicator[Hx-l;\beta](x) = \prod_{i=1}^{n_h} \frac{1}{1 + \exp(-\beta (H_i x - l_i))},
\end{equation}
where the parameter $\beta$ controls the rate of transition from 0 to 1 at the boundary of the polyhedron. Note that $\lim_{\beta \to \infty} \IIndicator[Hx-l;\beta](x) = \Indicator[Hx-l](x)$, which implies that (\ref{eq: positive indicator approx}) provides a good approximation of (\ref{eq: Positive Indicator}) for sufficiently large $\beta$. Let $f_0$ denote the Lipschitz continuous function associated with the convex object considered in this subsection. The convex object is modeled as
\begin{equation} \label{eq: convex object model}
    f(x) = \Indicator[Hx -l](x)f_0(x) = \lim_{\beta \to \infty} \IIndicator[Hx-l;\beta](x)f_0(x).
\end{equation}
If $H_i x < l_i$ for some row $i$, then the point $x$ lies outside the object of interest. In this case, the indicator function equals 0 by definition, and therefore $\Edge[f;\alpha](x) = 0$. When $x$ lies in the strict interior of the object ($Hx > l$ component-wise), $\Indicator[Hx-l](x) = 1$ and $\nabla \Indicator[Hx-l](x) = 0$ in the sharp limit. Applying the product rule to (\ref{eq: convex object model}) thus yields
\begin{equation}
\begin{split}
    \nabla f(x) &= \left(\nabla \Indicator[Hx -l](x)\right)f_0(x) \\
    &\quad + \Indicator[Hx -l](x)\nabla f_0(x) = \nabla f_0(x).
\end{split}
\end{equation}
Let $K_0$ be the Lipschitz constant of $f_0(x)$. When $Hx - l > 0$, since the inequality $1 - \exp \left(-\alpha\left\| \nabla f_0 (x) \right\|^2\right) \le 1 - \exp \left(-\alpha K_0^2 \right)$ holds and  $\Edge[f](x) \ge 0$ by definition, the edge-mapping function of $f$ is given by 
\begin{equation}
\begin{split}
    0 \leq \Edge[f](x) &= \lim_{\alpha \to 0}  1 - \exp \left(-\alpha\left\| \nabla f_0 (x) \right\|^2\right) \\
    &\leq \lim_{\alpha \to 0}  1 - \exp \left(-\alpha K_0^2 \right) = 0.
\end{split}
\end{equation}
Thus, in regions away from the object boundaries, the edge-mapping function of the image always evaluates to 0. 
Now consider the evaluation of the edge-mapping function on the boundary. If the polyhedron is expressed in minimal H-representation, that is, if there are no redundant inequalities describing the set, a boundary corresponding to row $i$ of $H$ and $l_i$ is denoted as $\Boundary_{H,i} = \left\{ x \;\middle|\; H_i x = l_i, \; H_j x \ge l_j \; \forall j \neq i \right\}$. At a point $x \in \Boundary_{H,i}$, the smoothed surrogate (\ref{eq: positive indicator approx}) gives $\IIndicator[Hx-l;\beta](x) \to \tfrac12$ as $\beta\to\infty$, while $\nabla \IIndicator[Hx-l;\beta](x)$ diverges in the direction $H_i^\top$. Applying the product rule to $f = \IIndicator\, f_0$ gives
\begin{equation} \label{eq: edge gradient}
    \begin{aligned}
        \nabla f(x) &= \left(\nabla \IIndicator[Hx-l;\beta](x)\right)f_0(x) \\
        &\quad + \IIndicator[Hx-l;\beta](x)\,\nabla f_0(x).
    \end{aligned}
\end{equation}
The bounded term $\IIndicator\,\nabla f_0$ contributes at most $K_0$ in norm uniformly in $\beta$, while a direct computation of the sigmoid derivative gives, on $\Boundary_{H,i}$,
\begin{equation} \label{eq: edge gradient evaluated}
    \begin{aligned}
        \nabla \IIndicator[Hx-l;\beta](x)\,f_0(x) &= \frac{\beta H_i^\top\, e^{-\beta(H_ix-l_i)}}{\left(1+e^{-\beta(H_ix-l_i)}\right)^2}\,f_0(x) \\
        &= \tfrac{1}{4}\,\beta H_i^\top f_0(x),
    \end{aligned}
\end{equation}
which diverges as $\beta \to \infty$ whenever $f_0(x)\neq 0$.

From (\ref{eq: edge gradient})--(\ref{eq: edge gradient evaluated}), the edge-mapping function evaluated on the boundary takes the form
\begin{equation} \label{eq: edge map alpha beta}
    \begin{aligned}
        \Edge[f;\alpha](x) &= 1 - \exp \left(-\alpha\left\| \nabla f (x) \right\|^2\right) \\
        &= 1 - \exp \Big(-\big\| \sqrt{\alpha}\,\IIndicator(x)\,\nabla f_0(x) \\
        &\qquad\quad +  \tfrac{1}{4} \sqrt{\alpha}\,\beta\, H_i^\top f_0(x) \big\|^2\Big).
    \end{aligned}
\end{equation}

The naive joint limit $\lim_{\alpha\to 0}\lim_{\beta\to\infty}$ is indeterminate because the term $\sqrt{\alpha}\beta$ has the form $0 \cdot \infty$. Recall that $\alpha$ controls the smoothing of the edge map within each component (so that regions where $\nabla f_0$ is bounded converge to zero), while $\beta$ controls the sharpness of the indicator at component boundaries. Resolving the indeterminacy requires specifying which effect dominates. We make the choice $\alpha = \beta^{-1}$, so that $\sqrt{\alpha}\beta = \sqrt{\beta}\to\infty$ at the boundary while $\sqrt{\alpha}\to 0$ inside each component. Under this scaling, evaluation at $x \in \Boundary_{H,i}$ with $f_0(x)\neq 0$ yields
\begin{equation}
\begin{multlined}
    \Edge[f](x) := \lim_{\beta \to \infty} \Edge\!\left[f;\beta^{-1}\right](x) \\
    = \lim_{\beta \to \infty} 1 - \exp \Big(-\big\| \tfrac{1}{\sqrt{\beta}}\,\IIndicator(x)\,\nabla f_0(x) \\
    + \tfrac{1}{4} \sqrt{\beta}\, H_i^\top f_0(x) \big\|^2\Big) = 1.
\end{multlined}
\end{equation}
Combining all the results above, we arrive at the following edge mapping for a convex object defined by any Lipschitz continuous function on the region constrained by the polyhedron $Hx \ge l$: 
\begin{equation} \label{eq: edge of image}
\begin{aligned}
    \Edge[f](x) &= \Edge[\Indicator[Hx -l](x)f_0(x)](x) \\
    &= \begin{cases}
        1 & \text{if } x \in \Boundary_{H,i} \text{ for some } i\leq n_h, \\
        0 & \text{otherwise.}
    \end{cases}
\end{aligned}
\end{equation}

Let $\measure(\Boundary_{H,i})$ denote the length of the boundary $\Boundary_{H,i}$ of the polyhedron $H x \ge l$. A straightforward way to compute this is to convert the H-representation of the polyhedron to its V-representation with the double description method\cite{fukuda1996}, obtain all vertices, identify the two vertices that lie on $\Boundary_{H,i}$, and then compute the Euclidean distance between these two vertices. 
The result in (\ref{eq: edge of image}) implies that the edge-mapping of the image equals 1 if and only if the point lies on the boundary separating the sample from the background. Consequently, if the line integral of the Radon transform intersects the boundary only at isolated points without overlapping, the integral evaluates to zero. In contrast, if the line integral overlaps with the boundary, the value of the integral is exactly the length of the boundary segment of the convex region.
Thus, the following expression is obtained for the Radon transform of the edge-mapping of a convex object.
\begin{equation} \label{eq: Radon of Edge}
    \Radon[\Edge[f]](\theta,r) = \begin{cases}
        \measure(\Boundary_{H,i}) & \text{if } \Boundary_{H,i} \subset L_{\theta,r} \\
        & \quad \text{for some } i\leq n_h, \\
        0 & \text{otherwise.}
    \end{cases}
\end{equation}

In order to formalize the relationship between the edge-mapping function and the Radon transform, the edge-mapping of the Radon transform must be computed and compared to the Radon transform of the edge-mapping shown in (\ref{eq: Radon of Edge}). Since the edge-mapping is defined in terms of the gradient in (\ref{eq: edge map}), the next step is to derive the gradient of the Radon transform with respect to the angle $\theta$ and the line distance $r$. We employ the reparameterization of the line $L_{\theta,r}$ in terms of the signed distance $z$ along the line from the projection of the origin:
\begin{equation} \label{eq:line_reparam}
\begin{aligned}
    L_{\theta,r} ={}& \Big\{ x \in \R^2 \;\Big|\; x(\theta,r,z) = \begin{bmatrix}
        r\cos(\theta) - z\sin(\theta) \\
        r\sin(\theta) + z\cos(\theta)
    \end{bmatrix} \\
    &\quad \forall z \in \R \Big\}.
\end{aligned}
\end{equation}
The derivative of each coordinate $x$ along the line $L_{\theta,r}$ is given by 
\begin{align}
    \frac{\partial x_1}{\partial \theta}(\theta,r,z) &= -r\sin(\theta) - z\cos(\theta) = -x_2(\theta,r,z), \\
    \frac{\partial x_2}{\partial \theta}(\theta,r,z) &= r\cos(\theta) - z\sin(\theta)  = x_1(\theta,r,z), \\
    \frac{\partial x_1}{\partial r}(\theta,r,z) &= \cos(\theta), \text{ and } \\
    \frac{\partial x_2}{\partial r}(\theta,r,z) &= \sin(\theta).
\end{align}
Hence, the derivative of the Radon transform with respect to the angle $\theta$ is expressed as
\begin{equation} \label{eq: dRadon_dtheta}
\begin{aligned}
    \frac{\partial }{\partial \theta}\Radon[f](\theta,r) &= \int_{-1}^{1} \frac{\partial}{\partial \theta}f(r\cos(\theta) - z\sin(\theta), \\
    &\qquad\qquad r\sin(\theta) + z\cos(\theta) ) \, dz \\
    & = \int_{-1}^{1} \frac{\partial f}{\partial x_1}\frac{\partial x_1}{\partial \theta} + \frac{\partial f}{\partial x_2}\frac{\partial x_2}{\partial \theta} \, dz \\
    & = \int_{-1}^{1} -x_2(\theta,r,z)\frac{\partial f}{\partial x_1} \\
    &\qquad\quad + x_1(\theta,r,z)\frac{\partial f}{\partial x_2}\, dz \\
    &= \Radon \left[-x_2 \frac{\partial f}{\partial x_1}+ x_1 \frac{\partial f}{\partial x_2}\right](\theta,r).
\end{aligned}    
\end{equation}

Consider the case where $\Boundary_{H,i} \not\subset L_{\theta,r}$ for all $i$. In this case, the line $L_{\theta,r}$ intersects the boundary of the object at at most two points, corresponding to parameter values $z^{\min}$ and $z^{\max}$ in the parameterization (\ref{eq:line_reparam}). Since the Radon transform is defined as a Lebesgue integral, and the Lebesgue integrals of a function over an open interval and its closure coincide, we can rewrite (\ref{eq: dRadon_dtheta}) as
\begin{equation}
    \frac{\partial }{\partial \theta}\Radon[f](\theta,r) = \int_{z\in (z^{\min},z^{\max})} -x_2 \frac{\partial f}{\partial x_1}+ x_1 \frac{\partial f}{\partial x_2} \, dz.
\end{equation}
Since $f(x) = \Indicator[Hx-l](x)\,f_0(x)$ with $f_0$ Lipschitz continuous with constant $K_0$, every $x(\theta,r,z)$ with $z\in(z^{\min},z^{\max})$ lies in the interior of the polyhedron, so $\Indicator[Hx-l](x)=1$ and $\nabla f(x) = \nabla f_0(x)$. The Lipschitz bound gives $\left|\partial_{x_1} f\right|\le K_0$ and $\left|\partial_{x_2} f\right|\le K_0$. Combined with $|x_1|, |x_2| \le 1$ on the unit disk, this yields
\begin{equation} \label{eq: dRadon_dtheta bound}
\begin{aligned}
    \left|\frac{\partial }{\partial \theta}\Radon[f](\theta,r)\right| &\leq \int_{z\in (z^{\min},z^{\max})} K_0\,(|x_1| + |x_2|)\, dz \\
    &\leq 2K_0\,(z^{\max} - z^{\min}).
\end{aligned}
\end{equation}

Since the object is defined in a unit disk of radius 1, $z^{\max} - z^{\min} \leq 2$, hence
\begin{equation}\label{eq: dRadon_dtheta upper bound}
    \left|\frac{\partial }{\partial \theta}\Radon[f](\theta,r)\right| \leq 4K_0.
\end{equation}

Similarly, the derivative of the Radon transform with respect to the line distance $r$ is expressed as
\begin{equation}\label{eq: dRadon_dr}
\begin{aligned}
    \frac{\partial }{\partial r}\Radon[f](\theta,r) &= \int_{-1}^{1} \frac{\partial}{\partial r}f(r\cos(\theta) - z\sin(\theta), \\
    &\qquad\qquad r\sin(\theta) + z\cos(\theta) ) \, dz \\
    & = \int_{-1}^{1} \frac{\partial f}{\partial x_1}\frac{\partial x_1}{\partial r} + \frac{\partial f}{\partial x_2}\frac{\partial x_2}{\partial r} \, dz \\
    & = \int_{-1}^{1} \cos(\theta)\frac{\partial f}{\partial x_1}+ \sin(\theta)\frac{\partial f}{\partial x_2}\, dz \\
    &= \Radon \left[\cos(\theta)\frac{\partial f}{\partial x_1}+ \sin(\theta)\frac{\partial f}{\partial x_2}\right](\theta,r).
\end{aligned}    
\end{equation}

Consider the case where $\Boundary_{H,i} \not\subset L_{\theta,r}$ for all $i$. Similarly to (\ref{eq: dRadon_dtheta bound}), we obtain the following upper bound on $\frac{\partial }{\partial r}\Radon[f](\theta,r)$:
\begin{equation} \label{eq: dRadon_dr upper bound}
    \begin{aligned}
        \left|\frac{\partial }{\partial r}\Radon[f](\theta,r)\right| &\leq \int_{z\in (z^{\min},z^{\max})} K_0\,(|\cos(\theta)| \\
        &\qquad\qquad\quad + |\sin(\theta)|) \, dz \\
        &\leq K_0\sqrt{2}\,(z^{\max} - z^{\min}) \\
        &\leq 2\sqrt{2}\,K_0.
    \end{aligned}
\end{equation}

The edge-mapping function of the Radon transform can be evaluated by substituting (\ref{eq: dRadon_dtheta}) and (\ref{eq: dRadon_dr}) into the edge-mapping definition (\ref{eq: edge map}), giving
\begin{equation}
    \begin{aligned}
    \Edge[\Radon[f]](\theta,r) &=  \lim_{\alpha \to 0} 1 - \exp \left(-\alpha\left\| \nabla \Radon[f](\theta,r) \right\|^2\right) \\
    &= \lim_{\alpha \to 0} 1 - \exp \Big(-\Big(\left(\sqrt{\alpha }\frac{\partial \Radon}{\partial \theta}\right)^2 \\
    &\qquad\qquad\quad + \left(\sqrt{\alpha } \frac{\partial \Radon}{\partial r}\right)^2\Big)  \Big)
    \end{aligned}
\end{equation}

Consider the case where $\Boundary_{H,i} \not\subset L_{\theta,r}$ for all $i$. An upper bound for $\Edge[\Radon[f]]$ is obtained by substituting (\ref{eq: dRadon_dtheta upper bound}) and (\ref{eq: dRadon_dr upper bound}) into the edge-mapping definition (\ref{eq: edge map}):
\begin{equation}
\begin{split}
    0 \leq  \Edge[\Radon[f]](\theta,r) &\leq  \lim_{\alpha \to 0} 1 - \exp \left(-\alpha\,\bigl(16 K_0^2 + 8 K_0^2\bigr) \right) \\
    &= 0.
\end{split}
\end{equation}

Consider the case where $\Boundary_{H,i} \subset L_{\theta,r}$ for some $i$. Since $\left(\sqrt{\alpha }\,\partial_\theta\Radon\right)^2 \ge 0$, dropping that contribution and substituting the boundary gradient (\ref{eq: edge gradient evaluated}) into (\ref{eq: dRadon_dr}) gives the lower bound
\begin{equation}
\begin{aligned}
&\Edge[\Radon[f]](\theta,r) \\
&\ge \lim_{\alpha \to 0} 1 - \exp \left(-\left(\sqrt{\alpha } \frac{\partial \Radon}{\partial r}\right)^2 \right)\\
&= \lim_{\alpha \to 0} 1 - \exp \Big(-\Big(\sqrt{\alpha}
\int_{z\in (z^{\min},z^{\max})} \\
&\qquad\qquad\quad [\cos(\theta) \sin(\theta)] \nabla f \, dz
\Big)^2 \Big) \\
&= \lim_{\alpha \to 0} \lim_{\beta \to \infty} 1 - \exp \Big(-\Big( \\
&\qquad
[\cos(\theta) \sin(\theta)]\int_{z\in (z^{\min},z^{\max})} \\
&\qquad\qquad \sqrt{\alpha}\,\IIndicator(x)\,\nabla f_0(x) \\
&\qquad\qquad + \tfrac{1}{4} \sqrt{\alpha}\,\beta\, H_i^{\top} f_0(x) \, dz
\Big)^2 \Big) \\
&\ge \lim_{\alpha \to 0} \lim_{\beta \to \infty} 1 - \exp \Big(-\Big( \\
&\qquad
\int_{z\in (z^{\min},z^{\max})} \tfrac{1}{4} \sqrt{\alpha}\,\beta \\
&\qquad\qquad \times [\cos(\theta) \sin(\theta)] H_i^{\top} f_0(x) \, dz
\Big)^2 \Big).
\end{aligned}
\end{equation}

Since $\Boundary_{H,i} \subset L_{\theta,r}$, the boundary segment and the beam $L_{\theta,r}$ are collinear, so their normals are parallel: the row $H_i$ is a scalar multiple of the line normal $[\cos(\theta)\;\sin(\theta)]$ (the normal of $L_{\theta,r}$ from its definition in Section~\ref{sec: Background}; the line's direction is $[-\sin(\theta)\;\cos(\theta)]$). Taking the rows of $H$ to be unit normals, which is without loss of generality by jointly rescaling each $H_i$ and $l_i$ so that the polyhedron $Hx\ge l$ is unchanged, and orienting the $(\theta,r)$ parameterization so the normals agree in sign, gives $H_i=[\cos(\theta)\;\sin(\theta)]$ and hence $[\cos(\theta) \sin(\theta)]\, H_i^{\top}=\cos^2(\theta)+\sin^2(\theta) =1$, yielding
\begin{equation}
\begin{aligned}
    &\Edge[\Radon[f]](\theta,r) \\
    &\ge \lim_{\alpha \to 0} \lim_{\beta \to \infty} 1 - \exp \Big(-\Big( \\
    &\qquad \int_{z\in (z^{\min},z^{\max})} \tfrac{1}{4} \sqrt{\alpha}\,\beta\, f_0(x) \, dz \Big)^2 \Big).
\end{aligned}
\end{equation}
Similar to (\ref{eq: edge map alpha beta}), the indeterminate form of the double limit of $\sqrt{\alpha}\beta$ is resolved by choosing $\alpha = \beta^{-1}$:
\begin{equation}
\begin{aligned}
    &\Edge[\Radon[f]](\theta,r) \\
    &\ge \lim_{\beta \to \infty} 1 - \exp \Big(-\Big( \\
    &\qquad \int_{z\in (z^{\min},z^{\max})} \tfrac{1}{4} \sqrt{\beta}\, f_0(x) \, dz \Big)^2 \Big) = 1.
\end{aligned}
\end{equation}
Since $\Edge[\Radon[f]](\theta,r) = 1 - \exp(-\|\nabla \Radon[f](\theta,r)\|^2) \le 1$ for all arguments, this lower bound is attained with equality, so $\Edge[\Radon[f]](\theta,r) = 1$ on such lines.

Overall, the value for the edge-mapping of the Radon transform is
\begin{equation} \label{eq: Edge of Radon}
     \Edge[\Radon[f]](\theta,r) = \left\{\begin{array}{l}
                                            1 \;  \text{if} \; \Boundary_{H,i}\subset L_{\theta,r}  \text{ for some } i\leq n_h, \\
                                            0 \; \text{otherwise.}
                                        \end{array}\right.
\end{equation}
Combining (\ref{eq: Radon of Edge}) and (\ref{eq: Edge of Radon}) and applying the sharp-indicator limit $\Indicator[\,\cdot\,] = \lim_{\gamma\to\infty}\Indicator[\,\cdot\,;\gamma]$ to the right-hand side of (\ref{eq: Radon of Edge}), since $\Radon[\Edge[f]](\theta,r)\in\{0,\measure(\Boundary_{H,i})\}$, applying $\Indicator$ maps the non-zero value to $1$ and leaves zero unchanged, yielding
\begin{equation} \label{eq: edge-radon equality}
    \Edge[\Radon[f]](\theta,r) = \Indicator[\Radon[\Edge[f]]](\theta,r),
\end{equation}
which establishes Theorem \ref{theorem: edge radon} for the convex case.

\section{Full Derivation of the Bayesian Sequential Design Baseline}
\label{appendix: bayesian baseline}
This appendix provides the full probabilistic derivation underlying the Bayesian sequential design baseline of Section~\ref{subsec: Bayesian Angle sOED}, comprising the Gaussian conditioning, the Kalman/GP posterior update, and the posterior predictive distribution from which the closed-form predictive covariance (\ref{eq: posterior covar candidate}) is obtained. We use the notation of Section~\ref{subsec: Bayesian Angle sOED} throughout.

Recalling that the Radon transform is linear in the sample for the fixed measurement matrix $\A_k$, at iteration $k$ the joint distribution of the latent sample and the noisy measurement is
\begin{equation}
\begin{aligned}
    &\left[\begin{array}{c}
        \sample_{k|k-1} \\
        \sinogram_{k|k-1}
    \end{array}\right]
    \sim
    \mathcal{N}\Big(
        \left[
            \begin{array}{c}
                \sampleMean_{k|k-1} \\
                \A_k \sampleMean_{k|k-1}
            \end{array}
        \right], \\
    &\qquad
        \left[
            \begin{array}{c c}
                \CoVar_{k|k-1} & \CoVar_{k|k-1} \A_k^\top \\
                \A_k \CoVar_{k|k-1} & \A_k \CoVar_{k|k-1} \A_k^\top + \CoVar_e
            \end{array}
        \right]
    \Big).
\end{aligned}
\end{equation}

After a new measurement is performed, the conditional distribution of the sample given the newly acquired sinogram is \(\sample_k \mid \sinogram_k \sim \mathcal{N}(\sampleMean_{k \mid k}, \CoVar_{k\mid k}) \), with the standard Kalman/GP update
\begin{align}
    \sampleMean_{k \mid k} &= \sampleMean_{k|k-1} + \CoVar_{k|k-1} \A_k^\top \nonumber \\
    &\quad \times \left[\A_k \CoVar_{k|k-1} \A_k^\top + \CoVar_e\right]^{-1} \nonumber \\
    &\quad\quad \times \left(\sinogram_k - \A_k \sampleMean_{k|k-1}\right),  \\
    \CoVar_{k\mid k} &= \CoVar_{k|k-1} - \CoVar_{k|k-1} \A_k^\top \nonumber \\
    &\quad \times \left[\A_k \CoVar_{k|k-1} \A_k^\top + \CoVar_e\right]^{-1} \nonumber \\
    &\quad\quad \times \A_k \CoVar_{k|k-1}.
\end{align}

For any candidate measurement set \(\BeamSet_i^c \subset \BeamSet\), the predicted measurement vector under the updated posterior is
\begin{equation}
    \widehat{\sinogram}_{k+1|k}(\BeamSet_i^c) = \A(\BeamSet_i^c)\,\sampleMean_{k|k}.
\end{equation}
As a motivating example, in the special (typical in practice) case where we scan at an evenly distributed set of beam positions along a single scan angle, the candidate set is
\begin{equation}
    \BeamSet_i^c = \left\{[\theta_i,r_1],[\theta_i,r_2],\ldots,[\theta_i,r_{n_r}]\right\}.
\end{equation}

Folding in the edge-prior covariance $\CoVar^c_{k|k}$ from (\ref{eq: edge-prior covar}) and treating the current posterior as the prior for the next acquisition, $\sample_{k+1|k} \sim \mathcal{N}(\sampleMean_{k|k},\,\CoVar^c_{k|k})$, the posterior predictive joint distribution over a candidate set $\BeamSet_i^c$ is
\begin{equation}
\begin{aligned}
    &\left[\begin{array}{c}
        \sample_{k+1|k}(\BeamSet_i^c) \\
        \sinogram_{k+1|k}(\BeamSet_i^c)
    \end{array}\right]
    \sim
    \mathcal{N}\Big(
        \left[
            \begin{array}{c}
                \sampleMean_{k|k} \\
                 \A(\BeamSet_i^c) \sampleMean_{k|k}
            \end{array}
        \right], \\
    &\setlength{\arraycolsep}{2pt}
        \left[
            \begin{array}{c c}
                \CoVar^c_{k|k} & \CoVar^c_{k|k} \A(\BeamSet_i^c)^\top \\
                \A(\BeamSet_i^c) \CoVar^c_{k|k} & \A(\BeamSet_i^c) \CoVar^c_{k|k} \A(\BeamSet_i^c)^\top + \CoVar^c_e
            \end{array}
        \right]
    \Big).
\end{aligned}
\end{equation}
Marginalizing this joint distribution yields the closed-form predictive posterior covariance $\CoVar^c_{k+1\mid k}(\BeamSet_i^c)$ stated in (\ref{eq: posterior covar candidate}).

\end{document}